\documentclass[11pt]{article}

\usepackage[letterpaper,margin=1in]{geometry}
\usepackage{times}
\usepackage{microtype}
\usepackage{authblk}
\usepackage{natbib}
\setcitestyle{authoryear,round,citesep={;},aysep={,},yysep={;}}

\usepackage{amsmath,amsfonts,bm}

\def\eqref#1{equation~\ref{#1}}

\def\1{\bm{1}}

\DeclareMathAlphabet{\mathsfit}{\encodingdefault}{\sfdefault}{m}{sl}
\SetMathAlphabet{\mathsfit}{bold}{\encodingdefault}{\sfdefault}{bx}{n}

\newcommand{\R}{\mathbb{R}}

\usepackage{hyperref}
\hypersetup{
  hidelinks,
  pdftitle={A Constrained Kuramoto Gradient-Flow System Can Perform High-Accuracy Finite-Time Inference},
  pdfauthor={Yi Cheng and Zongli Lin}
}
\usepackage{url}
\usepackage{graphicx}
\usepackage{booktabs}
\usepackage{tabularx}
\usepackage{array}
\newcolumntype{Y}{>{\centering\arraybackslash}X}

\newsavebox{\arxivtablebox}
\newenvironment{adjustbox}[1]{%
  \begin{lrbox}{\arxivtablebox}%
}{%
  \end{lrbox}%
  \ifdim\wd\arxivtablebox>\textwidth
    \resizebox{\textwidth}{!}{\usebox{\arxivtablebox}}%
  \else
    \usebox{\arxivtablebox}%
  \fi
}

\title{A Constrained Kuramoto Gradient-Flow System Can Perform High-Accuracy Finite-Time Inference}

\renewcommand{\Authands}{ and }
\renewcommand{\Affilfont}{\normalfont\small}
\author{Yi Cheng}
\author{Zongli Lin\thanks{Corresponding author.}}
\affil{The Charles L. Brown Department of Electrical and Computer Engineering,\\
University of Virginia, Charlottesville, VA 22904, U.S.A.\\
\texttt{zss7gw@virginia.edu} \qquad \texttt{zl5y@virginia.edu}}
\date{}

\newcommand{\TT}{\mathbb{T}}
\newcommand{\T}{{\rm T}}
\newcommand{\HH}{{\rm H}}
\newcommand{\OO}{{\rm O}}
\newcommand{\Z}{\mathbb{Z}}
\begin{document}

\maketitle

\begin{abstract}
A central question in physical inference is whether strongly constrained dynamical systems can realize accurate input--output maps through their own finite-time evolution. We study this question in Kuramoto phase networks, whose deterministic dynamics form an input-conditioned gradient flow and whose predictions are read directly from output oscillators. As a constructive training approach, we develop a two-stage teacher--student procedure. A neural teacher is first converted into an explicit phase trajectory whose terminal oscillator activations reproduce the teacher outputs, and the Kuramoto parameters are trained by matching the student vector field along this prescribed path. Because accurate teacher-forced path matching does not ensure accurate autonomous inference, we then differentiate through the autonomous finite-time rollout and directly align its terminal output with the neural target. The resulting oscillator system, with $74$ oscillators, reaches mean test accuracies of $96.711\%$ on MNIST and $86.399\%$ on Fashion-MNIST. This capability persists across neural-teacher architectures, matched system sizes, thermal perturbations, and integration-grid refinement. Together, these results provide a constructive demonstration that a strongly constrained, small-sized Kuramoto gradient-flow system can be trained for high-accuracy finite-time inference through a direct oscillator readout.
\end{abstract}

\section{Introduction}
\label{intro}

Physical neural networks and analog computing offer an alternative route to
artificial intelligence in which computation is performed directly by the
evolution of physical degrees of freedom rather than by emulating every
operation digitally
\citep{momeni2025training,kalinin2025analog}. Such systems can exploit
intrinsic nonlinear dynamics and parallel state evolution for tasks including
classification, signal processing, and combinatorial optimization
\citep{markovic2020physics,wright2022deep,
romera2018vowel,zhou2019self,moy20221}. A central question for physical
inference, however, is how much task-relevant computation can be realized when
the dynamical state, interaction structure, and readout are themselves
constrained by the underlying physical substrate. In the finite-time setting, this question becomes whether a compact physical dynamical system can be trained so that its natural transient evolution implements an accurate input--output map, rather than relying on flexible digital computation surrounding the dynamics. Constructive training procedures that can access such behavior under
physical constraints are therefore essential.

The Kuramoto model provides a particularly stringent setting in which to study
this question. It is a canonical model of synchronization in coupled
oscillators, with longstanding applications in neuroscience and other complex
systems
\citep{kuramoto2005self,acebron2005kuramoto,rodrigues2016kuramoto,
breakspear2010generative,cabral2011role}, while also having a direct connection
to hardware: under weak coupling, self-sustaining oscillator networks admit
phase-reduced dynamics of Kuramoto type
\citep{csaba2020coupled,wang2021solving,cheng2026delayed}. Related oscillator
platforms based on spin-torque devices, electronic oscillators, lasers, and
other technologies have been explored for neuromorphic inference and
optimization
\citep{torrejon2017neuromorphic,nikonov2020convolution,
cilasun2025coupled}. In the architecture considered here, the deterministic
Kuramoto dynamics form an input-conditioned gradient flow with symmetric
pairwise oscillator couplings; inputs act directly only on the hidden
subsystem, and classification is read directly from output-oscillator
states at a prescribed finite time. These restrictions make the Kuramoto
network not only a hardware-motivated substrate, but also a useful test of how
much inference can be realized within a compact, energy-descending dynamical
system.

Trainable oscillator networks have been studied using differentiable simulation,
equilibrium-based learning, and related approaches
\citep{rudner2024design,wang2024training,gower2025train,rageau2025training}.
Most closely related to our work,  \cite{whitelam2026training} developed a finite-time
teacher--student framework for classical nonlinear thermodynamic dynamics,
building on a teacher--student idea that traces back to autonomous quantum
thermal machines introduced in
\citep{lipka2024thermodynamic}.
This provides a natural starting point for asking whether a compact, strongly
constrained oscillator system can be trained to realize accurate finite-time
inference under its own autonomous dynamics.

To address this question, we develop a two-stage teacher--student procedure for
finite-time Kuramoto dynamics. We first convert a neural teacher into an
explicit phase trajectory whose terminal oscillator activations reproduce the
teacher outputs, and train the Kuramoto vector field along this prescribed
path. We find, however, that accurate teacher-forced path matching does not
ensure accurate autonomous inference: during deployment the student evolves
on states generated by its own dynamics, so local vector-field errors can
accumulate and shift the terminal output. We therefore introduce a second
stage that differentiates through the autonomous finite-time rollout and
directly optimizes the terminal physical readout. This stage trains the
input--output behavior of the actual autonomous dynamical system rather than
only its vector field along a teacher-supplied trajectory.


Our main contributions are:
\begin{itemize}
    \item We provide a constructive demonstration that a strongly constrained
    Kuramoto gradient-flow system with 74 evolving variables can be trained
    for high-accuracy finite-time classification using a direct oscillator
    readout, reaching $96.711\%$ on MNIST and $86.399\%$ on Fashion-MNIST at
    the principal operating point.

    \item We develop a Kuramoto-specific two-stage teacher--student training
    procedure. An explicit phase trajectory first provides path-based
    supervision of the physical vector field, after which autonomous endpoint
    optimization directly trains the finite-time input--output map generated
    by the student's own dynamics.

    \item We show that this capability is robust across neural-teacher
    architectures, matched teacher--student system sizes, thermal
    perturbations, and numerical-resolution changes. Additional trajectory
    analysis shows that successful output inference can coexist with
    substantial reorganization of the hidden dynamics.
\end{itemize}

\section{Background and Related Work}
\label{relt-work}

Physical dynamical systems have been trained through several approaches,
including equilibrium propagation, physics-aware training, and
backpropagation through time
\citep{scellier2016equilibrium,wright2022deep,rudner2024design}.
For Kuramoto and related oscillator networks, supervised learning has been
demonstrated using differentiable dynamical models and equilibrium propagation
\citep{rudner2024design,wang2024training,gower2025train,
rageau2025training}. These works establish that constrained physical
dynamics can support supervised computation under a range of training
paradigms. The present work builds most directly on the finite-time
teacher--student framework developed by Whitelam
\citep{whitelam2026training}.

In Whitelam's framework, a conventional neural network first serves as a
digital teacher. Its hidden and output activations are used to construct a
prescribed finite-time trajectory for a thermodynamic student governed by
overdamped Langevin dynamics. Rather than supervising only the terminal state,
the student parameters are optimized so that the prescribed teacher trajectory
has high probability under the student dynamics. Under an Euler--Maruyama
discretization, this trajectory likelihood reduces to a sum of local
transition costs, equivalently training the student drift field to reproduce
the teacher increments along the prescribed path. Training is performed
digitally, after which inference is
carried out by the student's own finite-time dynamics.

Our work adopts this teacher--student principle but applies it to a compact
Kuramoto system with periodic phase variables, symmetric pairwise couplings,
gradient-flow dynamics, and a direct oscillator readout. This requires a
different physical state representation, interaction structure, supervisory
trajectory, and readout, which we develop in Sec.~\ref{meth}. We further find
that path supervision alone does not reliably translate into accurate
autonomous finite-time behavior. This motivates an autonomous endpoint stage
that, together with path-based initialization, provides a constructive route
for training the constrained Kuramoto dynamics to realize accurate finite-time
inference.

\section{Methodology}
\label{meth}

\subsection{Thermodynamic Kuramoto Network}
\label{kuratonet}

We consider an input-conditioned network of $N=H+O$ oscillators, with $H$
hidden oscillators and $O$ output oscillators. The state of oscillator $i$ is
its phase $\theta_i(t)\in\TT$, where
$\TT:=\R/(2\pi\Z)$. Given an input $u\in\R^D$, the deterministic dynamics are
\begin{equation}
\dot{\theta}_i(t)
=
-\mu
\left(
b_i(u)\sin\theta_i
+
\sum_{j=1}^{N}
J_{ij}\sin(\theta_i-\theta_j)
\right),
\qquad
i=1,2,\ldots,N,
\label{eq:kuramoto_student}
\end{equation}
where $\mu>0$ sets the dynamical timescale, $b_i(u)$ is an input-dependent local field, $u$ is the input vector, and
$J_{ij}=J_{ji}$ is a symmetric coupling with $J_{ii}=0$.

\textbf{Network architecture.}
The input vector $u$ act only on the hidden oscillators, while output
oscillators indirectly receive input information through hidden--output coupling. With
the hidden variables ordered before the output variables, the local-field
vector and coupling matrix are
\begin{equation*}
b(u)
=
\begin{bmatrix}
W^\HH u+c^\HH\\
c^\OO
\end{bmatrix},
\qquad
J
=
\begin{bmatrix}
J^{\HH\HH} & J^{\HH\OO}\\
(J^{\HH\OO})^\T & 0
\end{bmatrix},
\end{equation*}
where
\[
W^\HH\in\R^{H\times D},
\qquad
c^\HH\in\R^H,
\qquad
c^\OO\in\R^O,
\qquad
J^{\HH\OO}\in\R^{H\times O},
\]
and
\[
J^{\HH\HH}=(J^{\HH\HH})^\T,
\qquad
\operatorname{diag}(J^{\HH\HH})=0,
\]
where $\HH$ and $\OO$ are superscripts denoting ``hidden layer'' and  ``output layer'', respectively. Thus, there are no direct input-to-output weights, no output--output
couplings, and no self-couplings. The trainable physical parameters are chosen as 
\begin{equation*}
\Theta
=
\left\{
W^\HH,c^\HH,c^\OO,J^{\HH\HH},J^{\HH\OO}
\right\}.
\end{equation*}
These choices define a deliberately restricted dynamical hypothesis class, where input dependence enters only through hidden local fields and oscillator--oscillator interactions are symmetric.
The deterministic dynamics in Eq.~\ref{eq:kuramoto_student} are gradient
dynamics of the input-conditioned energy
\begin{equation*}
E_\Theta(\theta;u)
=
-\sum_{i=1}^{N} b_i(u)\cos\theta_i
-\sum_{i<j}J_{ij}\cos(\theta_i-\theta_j),
\end{equation*}
since
\[
\dot{\theta}_i
=
-\mu\frac{\partial E_\Theta}{\partial\theta_i}.
\]
Consequently,
\[
\frac{{\rm d}E_\Theta}{{\rm d}t}
=
-\mu\sum_{i=1}^{N}
\left(
\frac{\partial E_\Theta}{\partial\theta_i}
\right)^2
\leq 0.
\]
Thus, despite being used for finite-time computation, the deterministic student remains an energy-descending gradient-flow system rather than a generic recurrent dynamical model. A stochastic thermodynamic extension is obtained by adding independent thermal
fluctuations,
\begin{equation*}
{\rm d}\theta_i
=
-\mu
\frac{\partial E_\Theta(\theta;u)}{\partial\theta_i}
\,{\rm d}t
+
\sqrt{2\mu k_{\rm B}T}\,{\rm d}W_i,
\end{equation*}
where $W_i$ are independent Wiener processes. In the principal training and
inference experiments, we set $k_{\rm B}T=0$, while stochastic inference is
examined separately at $k_{\rm B}T>0$.

\textbf{Initialization and finite-time readout.}
All oscillators are initialized at
\[
\theta_i(0)=\frac{\pi}{2},
\]
and we associate each oscillator with the bounded activation
\[
x_i(t)=\cos\theta_i(t),
\]
so that $x_i(0)=0$. Classification is performed from the output activations at
a prescribed observation time $t_{\rm f}$,
\begin{equation}
\widehat y
=
\arg\max_{o\in\{1,\ldots,O\}}
\cos\theta_o^\OO(t_{\rm f}).
\label{eq:kuramoto_readout}
\end{equation}
The network therefore performs finite-time dynamical inference rather than
waiting for convergence to equilibrium. The readout process shows that the classification of our model is obtained directly from the finite-time physical state without a trainable post-dynamical readout.


\subsection{Stage I: Teacher--Student Path Training}

\textbf{Neural teacher and transferred targets.}
We first train a conventional feedforward neural network, as a teacher, on the target
classification task and then keep its activations fixed. For each input
$u\in\R^D$, the transferred hidden and output activations are collected as
\begin{equation*}
a^{\rm TE}(u)
=
\begin{bmatrix}
a^{\rm TE,\HH}(u)\\
a^{\rm TE,\OO}(u)
\end{bmatrix}
\in[-1,1]^N,
\qquad
N=H+O,
\end{equation*}
where $\rm TE$ is the superscript denoting ``teacher'', the $H$ hidden activations define targets for the hidden oscillators, and
the $O$ output activations define the desired physical readout. The teacher is
frozen before training the Kuramoto student. Neural-teacher architectures,
training, and target construction are detailed in
Appendix~\ref{app:neural_teacher_training}.

\textbf{Explicit teacher trajectory.}
The neural teacher provides real-valued activations, whereas the Kuramoto
student evolves in phase space. We connect the two representations through the
physical activation $x_i=\cos\theta_i$. For a machine-safe transferred target
$\widetilde a_i^{\rm TE}\in(-1,1)$, the terminal teacher phase is chosen on
the principal branch as
\begin{equation*}
\theta_i^{\rm TE}(t_{\rm f})
=
\arccos\widetilde a_i^{\rm TE},
\end{equation*}
with the common initial condition
$\theta_i^{\rm TE}(0)=\pi/2$.

To construct the complete finite-time path, we use an uncoupled auxiliary
phase dynamics,
\begin{equation*}
\dot{\theta}_i^{\rm TE}
=
-\mu\beta_i^{\rm TE}\sin\theta_i^{\rm TE},
\qquad
\beta_i^{\rm TE}
=
\frac{
\operatorname{artanh}(\widetilde a_i^{\rm TE})
}{
\mu t_{\rm f}
}.
\end{equation*}
This yields the explicit activation and phase trajectories
\begin{equation*}
x_i^{\rm TE}(t)
=
\tanh
\left(
\frac{t}{t_{\rm f}}
\operatorname{artanh}
\left(
\widetilde a_i^{\rm TE}
\right)
\right),
\qquad
\theta_i^{\rm TE}(t)
=
\arccos x_i^{\rm TE}(t),
\end{equation*}
which satisfy
\[
x_i^{\rm TE}(0)=0,
\qquad
x_i^{\rm TE}(t_{\rm f})=\widetilde a_i^{\rm TE}.
\]
Thus, the transferred teacher neural activation $\widetilde a_i^{\rm TE}$ determines both the prescribed
finite-time endpoint and the complete supervisory phase trajectory. The full
derivation, principal-branch construction, and numerical treatment of targets
at $\pm1$ are provided in
Appendix~\ref{app:stage1_derivation}.

\textbf{Teacher-forced path objective.}
Let $f_\Theta(\theta,u)$ denote the deterministic Kuramoto vector field in
Eq.~\ref{eq:kuramoto_student}. For a discretized teacher trajectory
$\{\theta^{\rm TE}(t_k)\}_{k=0}^{K-1}$, with
$t_k=k\Delta t$ and $K\Delta t=t_{\rm f}$, the phase-increment residual for
training example $n$ is
\begin{equation*}
r_{n,k}
=
\theta_n^{\rm TE}(t_{k+1})
-
\theta_n^{\rm TE}(t_k)
-
\Delta t\,
f_\Theta
\left(
\theta_n^{\rm TE}(t_k),u_n
\right).
\end{equation*}
For a mini-batch of $B$ examples, Stage I minimizes
\begin{equation}
\mathcal L_{\rm path}
=
\frac{1}{BKN}
\sum_{n=1}^{B}
\sum_{k=0}^{K-1}
\|r_{n,k}\|_2^2.
\label{Lpath}
\end{equation}
The objective therefore constrains the student vector field across all
physical nodes and time intervals along the prescribed teacher trajectory.
Importantly, the vector field is evaluated on teacher-supplied states rather
than on states generated recursively by the student. 

All trainable physical parameters are initialized exactly at zero and
optimized jointly under $\mathcal L_{\rm path}$. The complete Kuramoto
training configuration, including parameter constraints, numerical
integration, initialization, and optimization settings, is given in
Appendix~\ref{app:kuramoto_training}.

The path objective also retains the thermodynamic likelihood interpretation of
teacher--student training. Specifically, under the stochastic Kuramoto extension, the local
Euler--Maruyama negative log-likelihood of the prescribed trajectory is,
up to a parameter-independent scale and additive constant, proportional to
$\mathcal L_{\rm path}$. The full derivation
and its periodic phase-space interpretation are given in
Appendix~\ref{app:stage1_derivation}.

Stage I constrains the student vector field only on the prescribed teacher
path. During autonomous inference, however, each subsequent state is generated
by the student itself. Local vector-field errors can therefore move the system
away from the states on which it was trained and accumulate over the
finite-time rollout. This teacher-forcing mismatch motivates the autonomous
endpoint fine-tuning introduced next.

\subsection{Stage II: Autonomous Endpoint Fine-Tuning}
To address this teacher-forcing mismatch, Stage II directly supervises the
output produced by an autonomous student rollout. For input $u_n$, the student
starts from the common initial phase
$\theta_n^{\rm S}(0)=(\pi/2,\pi/2,\ldots,\pi/2)$ and evolves recursively according
to
\begin{equation}
\theta_{n,k+1}^{\rm S}
=
\theta_{n,k}^{\rm S}
+
\Delta t\,
f_\Theta
\left(
\theta_{n,k}^{\rm S},u_n
\right),
\qquad
k=0,1,\ldots,K-1.
\label{discrettheta}
\end{equation}
where $\rm S$ is the superscript denoting ``student''. Unlike Stage I, every state on the right-hand side of
Eq.~\ref{discrettheta} is generated by the student rather than supplied by
the teacher.

At the terminal time, the physical output activation is
$\cos\theta_{n,o}^{\rm S,\OO}(t_{\rm f})$. We therefore define the autonomous
endpoint loss
\begin{equation*}
\mathcal L_{\rm end}
=
\frac{1}{BO}
\sum_{n=1}^{B}
\sum_{o=1}^{O}
\left(
\cos\theta_{n,o}^{\rm S,\OO}(t_{\rm f})
-
a_{n,o}^{\rm TE,\OO}
\right)^2.
\end{equation*}
Because the teacher output and the physical readout are represented in the
same bounded activation space, endpoint supervision can be applied directly
without introducing a phase-space distance or selecting between equivalent
phase representations.

Starting from the validation-selected Stage-I checkpoint, Stage II optimizes
\begin{equation*}
\mathcal L_{\rm Stage\,II}
=
\mathcal L_{\rm path}
+
\lambda_{\rm end}\mathcal L_{\rm end},
\end{equation*}
for the principal protocol. The path term retains its teacher-forced
definition from Eq.~\ref{Lpath}, whereas the endpoint term is evaluated
after the complete autonomous rollout. The two terms therefore provide
complementary supervision: $\mathcal L_{\rm path}$ constrains the local vector
field along the prescribed teacher path, while $\mathcal L_{\rm end}$ aligns
the terminal output produced by recursively generated student dynamics.

All physical parameters in $\Theta$ remain trainable during Stage II.
Gradients of $\mathcal L_{\rm end}$ are propagated through the complete
sequence in Eq.~\ref{discrettheta}, so the optimization accounts for how
parameter changes affect the terminal physical output through the full
finite-time evolution. Complete Stage-II optimization settings and checkpoint-selection rules are
provided in Appendix~\ref{app:kuramoto_training} and
Appendix~\ref{app:checkpoint_selection}, respectively.

After training, the teacher and both supervision objectives are removed;
inference uses the fixed learned parameters $\Theta^\star$ and the autonomous
finite-time readout of Eq.~\ref{eq:kuramoto_readout}.

\section{Experiments}
\label{others}

\paragraph{Experimental Setup and Evaluation Protocol.}
We evaluate the proposed method on MNIST and Fashion-MNIST. The principal
experiments use a $784$-$32$-$16$-$16$-$10$ neural teacher, whose transferred
activations are matched to a Kuramoto student with $H=64$ hidden and $O=10$
output oscillators ($N=74$). Main-text comparisons focus on autonomous test
top-1 accuracy, neural-teacher agreement, and output endpoint RMSE, with all
headline results obtained from validation-selected checkpoints and
deterministic finite-time inference. Complete dataset preprocessing,
neural-teacher training, Kuramoto-training settings, checkpoint-selection
rules, and metric definitions are provided in
Appendices~\ref{app:data_preprocessing}--\ref{app:evaluation_metrics}.

\paragraph{High-Accuracy Finite-Time Inference at the Principal Operating Point.}
We first ask whether the constrained Kuramoto system can be trained for
accurate autonomous finite-time inference. At the principal operating point,
the system contains 64 hidden and 10 output oscillators, for a total of 74
evolving variables, with classification read directly from the output
oscillators at the prescribed observation time. We evaluate ten paired
training-order runs per dataset. PATH is the validation-selected Stage-I
model; PATH-CONT continues optimizing $\mathcal L_{\rm path}$ from the same
checkpoint, whereas TWO-STAGE-END adds autonomous endpoint supervision through
$\mathcal L_{\rm path}+\mathcal L_{\rm end}$.

\begin{figure}[htbp!]
    \centering
    \includegraphics[width=0.9\linewidth]
    {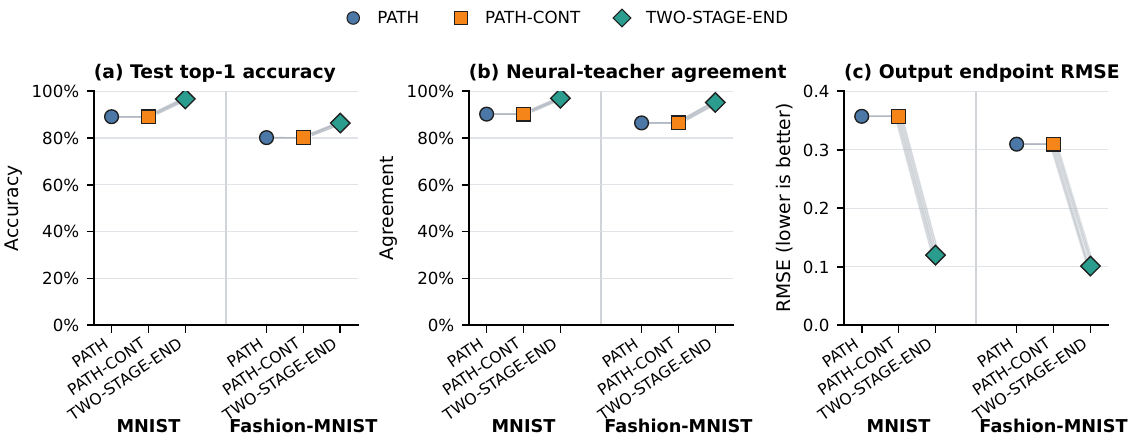}
    \caption{\textbf{High-accuracy finite-time inference in the
    74-variable Kuramoto system.}
    Test top-1 accuracy, neural-teacher agreement, and output endpoint RMSE on
    MNIST and Fashion-MNIST over ten paired training-order runs. Small points
    within the circles and the squares show individual runs, large markers show means, and error bars denote one
    sample standard deviation. PATH is the selected Stage-I model; PATH-CONT
    continues path-only optimization; TWO-STAGE-END adds autonomous endpoint
    supervision. Lower RMSE is better.}
    \label{fig:main_performance}
\end{figure}

Figure~\ref{fig:main_performance} shows that the resulting 74-variable system
reaches mean test top-1 accuracies of $96.711\%$ on MNIST and $86.399\%$ on
Fashion-MNIST. After path training alone, PATH reaches $89.112\%$ and
$80.172\%$, respectively. The improvement is positive in all ten paired runs
on both datasets, while PATH-CONT remains essentially unchanged, ruling out
additional path-only optimization as the explanation. TWO-STAGE-END also
increases neural-teacher agreement and substantially reduces output endpoint
RMSE. Detailed results and paired analyses are provided in
Appendices~\ref{app:multiseed_main_results} and
\ref{app:paired_improvements}.

Thus, path matching alone is insufficient to obtain the high-accuracy
finite-time inference of the fully trained autonomous system. Path training
provides a useful dynamical initialization, while endpoint supervision
optimizes behavior along states generated by the student itself. Together,
the two stages provide a constructive route to accurate inference in the
constrained Kuramoto model. Trajectory analysis further shows that this does
not require preservation of the complete hidden teacher trajectory
(Appendix~\ref{AppendixE}).

\paragraph{Generality Across Neural-Teacher Representations.}
We next ask whether the high-accuracy finite-time inference obtained by the
74-variable Kuramoto system depends on a particular neural-teacher
representation. On Fashion-MNIST, we evaluate 15 feedforward teachers with
different depths and hidden-width allocations while fixing the total number of
transferred hidden activations at $H=64$. The Kuramoto student therefore
remains fixed at $64$ hidden and $10$ output oscillators throughout the sweep;
only the teacher representation changes. The complete architecture set and
protocol are given in
Appendix~\ref{app:fifteen_teacher_architectures}.

\begin{figure}[htbp!]
    \centering
    \includegraphics[width=0.9\linewidth]
    {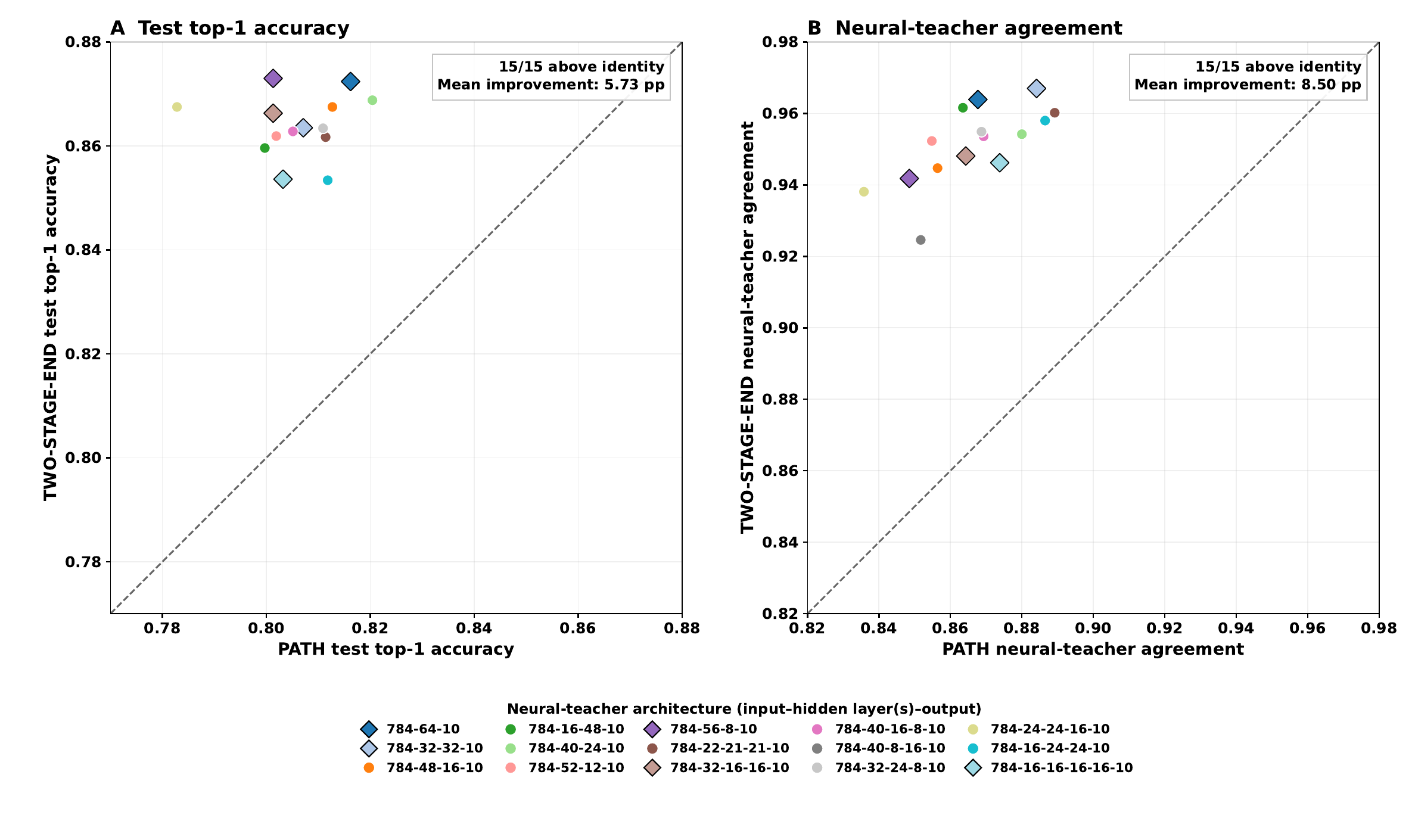}
    \caption{\textbf{Generality across neural-teacher representations.}
    PATH and TWO-STAGE-END performance across 15 Fashion-MNIST neural teachers
    with different hidden-layer depths and width allocations but fixed total
    transferred hidden width $H=64$. The Kuramoto system is fixed at 64 hidden
    and 10 output oscillators. Each point denotes one frozen teacher
    architecture; the dashed identity line indicates equal performance.
    TWO-STAGE-END improves autonomous top-1 accuracy and neural-teacher
    agreement for all 15 configurations.}
    \label{fig:teacher_architecture_robustness}
\end{figure}

As shown in Fig.~\ref{fig:teacher_architecture_robustness}, the two-stage
procedure improves autonomous top-1 accuracy for all $15/15$ teacher
architectures, with a mean gain of $5.73$ percentage points over PATH.
Neural-teacher agreement also increases and output endpoint RMSE decreases in
every case. Thus, the high-accuracy autonomous regime is not specific to the
$784$-$32$-$16$-$16$-$10$ teacher used at the principal operating point.
Complete results are reported in
Appendix~\ref{app:teacher_architecture_results}. This sweep is an auxiliary
single-run analysis across distinct teacher architectures; training-order
variability at the principal operating point is quantified separately by the
preceding paired-run experiment.

\paragraph{Roles of Path and Endpoint Supervision.}
To understand how the two-stage procedure reaches accurate autonomous
finite-time inference, we use an auxiliary six-method ablation to separate
the roles of additional optimization, path supervision, and endpoint
supervision (Appendix~\ref{app:full_protocol_ablation}). PATH-CONT controls
for additional path-only optimization, while matched zero-initialized and
PATH-pretrained variants probe the roles of the two objectives before and
after path-based initialization.

\begin{figure}[htbp!]
    \centering
    \includegraphics[width=0.9\linewidth]
    {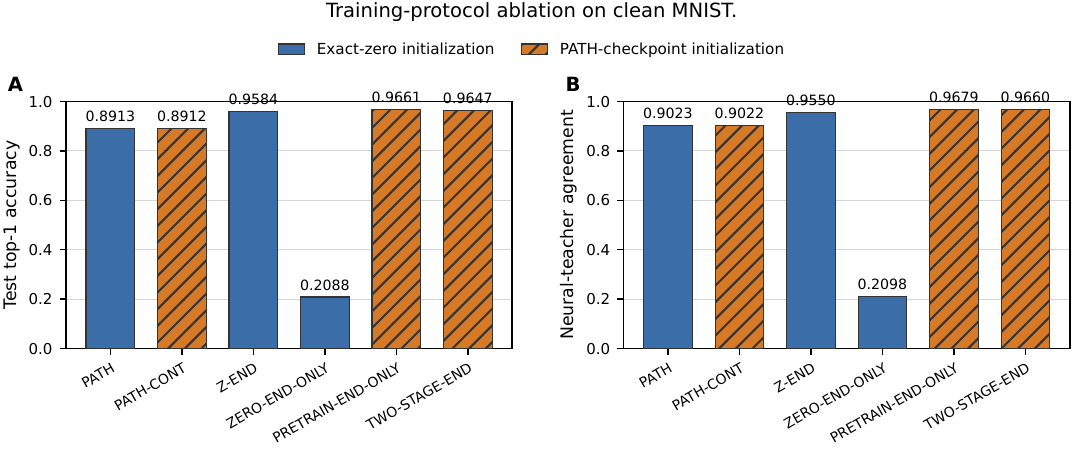}
    \caption{\textbf{Roles of path and endpoint supervision in the training
    procedure.}
    Test top-1 accuracy and neural-teacher agreement for six combinations of
    initialization and training objective. PATH-CONT controls for additional
    path-only optimization; ZERO-END-ONLY and Z-END compare direct training
    from exact zero without and with path supervision; and
    PRETRAIN-END-ONLY and TWO-STAGE-END compare endpoint continuation without
    and with continued path supervision.}
    \label{fig:protocol_ablation}
\end{figure}

As summarized in Fig.~\ref{fig:protocol_ablation}, PATH-CONT remains
essentially unchanged from PATH. When training begins from exact zero,
endpoint-only optimization reaches only $20.88\%$ top-1 accuracy on MNIST,
whereas joint path--endpoint training reaches $95.84\%$. After PATH
pretraining, however, endpoint-only and joint continuation perform nearly
identically. Under the tested exact-zero initialization and optimization
protocol, these results indicate distinct roles for the two stages: path
supervision establishes a useful dynamical initialization, while the major
post-pretraining improvement is driven by supervision of the autonomous
finite-time output. Complete results and endpoint-weight sensitivity are
given in Appendices~\ref{app:full_protocol_ablation} and
\ref{app:endpoint_weight_sensitivity}.

\begin{figure}[htbp!]
    \centering
    \includegraphics[width=0.8\textwidth]
    {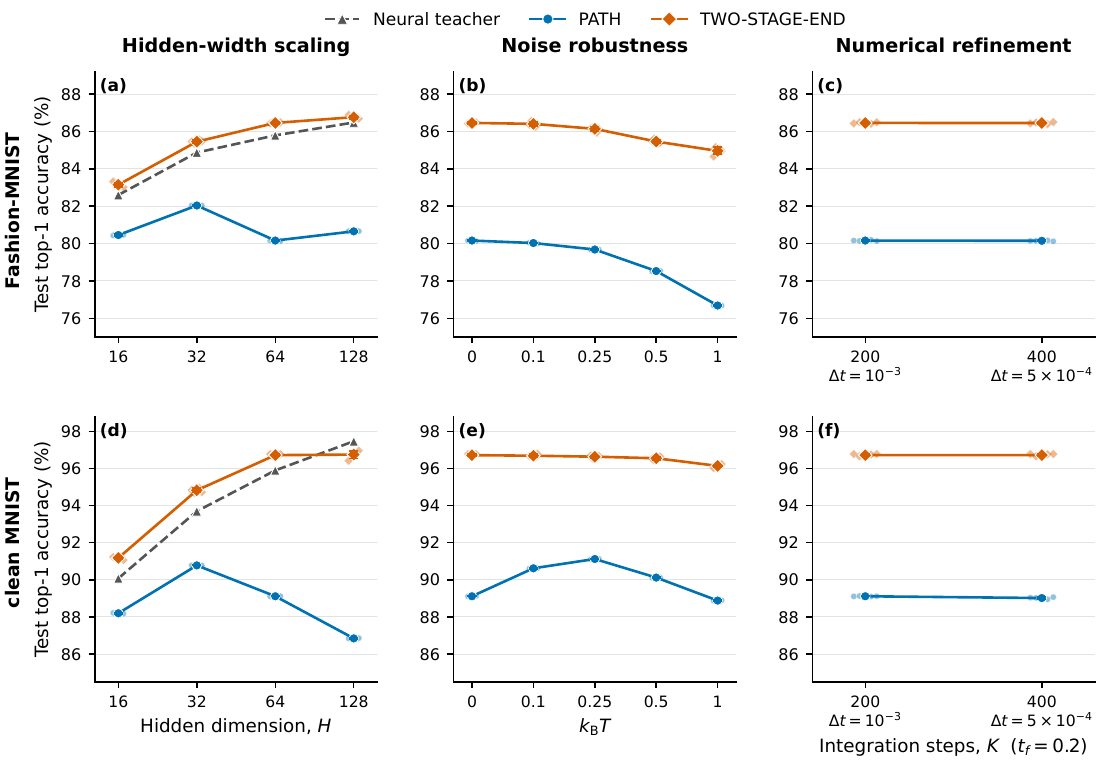}
    \caption{\textbf{Scaling and robustness of finite-time inference.}
    Fashion-MNIST is shown in the top row and MNIST in the bottom row.
    \textbf{(a,d)} Test top-1 accuracy under matched teacher--student scaling
    over $H=16,32,64,$ and $128$.
    \textbf{(b,e)} Test top-1 accuracy under thermal noise using
    $M_{\rm ens}=10$ continuous-output ensemble inference.
    \textbf{(c,f)} Deterministic evaluation of the same frozen $H=64$
    checkpoints using the nominal $K=200$ and refined $K=400$ integration
    grids at fixed $t_{\rm f}=0.2$.
    Student markers denote means over five training-order runs, error bars
    denote one sample standard deviation, and faint markers show individual
    runs.}
    \label{fig:scaling_noise_numerical}
\end{figure}

\paragraph{Scaling and Robustness of Finite-Time Inference.}
We now test whether accurate autonomous finite-time inference persists across
system size, thermal perturbations, and numerical resolution.
Figure~\ref{fig:scaling_noise_numerical} summarizes these three tests.

For matched teacher--student scaling
(Figs.~\ref{fig:scaling_noise_numerical}(a,d)), we vary the hidden dimension
over $H\in\{16,32,64,128\}$ while preserving a common three-hidden-layer
teacher family. Neural-teacher accuracy increases with $H$ on both datasets,
and the fully trained Kuramoto systems follow the same qualitative trend.
In contrast, PATH reaches its highest accuracy at $H=32$ and does not improve
systematically at larger widths. TWO-STAGE-END outperforms PATH at every
tested width, while its incremental benefit from increasing system size
diminishes at the largest widths: from $H=64$ to $H=128$, mean accuracy
changes by only $0.030$ percentage points on MNIST and $0.308$ on
Fashion-MNIST. Thus, the principal $H=64$ result is part of a broader matched
family rather than an isolated operating point. Complete results and the
matched scaling construction are given in
Appendices~\ref{app:matched_scaling_protocol} and
\ref{app:hidden_width_scaling_results}.

For stochastic inference
(Figs.~\ref{fig:scaling_noise_numerical}(b,e)), we add thermal noise and
average the continuous terminal outputs of $M_{\rm ens}=10$ trajectories
before classification. Across
$k_{\rm B}T\in\{0,0.1,0.25,0.5,1\}$, the fully trained system remains more
accurate than PATH at every nonzero temperature for all five paired
training-order runs on both datasets. At $k_{\rm B}T=1$, ensemble accuracy
remains $96.142\%$ on MNIST and $84.958\%$ on Fashion-MNIST, compared with
$88.884\%$ and $76.684\%$ for PATH. Continuous-output averaging further
mitigates degradation of individual noisy trajectories. The complete
stochastic protocol and temperature-dependent results are reported in
Appendices~\ref{app:stochastic_protocol} and
\ref{app:temperature_robustness}.

Finally, Figs.~\ref{fig:scaling_noise_numerical}(c,f) test numerical
sensitivity by re-evaluating the same frozen $H=64$ checkpoints with
$K=200$ and $K=400$ explicit-Euler steps at fixed $t_{\rm f}=0.2$, without
retraining or checkpoint reselection. The largest mean top-1 change among the
four dataset--method groups is only $0.096$ percentage points, while the
performance separation between PATH and TWO-STAGE-END is preserved.
Continuous terminal outputs are similarly stable and all audited evaluations
remain finite. The reported finite-time inference is therefore stable under
the tested twofold integration-grid refinement; complete numerical-safety
results are provided in Appendix~\ref{AppendixH}.

\paragraph{Cross-Platform Comparison.}
We compare the proposed method with recent trainable oscillator, Ising-machine,
and quantum-annealing approaches
\citep{rageau2025training,gower2025train,wang2024training,
laydevant2024training,fan2025equilibrium}.
The principal Kuramoto system uses $74$ logical dynamical variables and
$2{,}656$ symmetric couplers, giving the smallest complete logical state count
among these closely related operating points. We also place this state scale
in broader descriptive context with representative energy-based, equilibrium,
neural-ODE, reservoir, analog-computing, and spiking classifiers. We note that these
model-level comparisons are not hardware-footprint or controlled efficiency
rankings. See
Appendices~\ref{app:cross_platform_scope}--
\ref{app:broader_dynamical_context} for detailed resource, protocol comparisons, and broader-context.

\section{Conclusion}

We showed that a compact, strongly constrained Kuramoto gradient-flow system
can be trained for high-accuracy finite-time inference through a direct
oscillator readout. At the principal operating point, the model contains only
74 evolving variables. A two-stage teacher--student procedure provides a
constructive route to this regime: path supervision first establishes useful
dynamics, while autonomous endpoint supervision subsequently optimizes the
finite-time input--output behavior generated by the student's own evolution.
The resulting capability persists across neural-teacher representations,
matched system sizes, thermal perturbations, and integration-grid refinement.

These results suggest that accurate physical inference can be realized within
a small dynamical state space despite strong constraints on the underlying
dynamics, connectivity, and readout. The present study remains a system-level
numerical demonstration; experimental realization will require incorporating
programmable hardware constraints and device nonidealities. Extending the
approach to more general phase-native objectives and physical oscillator
implementations is an important direction for future work.


\newpage

\bibliography{references}

@article{kalinin2025analog,
  title={Analog optical computer for AI inference and combinatorial optimization},
  author={Kalinin, Kirill P and Gladrow, Jannes and Chu, Jiaqi and Clegg, James H and Cletheroe, Daniel and Kelly, Douglas J and Rahmani, Babak and Brennan, Grace and Canakci, Burcu and Falck, Fabian and others},
  journal={Nature},
  volume={645},
  number={8080},
  pages={354--361},
  year={2025},
  publisher={Nature Publishing Group UK London}
}

@article{momeni2025training,
  title={Training of physical neural networks},
  author={Momeni, Ali and Rahmani, Babak and Scellier, Benjamin and Wright, Logan G and McMahon, Peter L and Wanjura, Clara C and Li, Yuhang and Skalli, Anas and Berloff, Natalia G and Onodera, Tatsuhiro and others},
  journal={Nature},
  volume={645},
  number={8079},
  pages={53--61},
  year={2025},
  publisher={Nature Publishing Group UK London}
}

@article{markovic2020physics,
  title={Physics for neuromorphic computing},
  author={Markovi{\'c}, Danijela and Mizrahi, Alice and Querlioz, Damien and Grollier, Julie},
  journal={Nature Reviews Physics},
  volume={2},
  number={9},
  pages={499--510},
  year={2020},
  publisher={Nature Publishing Group UK London}
}

@article{wright2022deep,
  title={Deep physical neural networks trained with backpropagation},
  author={Wright, Logan G and Onodera, Tatsuhiro and Stein, Martin M and Wang, Tianyu and Schachter, Darren T and Hu, Zoey and McMahon, Peter L},
  journal={Nature},
  volume={601},
  number={7894},
  pages={549--555},
  year={2022},
  publisher={Nature Publishing Group UK London}
}

@article{romera2018vowel,
  title={Vowel recognition with four coupled spin-torque nano-oscillators},
  author={Romera, Miguel and Talatchian, Philippe and Tsunegi, Sumito and Abreu Araujo, Flavio and Cros, Vincent and Bortolotti, Paolo and Trastoy, Juan and Yakushiji, Kay and Fukushima, Akio and Kubota, Hitoshi and others},
  journal={Nature},
  volume={563},
  number={7730},
  pages={230--234},
  year={2018},
  publisher={Nature Publishing Group UK London}
}

@article{zhou2019self,
  title={Self-learning photonic signal processor with an optical neural network chip},
  author={Zhou, Hailong and Zhao, Yuhe and Wang, Xu and Gao, Dingshan and Dong, Jianji and Zhang, Xinliang},
  journal={arXiv preprint arXiv:1902.07318},
  year={2019}
}

@article{moy20221,
  title={A 1,968-node coupled ring oscillator circuit for combinatorial optimization problem solving},
  author={Moy, William and Ahmed, Ibrahim and Chiu, Po-wei and Moy, John and Sapatnekar, Sachin S and Kim, Chris H},
  journal={Nature Electronics},
  volume={5},
  number={5},
  pages={310--317},
  year={2022},
  publisher={Nature Publishing Group UK London}
}

@inproceedings{kuramoto2005self,
  title={Self-entrainment of a population of coupled non-linear oscillators},
  author={Kuramoto, Yoshiki},
  booktitle={International symposium on mathematical problems in theoretical physics: January 23--29, 1975, kyoto university, kyoto/Japan},
  pages={420--422},
  year={2005},
  organization={Springer}
}

@article{acebron2005kuramoto,
  title={The Kuramoto model: A simple paradigm for synchronization phenomena},
  author={Acebr{\'o}n, Juan A and Bonilla, Luis L and P{\'e}rez Vicente, Conrad J and Ritort, F{\'e}lix and Spigler, Renato},
  journal={Reviews of modern physics},
  volume={77},
  number={1},
  pages={137--185},
  year={2005},
  publisher={APS}
}

@article{rodrigues2016kuramoto,
  title={The Kuramoto model in complex networks},
  author={Rodrigues, Francisco A and Peron, Thomas K DM and Ji, Peng and Kurths, J{\"u}rgen},
  journal={Physics Reports},
  volume={610},
  pages={1--98},
  year={2016},
  publisher={Elsevier}
}

@article{csaba2020coupled,
  title={Coupled oscillators for computing: A review and perspective},
  author={Csaba, Gyorgy and Porod, Wolfgang},
  journal={Applied physics reviews},
  volume={7},
  number={1},
  year={2020},
  publisher={AIP Publishing}
}

@article{wang2021solving,
  title={Solving combinatorial optimisation problems using oscillator based Ising machines},
  author={Wang, Tianshi and Wu, Leon and Nobel, Parth and Roychowdhury, Jaijeet},
  journal={Natural Computing},
  volume={20},
  number={2},
  year={2021}
}

@article{cheng2026delayed,
  title={Delayed Coupling Restores {I}sing Phase Dynamics in Physical Oscillator Networks},
  author={Cheng, Yi and Dai, Liangtao and Stan, Mircea R and Lin, Zongli},
  journal={arXiv preprint arXiv:2607.16634},
  year={2026}
}

@article{torrejon2017neuromorphic,
  title={Neuromorphic computing with nanoscale spintronic oscillators},
  author={Torrejon, Jacob and Riou, Mathieu and Araujo, Flavio Abreu and Tsunegi, Sumito and Khalsa, Guru and Querlioz, Damien and Bortolotti, Paolo and Cros, Vincent and Yakushiji, Kay and Fukushima, Akio and others},
  journal={Nature},
  volume={547},
  number={7664},
  pages={428--431},
  year={2017},
  publisher={Nature Publishing Group UK London}
}

@article{nikonov2020convolution,
  title={Convolution inference via synchronization of a coupled {C}{M}{O}{S} oscillator array},
  author={Nikonov, Dmitri E and Kurahashi, Peter and Ayers, James S and Li, Hai and Kamgaing, Telesphor and Dogiamis, Georgios C and Lee, Hyung-Jin and Fan, Yongping and Young, IA},
  journal={IEEE Journal on Exploratory Solid-State Computational Devices and Circuits},
  volume={6},
  number={2},
  pages={170--176},
  year={2020},
  publisher={IEEE}
}

@article{cilasun2025coupled,
  title={A coupled-oscillator-based {I}sing chip for combinatorial optimization},
  author={C{\i}lasun, H{\"u}srev and Moy, William and Zeng, Ziqing and Islam, Tahmida and Lo, Hao and Vanasse, Alex and Tan, Megan and Anees, Mohammad and S, Ramprasath and Kumar, Abhimanyu and others},
  journal={Nature Electronics},
  volume={8},
  number={6},
  pages={537--546},
  year={2025},
  publisher={Nature Publishing Group UK London}
}

@article{rudner2024design,
  title={Design of oscillatory neural networks by machine learning},
  author={Rudner, Tamas and Porod, Wolfgang and Csaba, Gyorgy},
  journal={Frontiers in Neuroscience},
  volume={18},
  pages={1307525},
  year={2024},
  publisher={Frontiers Media SA}
}

@inproceedings{gower2025train,
  title={How to Train an Oscillator Ising Machine using Equilibrium Propagation},
  author={Gower, Alex},
  booktitle={2025 International Conference on Neuromorphic Systems (ICONS)},
  pages={229--234},
  year={2025},
  organization={IEEE}
}

@article{rageau2025training,
  title={Training and synchronizing oscillator networks with Equilibrium Propagation},
  author={Rageau, Th{\'e}ophile and Grollier, Julie},
  journal={Neuromorphic Computing and Engineering},
  volume={5},
  number={3},
  pages={034008},
  year={2025},
  publisher={IOP Publishing}
}

@article{wang2024training,
  title={Training coupled phase oscillators as a neuromorphic platform using equilibrium propagation},
  author={Wang, Qingshan and Wanjura, Clara C and Marquardt, Florian},
  journal={Neuromorphic Computing and Engineering},
  volume={4},
  number={3},
  pages={034014},
  year={2024},
  publisher={IOP Publishing}
}

@article{
whitelam2026training,
author = {Stephen Whitelam },
title = {Training thermodynamic computers by gradient descent},
journal = {Proceedings of the National Academy of Sciences},
volume = {123},
number = {14},
pages = {e2528413123},
year = {2026},
doi = {10.1073/pnas.2528413123},
URL = {https://www.pnas.org/doi/abs/10.1073/pnas.2528413123},
eprint = {https://www.pnas.org/doi/pdf/10.1073/pnas.2528413123}
}

@article{scellier2016equilibrium,
  title={Equilibrium propagation: {B}ridging the gap between energy-based models and backpropagation},
  author={Scellier, Benjamin and Bengio, Yoshua},
  journal={arXiv preprint arXiv:1602.05179},
  year={2016}
}

@article{cugliandolo2017rules,
  title={Rules of calculus in the path integral representation of white noise Langevin equations: the Onsager--Machlup approach},
  author={Cugliandolo, Leticia F and Lecomte, Vivien},
  journal={Journal of Physics A: Mathematical and Theoretical},
  volume={50},
  number={34},
  pages={345001},
  year={2017},
  publisher={IOP Publishing}
}

@article{laydevant2024training,
  title={Training an {I}sing machine with equilibrium propagation},
  author={Laydevant, J{\'e}r{\'e}mie and Markovi{\'c}, Danijela and Grollier, Julie},
  journal={Nature Communications},
  volume={15},
  number={1},
  pages={3671},
  year={2024},
  publisher={Nature Publishing Group UK London}
}

@inproceedings{fan2025equilibrium,
  title={Equilibrium Propagation Training Neural Network Based on Coherent {I}sing Machine},
  author={Fan, Chenrui and Lu, Bo and Wang, Chuan},
  booktitle={2025 Photonics \& Electromagnetics Research Symposium-Fall (PIERS-Fall)},
  pages={1--6},
  year={2025},
  organization={IEEE}
}

@article{breakspear2010generative,
  title={Generative models of cortical oscillations: neurobiological implications of the Kuramoto model},
  author={Breakspear, Michael and Heitmann, Stewart and Daffertshofer, Andreas},
  journal={Frontiers in human neuroscience},
  volume={4},
  pages={190},
  year={2010},
  publisher={Frontiers Research Foundation}
}

@article{cabral2011role,
  title={Role of local network oscillations in resting-state functional connectivity},
  author={Cabral, Joana and Hugues, Etienne and Sporns, Olaf and Deco, Gustavo},
  journal={Neuroimage},
  volume={57},
  number={1},
  pages={130--139},
  year={2011},
  publisher={Elsevier}
}

@inproceedings{
sittoni2024subhomogeneous,
title={Subhomogeneous Deep Equilibrium Models},
author={Pietro Sittoni and Francesco Tudisco},
booktitle={Forty-first International Conference on Machine Learning},
year={2024},
url={https://openreview.net/forum?id=YBXwr7wF7i}
}

@article{yi2023nmode,
  title={nmODE: neural memory ordinary differential equation: Z. Yi},
  author={Yi, Zhang},
  journal={Artificial Intelligence Review},
  volume={56},
  number={12},
  pages={14403--14438},
  year={2023},
  publisher={Springer}
}

@article{lopezortiz2024exploring,
  title={Exploring deep echo state networks for image classification: A multi-reservoir approach},
  author={L{\'o}pez-Ortiz, Enrique J and Perea-Trigo, Marina and Soria-Morillo, Luis Miguel and Sancho-Caparrini, Fernando and Vegas-Olmos, JJ},
  journal={Neural Computing and Applications},
  volume={36},
  number={20},
  pages={11901--11918},
  year={2024},
  publisher={Springer}
}

@article{bacho2023exploring,
  title={Exploring trade-offs in spiking neural networks},
  author={Bacho, Florian and Chu, Dominique},
  journal={Neural Computation},
  volume={35},
  number={10},
  pages={1627--1656},
  year={2023},
  publisher={MIT Press One Rogers Street, Cambridge, MA 02142-1209, USA journals-info~…}
}

@inproceedings{cai2025oscnet,
  title={OscNet v1. 5: energy efficient hopfield network on CMOS oscillators for image classification},
  author={Cai, Wenxiao and Li, Zongru and Wang, Iris and Wang, Yu-Neng and Lee, Thomas H},
  booktitle={2025 IEEE/CVF International Conference on Computer Vision Workshops (ICCVW)},
  pages={4792--4800},
  year={2025},
  organization={IEEE}
}

@article{abernot2023training,
  title={Training energy-based single-layer Hopfield and oscillatory networks with unsupervised and supervised algorithms for image classification},
  author={Abernot, Madeleine and Todri-Sanial, Aida},
  journal={Neural Computing and Applications},
  volume={35},
  number={25},
  pages={18505--18518},
  year={2023},
  publisher={Springer}
}

@article{gemo2026leveraging,
  title={Leveraging phase model reduction and equilibrium propagation for neuromorphic design optimization of oscillatory neural networks},
  author={Gemo, Emanuele and Bonnin, Michele and Corinto, Fernando},
  journal={Neuromorphic Computing and Engineering},
  volume={6},
  number={2},
  pages={024016},
  year={2026},
  publisher={IOP Publishing}
}

@article{moayed2026design,
  title={Design of Oscillatory Neural Networks Using Machine-Learned Templates},
  author={Moayed, Mitra and Csaba, Gyorgy},
  journal={Electronics},
  volume={15},
  number={13},
  pages={2897},
  year={2026},
  publisher={MDPI}
}

@article{lipka2024thermodynamic,
  title={Thermodynamic computing via autonomous quantum thermal machines},
  author={Lipka-Bartosik, Patryk and Perarnau-Llobet, Mart{\'\i} and Brunner, Nicolas},
  journal={Science advances},
  volume={10},
  number={36},
  pages={eadm8792},
  year={2024},
  publisher={American Association for the Advancement of Science}
}
\bibliographystyle{plainnat}

\appendix
\section{Experimental Details and Reproducibility}
\label{AppendixA}

\subsection{Datasets, Splits, and Preprocessing}
\label{app:data_preprocessing}

We evaluate the method on MNIST and Fashion-MNIST, each containing
$60{,}000$ official training images and $10{,}000$ official test images.
Each example is a $28\times28$ grayscale image belonging to one of ten
classes. For both datasets, the official training set is divided into
$50{,}000$ training and $10{,}000$ validation examples using a fixed
stratified split that is held unchanged throughout all subsequent experiments.
The official test set is kept separate and used only for final evaluation.

Each image is flattened into $D=784$ components and rescaled from integer
pixel values to
\[
p_n\in[0,1]^D.
\]
For dataset $\mathcal D$, let $\mu_{\mathcal D}$ and
$\sigma_{\mathcal D}$ denote the scalar mean and population standard deviation
computed from all pixel values in the $50{,}000$-example training subset.
Each input is standardized as
\begin{equation*}
    z_n
    =
    \frac{
        p_n-\mu_{\mathcal D}\mathbf 1
    }{
        \sigma_{\mathcal D}
    },
\end{equation*}
where $\mathbf 1\in\mathbb R^D$ is the all-ones vector. The same
$\mu_{\mathcal D}$ and $\sigma_{\mathcal D}$ are applied to the training,
validation, and test splits.

The standardized input is then rescaled to a prescribed reference
$\ell_2$ norm,
\begin{equation*}
    u_n
    =
    \rho
    \frac{
        z_n
    }{
        \lVert z_n\rVert_2
    },
\end{equation*}
so that
\begin{equation*}
    \lVert u_n\rVert_2=\rho.
\end{equation*}
This normalization controls the overall magnitude of the vector entering the
input-dependent Kuramoto fields while preserving the direction of the
standardized image in input space.

For MNIST, a single reference norm $\rho_{\rm M}$ derived from the training
subset is used for training, validation, and test inputs. Fashion-MNIST
retains the source-domain scaling convention used in the experimental
pipeline: training and validation examples use
$\rho_{\rm F,train}$, whereas official-test examples use
$\rho_{\rm F,test}$. The latter is computed only from the test input values
and does not use class labels or model predictions. This distinction is kept
fixed throughout all Fashion-MNIST teacher and Kuramoto experiments.
The preprocessing constants are summarized in
Table~\ref{tab:dataset_preprocessing}. For each dataset, the frozen neural teacher and the Kuramoto student receive
the same processed input vector $u_n$.

\begin{table*}[htbp!]
    \centering
    \caption{\textbf{Dataset splits and preprocessing used in the main
    experiments.}
    Both datasets use the same stratified train--validation split and scalar
    standardization. MNIST uses one training-derived reference
    $\ell_2$ norm for all splits, whereas Fashion-MNIST retains separate
    label-independent reference norms for the training/validation and
    official-test input domains.}
    \label{tab:dataset_preprocessing}
    \small
    \begin{tabularx}{\textwidth}{
        @{}>{\raggedright\arraybackslash}p{0.26\textwidth}XX@{}}
        \toprule
        Item & MNIST & Fashion-MNIST \\
        \midrule

        Training examples
        & $50{,}000$
        & $50{,}000$ \\

        Validation examples
        & $10{,}000$
        & $10{,}000$ \\

        Test examples
        & $10{,}000$
        & $10{,}000$ \\

        Train--validation split
        & Fixed stratified split
        & Fixed stratified split \\

        $\mu_{\mathcal D}$
        & $0.131$
        & $0.286$ \\

        $\sigma_{\mathcal D}$
        & $0.308$
        & $0.353$ \\

        Training reference norm
        & $\rho_{\rm M}=48.902$
        & $\rho_{\rm F,train}=45.662$ \\

        Validation reference norm
        & $48.902$
        & $45.662$ \\

        Test reference norm
        & $48.902$
        & $\rho_{\rm F,test}=44.585$ \\

        Reference-norm convention
        & Training-derived value reused for all splits
        & Training-domain value for training/validation;
          label-independent test-domain value for test \\

        Data augmentation
        & None
        & None \\

        \bottomrule
    \end{tabularx}
\end{table*}

\subsection{Neural-Teacher Training}
\label{app:neural_teacher_training}

For both datasets, a feedforward neural network is trained first and then
held fixed throughout the subsequent construction and training of the
Kuramoto student. The principal teacher has architecture
\[
784\text{-}32\text{-}16\text{-}16\text{-}10.
\]
For an input $u\in\mathbb R^D$, the three hidden activations are
\begin{align*}
    h^{(1)}
    &=
    \tanh\!\left(
        W^{(1)}u+q^{(1)}
    \right),\\
    h^{(2)}
    &=
    \tanh\!\left(
        W^{(2)}h^{(1)}+q^{(2)}
    \right),\\
    h^{(3)}
    &=
    \tanh\!\left(
        W^{(3)}h^{(2)}+q^{(3)}
    \right),
\end{align*}
followed by the output logits
\begin{equation*}
    \zeta^{\rm TE,\OO}
    =
    W^{(4)}h^{(3)}+q^{(4)}.
\end{equation*}
Here, the parenthesized superscript denotes the neural-network layer index.

The hidden representation transferred to the physical student is obtained by
concatenating all three hidden layers,
\begin{equation*}
    a^{\rm TE,\HH}
    =
    \begin{pmatrix}
        h^{(1)}\\
        h^{(2)}\\
        h^{(3)}
    \end{pmatrix}
    \in[-1,1]^H,
    \qquad
    H=32+16+16=64.
\end{equation*}
Thus, each transferred hidden activation is associated one-to-one with one
hidden oscillator.

Because the raw output logits are unbounded, the output representation used by
the Kuramoto student is defined through the signed-softmax transformation
\begin{equation*}
    a_o^{\rm TE,\OO}
    =
    2
    \frac{
        \exp\!\left(\zeta_o^{\rm TE,\OO}\right)
    }{
        \sum_{c=1}^{O}
        \exp\!\left(\zeta_c^{\rm TE,\OO}\right)
    }
    -1,
    \qquad
    o=1,2,\ldots,O,
\end{equation*}
with $O=10$. The complete transferred state is therefore
\begin{equation*}
    a^{\rm TE}
    =
    \begin{pmatrix}
        a^{\rm TE,\HH}\\
        a^{\rm TE,\OO}
    \end{pmatrix}
    \in[-1,1]^N,
    \qquad
    N=H+O=74.
\end{equation*}
The signed-softmax transformation is monotonic in each logit relative to the
common normalization and therefore preserves the predicted class,
\begin{equation*}
    \arg\max_o \zeta_o^{\rm TE,\OO}
    =
    \arg\max_o a_o^{\rm TE,\OO}.
\end{equation*}
It consequently provides a bounded output representation compatible with the
physical readout $x_o=\cos\theta_o$ without altering the teacher's hard
classification.

The teacher is optimized on the raw logits using the mean multiclass
cross-entropy
\begin{equation*}
    \mathcal L_{\rm CE}
    =
    -\frac{1}{B}
    \sum_{n=1}^{B}
    \log
    \frac{
        \exp\!\left(
            \zeta_{n,y_n}^{\rm TE,\OO}
        \right)
    }{
        \sum_{c=1}^{O}
        \exp\!\left(
            \zeta_{n,c}^{\rm TE,\OO}
        \right)
    }.
\end{equation*}
Only the training subset is used for optimization, and checkpoint selection
uses the disjoint validation subset defined in
Appendix~\ref{app:data_preprocessing}. The official test set is evaluated
only after the teacher checkpoint has been fixed.

For MNIST, five prespecified teacher initializations are trained independently.
A validation-selected checkpoint is first identified for each run, after which
the five teacher candidates are compared using validation top-1 accuracy, with
validation cross-entropy used as a secondary criterion. For Fashion-MNIST, a
single prespecified teacher training run is used, with checkpoint selection by
minimum validation cross-entropy. These procedures yield frozen teachers with
test top-1 accuracies of $95.90\%$ on MNIST and $85.79\%$ on Fashion-MNIST.
The corresponding training configurations are summarized in
Table~\ref{tab:teacher_training}.

After teacher selection, its parameters remain fixed and the transferred
hidden and output activations are used unchanged as targets throughout
Kuramoto training. The numerical endpoint treatment required by the subsequent
$\operatorname{artanh}$-based phase-trajectory construction is described
separately in Appendix~\ref{app:stage1_derivation}.

\begin{table*}[htbp!]
    \centering
    \caption{\textbf{Neural-teacher configuration and training protocol.}
    Both datasets use the same principal architecture and optimization
    settings. Teacher checkpoints are selected using validation data only
    before official-test evaluation.}
    \label{tab:teacher_training}
    \small
    \begin{tabularx}{\textwidth}{
        @{}>{\raggedright\arraybackslash}p{0.29\textwidth}XX@{}}
        \toprule
        Item & MNIST & Fashion-MNIST \\
        \midrule

        Architecture
        & $784$--$32$--$16$--$16$--$10$
        & $784$--$32$--$16$--$16$--$10$ \\

        Trainable parameters
        & $26{,}090$
        & $26{,}090$ \\

        Transferred states
        & $64$ hidden; $10$ signed-softmax outputs
        & $64$ hidden; $10$ signed-softmax outputs \\

        Training loss
        & Mean cross-entropy on raw logits
        & Mean cross-entropy on raw logits \\

        Optimizer
        & SGD
        & SGD \\

        Learning rate
        & $0.2$
        & $0.2$ \\

        Weight decay
        & $10^{-4}$
        & $10^{-4}$ \\

        Batch size
        & $256$
        & $256$ \\

        Training budget
        & $300$ epochs
        & $300$ epochs \\

        Teacher-training runs
        & Five prespecified initializations
        & One prespecified initialization \\

        Checkpoint selection
        & Validation top-1, then validation CE
        & Minimum validation CE \\

        Validation top-1
        & $95.63\%$
        & $86.81\%$ \\

        Test top-1
        & $95.90\%$
        & $85.79\%$ \\

        \bottomrule
    \end{tabularx}
\end{table*}

\subsection{Teacher-Trajectory Construction and Path-Likelihood Interpretation}
\label{app:stage1_derivation}

This section develops the phase trajectory used for Stage-I supervision and
relates the resulting phase-increment objective to the trajectory likelihood
of a thermally perturbed Kuramoto model. The teacher trajectory is an
auxiliary supervisory construction: it specifies the states and increments at
which the student vector field is trained, but it is not an additional
dynamical component of the Kuramoto student.

\paragraph{Analytic teacher-trajectory construction.}

Let $a_i^{\rm TE}\in[-1,1]$ denote one component of the transferred neural
representation defined in Appendix~\ref{app:neural_teacher_training}. The
Kuramoto representation uses the physical activation
\begin{equation*}
    x_i=\cos\theta_i, \; x_i = a_i^{\rm TE}
\end{equation*}
so each transferred activation corresponds to a phase on the principal branch
$[0,\pi]$.

Because the construction below contains
$\operatorname{artanh}(a_i^{\rm TE})$, activations represented numerically
at exactly $\pm1$ require a finite-precision endpoint treatment. We define
\begin{equation*}
    \alpha_{\rm fp}
    =
    \operatorname{nextafter}_{\texttt{float64}}(1,0),
    \qquad
    \widetilde a_i^{\rm TE}
    =
    \operatorname{clip}
    \left(
        a_i^{\rm TE},
        -\alpha_{\rm fp},
        \alpha_{\rm fp}
    \right).
\end{equation*}
Targets already lying strictly inside $(-1,1)$ are unchanged. This operation
only regularizes the singular endpoints of $\operatorname{artanh}$ and does
not alter the teacher hard prediction.

All teacher phases begin from the same initial state used by the Kuramoto
student,
\begin{equation*}
    \theta_i^{\rm TE}(0)=\frac{\pi}{2},
    \qquad
    x_i^{\rm TE}(0)=0,
\end{equation*}
and the desired terminal phase is
\begin{equation*}
    \theta_i^{\rm TE}(t_{\rm f})
    =
    \arccos\widetilde a_i^{\rm TE}.
\end{equation*}

To construct the complete path between these endpoints, we introduce an
independent auxiliary guide dynamics for each transferred activation,
\begin{equation}
    \dot{\theta}_i^{\rm TE}
    =
    -\mu\beta_i^{\rm TE}\sin\theta_i^{\rm TE},
    \label{eq:app_teacher_guide_phase}
\end{equation}
where $\beta_i^{\rm TE}$ is a guide field used only to define the supervisory
trajectory and is distinct from the trainable local field $b_i(u)$ of the
Kuramoto student. Under
$x_i^{\rm TE}=\cos\theta_i^{\rm TE}$,
Eq.~\ref{eq:app_teacher_guide_phase} becomes
\begin{equation*}
    \dot{x}_i^{\rm TE}
    =
    \mu\beta_i^{\rm TE}
    \left(
        1-\left(x_i^{\rm TE}\right)^2
    \right).
\end{equation*}
With $x_i^{\rm TE}(0)=0$, its solution is
\begin{equation*}
    x_i^{\rm TE}(t)
    =
    \tanh\left(\mu\beta_i^{\rm TE}t\right).
\end{equation*}
Imposing
$x_i^{\rm TE}(t_{\rm f})=\widetilde a_i^{\rm TE}$
therefore gives
\begin{equation*}
    \beta_i^{\rm TE}
    =
    \frac{
        \operatorname{artanh}
        \left(\widetilde a_i^{\rm TE}\right)
    }{
        \mu t_{\rm f}
    }.
\end{equation*}
Substitution yields the explicit activation trajectory
\begin{equation*}
    x_i^{\rm TE}(t)
    =
    \tanh
    \left(
        \frac{t}{t_{\rm f}}
        \operatorname{artanh}
        \left(\widetilde a_i^{\rm TE}\right)
    \right),
    \qquad
    0\leq t\leq t_{\rm f},
\end{equation*}
and the corresponding phase trajectory
\begin{equation*}
    \theta_i^{\rm TE}(t)
    =
    \arccos
    \left(
        \tanh
        \left(
            \frac{t}{t_{\rm f}}
            \operatorname{artanh}
            \left(\widetilde a_i^{\rm TE}\right)
        \right)
    \right).
\end{equation*}
Hence,
\begin{equation*}
    x_i^{\rm TE}(0)=0,
    \qquad
    x_i^{\rm TE}(t_{\rm f})=\widetilde a_i^{\rm TE}.
\end{equation*}

The principal branch of $\arccos$ keeps the complete teacher path within
$[0,\pi]$. Positive targets move monotonically from $\pi/2$ toward $0$,
negative targets move toward $\pi$, and a zero target remains around $\pi/2$.
The construction therefore determines both the finite-time endpoint and the
complete supervisory phase trajectory directly from the transferred neural
activation. Importantly, the independent guide dynamics specify only the
teacher path; they do not restrict the topology of the interacting Kuramoto
student.

\paragraph{Teacher-forced phase-increment objective.}

Let $f_\Theta(\theta,u)\in\mathbb R^N$ denote the deterministic Kuramoto
vector field. For
\begin{equation*}
    t_k=k\Delta t,
    \qquad
    k=0,\ldots,K,
    \qquad
    K\Delta t=t_{\rm f},
\end{equation*}
the teacher increment for training example $n$ is
\begin{equation*}
    \Delta\theta_{n,k}^{\rm TE}
    =
    \theta_n^{\rm TE}(t_{k+1})
    -
    \theta_n^{\rm TE}(t_k).
\end{equation*}
Stage I evaluates the student vector field at the prescribed teacher state,
giving the locally predicted increment
\begin{equation*}
    \Delta\theta_{n,k}^{\rm S}
    =
    \Delta t\,
    f_\Theta
    \left(
        \theta_n^{\rm TE}(t_k),
        u_n
    \right).
\end{equation*}
The phase-increment residual is therefore
\begin{equation*}
    r_{n,k}
    =
    \theta_n^{\rm TE}(t_{k+1})
    -
    \theta_n^{\rm TE}(t_k)
    -
    \Delta t\,
    f_\Theta
    \left(
        \theta_n^{\rm TE}(t_k),
        u_n
    \right).
\end{equation*}
For a mini-batch of $B$ examples, the Stage-I objective is
\begin{equation}
    \mathcal L_{\rm path}
    =
    \frac{1}{BKN}
    \sum_{n=1}^{B}
    \sum_{k=0}^{K-1}
    \left\|r_{n,k}\right\|_2^2.
    \label{eq:app_path_loss}
\end{equation}

Eq.~\ref{eq:app_path_loss} constrains the student vector field across
all physical degrees of freedom and throughout the prescribed finite-time
trajectory. The important distinction is that
$f_\Theta$ is evaluated at
$\theta_n^{\rm TE}(t_k)$ rather than at a recursively generated student
state. 

\paragraph{Trajectory-likelihood interpretation.}

The squared phase-increment residual admits a probabilistic interpretation
through the thermally perturbed Kuramoto dynamics
\begin{equation*}
    {\rm d}\theta_i
    =
    f_{\Theta,i}(\theta,u)\,{\rm d}t
    +
    \sqrt{2\mu k_{\rm B}T}\,{\rm d}W_i,
    \qquad T>0,
\end{equation*}
where the $W_i$ are independent Wiener processes. In the local unwrapped
coordinates used to represent the numerical phase increments, one
Euler--Maruyama step is
\begin{equation*}
    \theta_{k+1}
    =
    \theta_k
    +
    \Delta t\,
    f_\Theta(\theta_k,u)
    +
    \sqrt{2\mu k_{\rm B}T\Delta t}\,
    \xi_k,
    \qquad
    \xi_k\sim\mathcal N(0,I_N).
\end{equation*}
Its conditional transition density is
\begin{equation*}
    p_\Theta^{\rm loc}
    (\theta_{k+1}\mid\theta_k,u)
    =
    \frac{1}
    {\left(4\pi\mu k_{\rm B}T\Delta t\right)^{N/2}}
    \exp
    \left(
        -
        \frac{
            \left\|
                \theta_{k+1}
                -
                \theta_k
                -
                \Delta t\,f_\Theta(\theta_k,u)
            \right\|_2^2
        }{
            4\mu k_{\rm B}T\Delta t
        }
    \right).
\end{equation*}

Because phase variables are defined modulo $2\pi$, the exact transition
density on $\TT^N$ is a wrapped density and its negative log-likelihood is
not generally a single quadratic residual. Stage I, however, evaluates the
student vector field only along the prescribed teacher trajectory. This
trajectory is represented by a continuous lift on the principal branch
$[0,\pi]$, so its successive phase increments are unambiguous in the local
coordinates used by the numerical integrator. The likelihood interpretation
below therefore refers to the local Euler--Maruyama model for these prescribed
teacher transitions, rather than to the exact wrapped transition kernel on
$\TT^N$.

For the teacher trajectory
\begin{equation*}
    \omega_n^{\rm TE}
    =
    \left\{
        \theta_n^{\rm TE}(t_k)
    \right\}_{k=0}^{K},
\end{equation*}
define the corresponding local path likelihood as
\begin{equation*}
    P_\Theta^{\rm loc}
    \left(
        \omega_n^{\rm TE}
        \mid
        \theta_n^{\rm TE}(t_0),u_n
    \right)
    =
    \prod_{k=0}^{K-1}
    p_\Theta^{\rm loc}
    \left(
        \theta_n^{\rm TE}(t_{k+1})
        \mid
        \theta_n^{\rm TE}(t_k),u_n
    \right).
\end{equation*}
Substituting the teacher increments gives
\begin{align*}
    -\log
    P_\Theta^{\rm loc}
    \left(
        \omega_n^{\rm TE}
        \mid
        \theta_n^{\rm TE}(t_0),u_n
    \right)
    &=
    \frac{1}
    {4\mu k_{\rm B}T\Delta t}
    \sum_{k=0}^{K-1}
    \left\|r_{n,k}\right\|_2^2
    \nonumber\\
    &\quad+
    \frac{KN}{2}
    \log
    \left(
        4\pi\mu k_{\rm B}T\Delta t
    \right).
\end{align*}
The second term is independent of the trainable parameters $\Theta$.
Consequently, for fixed $\mu$, $T>0$, and $\Delta t$, maximizing this local
trajectory likelihood is equivalent to minimizing the summed squared
phase-increment residual.

After normalization over examples, time intervals, and physical degrees of
freedom,
\begin{equation*}
    \overline{\mathcal L}_{\rm NLL}^{\rm loc}
    =
    \frac{
        \mathcal L_{\rm path}
    }{
        4\mu k_{\rm B}T\Delta t
    }
    +
    \frac{1}{2}
    \log
    \left(
        4\pi\mu k_{\rm B}T\Delta t
    \right).
\end{equation*}
Thus, up to a positive multiplicative factor and an additive constant,
$\mathcal L_{\rm path}$ is the discrete negative log-likelihood of the
prescribed trajectory under the local Euler--Maruyama transition model. In
this sense, it is also a discrete Onsager--Machlup-type path objective for the
chosen stochastic discretization \citep{cugliandolo2017rules}.

This probabilistic construction provides an interpretation of Stage I. The actual
Stage-I training protocol minimizes the deterministic, unweighted
$\mathcal L_{\rm path}$ at $k_{\rm B}T=0$. Introducing $T>0$ above only
exposes the connection between phase-increment matching and thermodynamic
trajectory likelihood; for any fixed positive $T$, the multiplicative factor
relating the two objectives is independent of $\Theta$.

Finally, Stage-I supervision constrains the student vector field only at
states on the prescribed teacher path. During autonomous inference, the state
at each subsequent time is generated by the student itself. A local
vector-field discrepancy can therefore move the trajectory away from the
supervised path, after which the learned field is evaluated at off-trajectory
states. This teacher-forcing mismatch provides the motivation for the
autonomous endpoint-fine-tuning stage introduced in the main text.

\subsection{Kuramoto Model and Training Configuration}
\label{app:kuramoto_training}

This section specifies the physical Kuramoto architecture and the numerical
configuration used in the principal MNIST and Fashion-MNIST experiments. The
teacher-trajectory construction and Stage-I path objective are derived
separately in Appendix~\ref{app:stage1_derivation}, while checkpoint-selection
rules and alternative training protocols are described in
Appendix~\ref{app:checkpoint_selection}.

\paragraph{Physical network.}

The Kuramoto student contains $H$ hidden oscillators and $O$ output
oscillators, with
\[
    N=H+O.
\]
For processed input $u_n\in\mathbb R^D$, sample-dependent fields act only on
the hidden oscillators,
\begin{equation*}
    b^{\HH}(u_n)
    =
    W^{\HH}u_n+c^{\HH},
    \qquad
    b^{\OO}
    =
    c^{\OO},
\end{equation*}
where
\[
    W^{\HH}\in\mathbb R^{H\times D},
    \qquad
    c^{\HH}\in\mathbb R^H,
    \qquad
    c^{\OO}\in\mathbb R^O.
\]
Thus, the image directly modulates only the hidden local fields; the output
oscillators receive input information through their dynamical coupling to the
hidden subsystem.

The symmetric coupling matrix is
\begin{equation*}
    J
    =
    \begin{pmatrix}
        J^{\HH\HH} & J^{\HH\OO}\\
        (J^{\HH\OO})^\T & 0
    \end{pmatrix},
\end{equation*}
with
\begin{equation*}
    J^{\HH\HH}
    =
    (J^{\HH\HH})^\T,
    \qquad
    \operatorname{diag}(J^{\HH\HH})=0,
    \qquad
    J^{\HH\OO}\in\mathbb R^{H\times O}.
\end{equation*}
There are therefore no self-couplings, no output--output couplings, and no
direct input-to-output weights. The trainable physical parameter set is
\begin{equation*}
    \Theta
    =
    \left\{
        W^{\HH},
        c^{\HH},
        c^{\OO},
        J^{\HH\HH},
        J^{\HH\OO}
    \right\}.
\end{equation*}

At the principal operating point,
\[
    D=784,
    \qquad
    H=64,
    \qquad
    O=10,
    \qquad
    N=74.
\]
The model contains $52{,}906$ trainable parameters, including $50{,}176$
input coefficients, $74$ local-field biases, $2{,}016$ independent
hidden--hidden couplings, and $640$ hidden--output couplings. The evolving
oscillator network therefore contains $2{,}656$ independently programmable
symmetric dynamical couplings.

\paragraph{Deterministic dynamics and energy structure.}

For the $n$-th input, oscillator $i$ evolves according to
\begin{equation*}
    \dot{\theta}_{n,i}^{\rm S}
    =
    f_{\Theta,i}
    \left(
        \theta_n^{\rm S},u_n
    \right)
    =
    -\mu
    \left[
        b_i(u_n)\sin\theta_{n,i}^{\rm S}
        +
        \sum_{j\neq i}
        J_{ij}
        \sin
        \left(
            \theta_{n,i}^{\rm S}
            -
            \theta_{n,j}^{\rm S}
        \right)
    \right].
\end{equation*}
Separating the hidden and output blocks gives, for
$h=1,\ldots,H$,
\begin{align*}
\dot{\theta}_{n,h}^{\rm S,\HH}
=
-\mu\Bigg[
&
\left(
\sum_{d=1}^{D}
W_{hd}^{\HH}u_{n,d}
+
c_h^{\HH}
\right)
\sin\theta_{n,h}^{\rm S,\HH}
\nonumber\\
&+
\sum_{\substack{h'=1\\h'\neq h}}^{H}
J_{hh'}^{\HH\HH}
\sin
\left(
\theta_{n,h}^{\rm S,\HH}
-
\theta_{n,h'}^{\rm S,\HH}
\right)
\nonumber\\
&+
\sum_{o=1}^{O}
J_{ho}^{\HH\OO}
\sin
\left(
\theta_{n,h}^{\rm S,\HH}
-
\theta_{n,o}^{\rm S,\OO}
\right)
\Bigg],
\end{align*}
and, for $o=1,2,\ldots,O$,
\begin{equation*}
\dot{\theta}_{n,o}^{\rm S,\OO}
=
-\mu
\left[
c_o^{\OO}
\sin\theta_{n,o}^{\rm S,\OO}
+
\sum_{h=1}^{H}
J_{ho}^{\HH\OO}
\sin
\left(
\theta_{n,o}^{\rm S,\OO}
-
\theta_{n,h}^{\rm S,\HH}
\right)
\right].
\end{equation*}
These equations make explicit that the output subsystem receives no direct
sample-dependent field and communicates with the hidden subsystem only through
the symmetric $J^{\HH\OO}$ interactions.

Because the couplings are symmetric, the deterministic dynamics form the
gradient flow of the input-conditioned energy
\begin{equation*}
    E_{\Theta}(\theta;u)
    =
    -
    \sum_{i=1}^{N}
    b_i(u)\cos\theta_i
    -
    \sum_{i<j}
    J_{ij}\cos(\theta_i-\theta_j),
\end{equation*}
with
\begin{equation*}
    \dot{\theta}
    =
    -\mu\nabla_{\theta}E_{\Theta}.
\end{equation*}
For a fixed input,
\begin{equation*}
    \frac{{\rm d}E_{\Theta}}{{\rm d}t}
    =
    -\mu
    \left\|
        \nabla_{\theta}E_{\Theta}
    \right\|_2^2
    \leq0.
\end{equation*}
The principal training and inference experiments use deterministic dynamics
with $k_{\rm B}T=0$. No natural-frequency terms, phase lags, higher harmonics,
or nonsymmetric interactions are introduced.

\paragraph{Initialization, numerical integration, and readout.}

Every autonomous student rollout begins from
\begin{equation*}
    \theta_{n,i}^{\rm S}(0)
    =
    \frac{\pi}{2},
    \qquad
    i=1,\ldots,N.
\end{equation*}
The corresponding physical activations
$x_{n,i}^{\rm S}=\cos\theta_{n,i}^{\rm S}$ therefore start from zero.

The deterministic dynamics are integrated using explicit Euler,
\begin{equation*}
    \theta_{n,k+1}^{\rm S}
    =
    \theta_{n,k}^{\rm S}
    +
    \Delta t\,
    f_{\Theta}
    \left(
        \theta_{n,k}^{\rm S},u_n
    \right),
    \qquad
    k=0,1,\ldots,K-1,
\end{equation*}
with
\[
    t_k=k\Delta t,
    \qquad
    K\Delta t=t_{\rm f}.
\]
The principal experiments use
\[
    \mu=1,
    \qquad
    t_{\rm f}=0.2,
    \qquad
    K=200,
    \qquad
    \Delta t=10^{-3}.
\]
Sensitivity to the integration grid is
examined separately in Appendix~\ref{AppendixH}.

Classification is read directly from the output oscillator activations at the
fixed observation time,
\begin{equation*}
    \widehat y_n^{\rm S}
    =
    \arg\max_{o\in\{1,\ldots,O\}}
    \cos\theta_{n,o}^{\rm S,\OO}(t_{\rm f}).
\end{equation*}
No additional trainable readout layer is introduced.

\paragraph{Stage-I and Stage-II optimization.}

Stage-I teacher trajectories and the corresponding teacher-forced
phase-increment loss $\mathcal L_{\rm path}$ are defined in
Appendix~\ref{app:stage1_derivation}. All trainable physical parameters are
initialized exactly at zero,
\begin{equation*}
    W^{\HH}=0,
    \qquad
    c^{\HH}=0,
    \qquad
    c^{\OO}=0,
    \qquad
    J^{\HH\HH}=0,
    \qquad
    J^{\HH\OO}=0,
\end{equation*}
and Stage I optimizes $\mathcal L_{\rm path}$ jointly over all components of
$\Theta$.

Stage II begins from the validation-selected Stage-I checkpoint and rolls the
student autonomously over the full observation interval. The endpoint loss is
\begin{equation*}
    \mathcal L_{\rm end}
    =
    \frac{1}{BO}
    \sum_{n=1}^{B}
    \sum_{o=1}^{O}
    \left(
        \cos
        \theta_{n,o}^{\rm S,\OO}(t_{\rm f})
        -
        a_{n,o}^{\rm TE,\OO}
    \right)^2,
\end{equation*}
and the principal TWO-STAGE-END objective is
\begin{equation*}
    \mathcal L_{\rm Stage\,II}
    =
    \mathcal L_{\rm path}
    +
    \lambda_{\rm end}\mathcal L_{\rm end},
    \qquad
    \lambda_{\rm end}=1.
\end{equation*}
Gradients of $\mathcal L_{\rm end}$ propagate through the complete autonomous
rollout, and all components of $\Theta$ remain trainable. The numerical
optimization settings used in the principal protocol are summarized in
Table~\ref{tab:kuramoto_training_configuration}; validation-based checkpoint
selection and alternative training variants are described in
Appendix~\ref{app:checkpoint_selection}.

\begin{table*}[htbp!]
    \centering
    \caption{\textbf{Physical, numerical, and optimization configuration for
    the principal Kuramoto experiments.}
    The same $H=64$, $O=10$ architecture and numerical integration are used
    for the principal MNIST and Fashion-MNIST experiments.}
    \label{tab:kuramoto_training_configuration}
    \small
    \begin{tabularx}{\textwidth}{
        @{}>{\raggedright\arraybackslash}p{0.32\textwidth}X@{}}
        \toprule
        Item & Configuration \\
        \midrule

        Hidden / output oscillators
        & $H=64$, $O=10$, $N=74$ \\

        Input dimension
        & $D=784$ \\

        Trainable physical parameters
        & $52{,}906$ \\

        symmetric dynamical couplings
        & $2{,}656$ \\

        Mobility
        & $\mu=1$ \\

        Training temperature
        & $k_{\rm B}T=0$ \\

        Initial phase
        & $\theta_i^{\rm S}(0)=\pi/2$ \\

        Observation time
        & $t_{\rm f}=0.2$ \\

        Numerical integration
        & Explicit Euler; $K=200$; $\Delta t=10^{-3}$ \\

        Stage-I initialization
        & Exact zero for all trainable physical parameters \\

        Stage-I optimization
        & SGD; learning rate $3000$; batch size $256$; $300$ epochs \\

        Stage-I objective
        & $\mathcal L_{\rm path}$ \\

        Stage-II initialization
        & Validation-selected Stage-I checkpoint; fresh optimizer \\

        Stage-II optimization
        & SGD; learning rate $10$; batch size $256$; $20$ epochs \\

        Stage-II objective
        & $\mathcal L_{\rm path}
        +\lambda_{\rm end}\mathcal L_{\rm end}$,
        $\lambda_{\rm end}=1$ \\

        \bottomrule
    \end{tabularx}
\end{table*}

\subsection{Evaluation Metrics and Statistical Procedures}
\label{app:evaluation_metrics}

Let $\mathcal D_{\rm eval}$ denote an evaluation set containing
$M=|\mathcal D_{\rm eval}|$ examples. For example $n$, the autonomous
Kuramoto output score is
\begin{equation*}
    s_{n,o}^{\rm S,\OO}
    =
    \cos\theta_{n,o}^{\rm S,\OO}(t_{\rm f}),
    \qquad
    o=1,\ldots,O,
\end{equation*}
and the corresponding frozen neural-teacher target is
$a_{n,o}^{\rm TE,\OO}$. The student and teacher predictions are
\begin{equation*}
    \widehat y_n^{\rm S}
    =
    \arg\max_o s_{n,o}^{\rm S,\OO},
    \qquad
    \widehat y_n^{\rm TE}
    =
    \arg\max_o a_{n,o}^{\rm TE,\OO}.
\end{equation*}
All official-test metrics are evaluated only after checkpoint selection has
been completed using validation data, as described in
Appendix~\ref{app:checkpoint_selection}.

\paragraph{Classification and teacher agreement.}

Top-$k$ accuracy is defined as
\begin{equation*}
    A_{\rm top\text{-}k}
    =
    \frac{1}{M}
    \sum_{n=1}^{M}
    \mathbf 1
    \left[
        y_n
        \in
        \operatorname{TopK}
        \left(
            s_n^{\rm S,\OO}
        \right)
    \right],
\end{equation*}
where $s_n^{\rm S,\OO}\in\mathbb R^O$ is the vector of terminal output
scores. We report top-1 accuracy as the principal task metric.

Neural-teacher agreement measures sample-wise consistency between the student
and the frozen teacher,
\begin{equation*}
    A_{\rm agree}
    =
    \frac{1}{M}
    \sum_{n=1}^{M}
    \mathbf 1
    \left[
        \widehat y_n^{\rm S}
        =
        \widehat y_n^{\rm TE}
    \right].
\end{equation*}
Teacher agreement is distinct from ground-truth accuracy: two models can have
similar aggregate accuracy while differing in the examples they classify
correctly.

\paragraph{Continuous endpoint fidelity.}

The principal continuous measure of terminal teacher--student alignment is the
output endpoint RMSE,
\begin{equation*}
    R_{\rm end}^{\OO}
    =
    \left(
        \frac{1}{MO}
        \sum_{n=1}^{M}
        \sum_{o=1}^{O}
        \left(
            s_{n,o}^{\rm S,\OO}
            -
            a_{n,o}^{\rm TE,\OO}
        \right)^2
    \right)^{1/2}.
\end{equation*}
Unlike classification accuracy, this quantity compares the complete continuous
output representation rather than only its largest component.

We additionally use the Pearson correlation between the flattened student and
teacher output arrays,
\begin{equation*}
    \rho_{\rm end}^{\OO}
    =
    \frac{
        \sum_{n,o}
        \left(
            s_{n,o}^{\rm S,\OO}-\bar s
        \right)
        \left(
            a_{n,o}^{\rm TE,\OO}-\bar a
        \right)
    }{
        \left(
            \sum_{n,o}
            \left(
                s_{n,o}^{\rm S,\OO}-\bar s
            \right)^2
        \right)^{1/2}
        \left(
            \sum_{n,o}
            \left(
                a_{n,o}^{\rm TE,\OO}-\bar a
            \right)^2
        \right)^{1/2}
    },
\end{equation*}
where
\[
    \bar s
    =
    \frac{1}{MO}\sum_{n,o}s_{n,o}^{\rm S,\OO},
    \qquad
    \bar a
    =
    \frac{1}{MO}\sum_{n,o}a_{n,o}^{\rm TE,\OO}.
\]
Endpoint RMSE measures absolute discrepancy, whereas correlation measures
global linear alignment; the two quantities are therefore complementary.

\paragraph{Trajectory fidelity.}

For an autonomous student rollout, define
\begin{equation*}
    x_{n,i}^{\rm S}(t_k)
    =
    \cos\theta_{n,i}^{\rm S}(t_k),
    \qquad
    x_{n,i}^{\rm TE}(t_k)
    =
    \cos\theta_{n,i}^{\rm TE}(t_k).
\end{equation*}
For a set of dynamical variables $\mathcal I$, the activation-space trajectory
RMSE is
\begin{equation*}
    R_{{\rm traj},x}^{\mathcal I}
    =
    \left(
        \frac{1}
        {M(K+1)|\mathcal I|}
        \sum_{n=1}^{M}
        \sum_{k=0}^{K}
        \sum_{i\in\mathcal I}
        \left(
            x_{n,i}^{\rm S}(t_k)
            -
            x_{n,i}^{\rm TE}(t_k)
        \right)^2
    \right)^{1/2}.
\end{equation*}

\paragraph{Statistics across training-order runs.}

The principal repeated-training experiments vary the minibatch ordering while
holding the dataset split, frozen neural teacher, physical architecture, and
parameter initialization fixed. For a metric $g$ evaluated over $R$
training-order runs, we report
\begin{equation*}
    \bar g
    =
    \frac{1}{R}
    \sum_{r=1}^{R}g_r,
    \qquad
    s_g
    =
    \left(
        \frac{1}{R-1}
        \sum_{r=1}^{R}
        (g_r-\bar g)^2
    \right)^{1/2},
\end{equation*}
as mean $\pm$ sample standard deviation.

\paragraph{Paired effects and confidence intervals.}

For methods $A$ and $B$ evaluated under matched training-order runs, the
paired effect for metrics in which larger values are favorable is
\begin{equation*}
    d_r
    =
    g_{A,r}-g_{B,r},
    \qquad
    r=1,2,\ldots,R.
\end{equation*}

Let $\bar d$ and $s_d$ denote the mean and sample standard deviation of the
paired effects. The two-sided $95\%$ Student-$t$ interval is
\begin{equation*}
    \bar d
    \pm
    t_{0.975,R-1}
    \frac{s_d}{\sqrt R}.
\end{equation*}


Specialized inference protocols and diagnostics are defined with the
experiments in which they are used. Stochastic single-trajectory and
ensemble readouts are defined in Appendix~\ref{app:stochastic_protocol},
while integration-grid sensitivity and numerical-safety checks are
described in Appendix~\ref{AppendixH}.

\subsection{Training Variants and Checkpoint Selection}
\label{app:checkpoint_selection}

This section defines the validation criteria used to select Kuramoto
checkpoints and the training variants used to isolate the roles of path
supervision, autonomous endpoint supervision, and training order. All
checkpoint and protocol choices are determined using training and validation
data only; the official test set is evaluated only after the corresponding
checkpoint has been fixed.

\paragraph{Validation-based checkpoint selection.}

Stage-I PATH checkpoints are selected by minimizing the validation output
endpoint RMSE,
\begin{equation*}
    e_{\rm PATH}^{\star}
    =
    \arg\min_e
    R_{\rm end}^{\rm val}(e),
\end{equation*}
where $R_{\rm end}^{\rm val}(e)$ is computed from an autonomous validation
rollout after epoch $e$.

Endpoint-supervised continuation and direct-training variants are selected
using the validation joint score
\begin{equation*}
    S_{\rm joint}^{\rm val}(e)
    =
    \frac{1}{2}
    \left(
        A_{\rm top1}^{\rm val}(e)
        +
        A_{\rm agree}^{\rm val}(e)
    \right),
\end{equation*}
where $A_{\rm top1}^{\rm val}$ is the validation top-1 accuracy and
$A_{\rm agree}^{\rm val}$ is the agreement between the Kuramoto student and
the frozen neural teacher. The selected checkpoint maximizes
$S_{\rm joint}^{\rm val}$. Any exact ties are resolved deterministically using
validation quantities only. No official-test metric enters either checkpoint
criterion.

\paragraph{Paper-main continuation comparison.}

The principal comparison after Stage-I path training is between PATH-CONT and
TWO-STAGE-END. For each paired training-order run, both methods start from the
same validation-selected PATH checkpoint and use matched continuation
conditions; they differ only in the continuation objective,
\begin{align*}
    \mathcal L_{\rm PATH\mbox{-}CONT}
    &=
    \mathcal L_{\rm path},
    \\
    \mathcal L_{\rm TWO\mbox{-}STAGE\mbox{-}END}
    &=
    \mathcal L_{\rm path}
    +
    \mathcal L_{\rm end}.
\end{align*}
PATH-CONT therefore controls for the effect of additional optimization after
Stage I, whereas TWO-STAGE-END isolates the effect of introducing autonomous
endpoint supervision. The principal continuation uses the optimization
configuration summarized in
Table~\ref{tab:kuramoto_training_configuration}, and the multi-seed comparison
uses matched minibatch orderings within each paired run.

\paragraph{Auxiliary training variants.}

Four additional variants are used to distinguish the roles of initialization,
path supervision, and training order:

\begin{description}

    \item[\textbf{PRETRAIN-END-ONLY}]
    starts from the validation-selected PATH checkpoint and optimizes
    $\mathcal L_{\rm end}$ alone. Comparison with TWO-STAGE-END tests whether
    the path objective must remain active after path pretraining.

    \item[\textbf{Z-END}]
    starts from exact-zero physical parameters and optimizes
    $\mathcal L_{\rm path}+\mathcal L_{\rm end}$ from the beginning.
    Comparison with TWO-STAGE-END probes the effect of staged path
    pretraining versus direct joint optimization.

    \item[\textbf{ZERO-END-ONLY}]
    starts from the same exact-zero physical initialization as Z-END but
    optimizes only $\mathcal L_{\rm end}$. The Z-END versus
    ZERO-END-ONLY comparison therefore isolates the role of path supervision
    when training begins from zero.

    \item[\textbf{PATH-CONT}]
    continues $\mathcal L_{\rm path}$ from the selected PATH checkpoint and
    serves as the matched additional-optimization control described above.

\end{description}

Together with PATH and TWO-STAGE-END, these variants form the six-method
training-protocol ablation analyzed in
Appendix~\ref{app:full_protocol_ablation}. That experiment is an auxiliary
single-run study and uses a continuation configuration distinct from the final
multi-seed principal protocol; its matched optimization settings are specified
with the ablation results rather than repeated here.

\paragraph{Endpoint-objective weighting.}

The dependence on the relative weight of autonomous endpoint supervision is
examined separately by varying $\lambda_{\rm end}$ in
\begin{equation*}
    \mathcal L
    =
    \mathcal L_{\rm path}
    +
    \lambda_{\rm end}\mathcal L_{\rm end}.
\end{equation*}
The corresponding matched sensitivity experiment and numerical results are
reported in Appendix~\ref{app:endpoint_weight_sensitivity}.

\section{Path Matching and Autonomous Inference}
\label{AppendixB}

\subsection{Multi-Seed Autonomous-Inference Results}
\label{app:multiseed_main_results}

We first evaluate the principal Kuramoto training protocol across ten paired
training-order runs on MNIST and Fashion-MNIST. The dataset split,
frozen neural teacher, Kuramoto architecture, parameter initialization, and
numerical configuration are fixed across runs, while the minibatch ordering is
varied. PATH-CONT and TWO-STAGE-END are continued from the same
validation-selected PATH checkpoint under the matched protocol defined in
Appendix~\ref{app:checkpoint_selection}; their defining difference is whether
autonomous endpoint supervision is introduced during continuation. All
official-test results are evaluated only after validation-based checkpoint
selection. Reported uncertainties are sample standard deviations across the
ten training-order runs.

Table~\ref{tab:appendix_B1_multiseed_main_results_lr10} summarizes the
autonomous-inference results. On MNIST, PATH reaches
$89.112\pm0.020\%$ top-1 accuracy, while PATH-CONT remains essentially
unchanged at $89.112\pm0.016\%$. Introducing autonomous endpoint supervision
with TWO-STAGE-END increases the accuracy to $96.711\pm0.065\%$. This
improvement is accompanied by substantially stronger agreement with the
neural teacher and a reduction of the output endpoint RMSE from
$0.357101\pm0.000055$ for PATH to $0.119756\pm0.001362$ for
TWO-STAGE-END.

Fashion-MNIST shows the same qualitative behavior. PATH and PATH-CONT obtain
$80.172\pm0.024\%$ and $80.171\pm0.014\%$ top-1 accuracy, respectively,
whereas TWO-STAGE-END reaches $86.399\pm0.184\%$. The output endpoint RMSE
simultaneously decreases from $0.309501\pm0.000062$ to
$0.101118\pm0.002507$, and neural-teacher agreement increases from
$86.464\pm0.021\%$ to $95.206\pm0.156\%$. Top-2 accuracy follows the same
overall pattern. Thus, on both datasets, continued path-only optimization
produces negligible changes in autonomous inference, whereas endpoint
supervision substantially improves both task performance and continuous
teacher--student endpoint alignment.

\begin{table}[htbp!]
\centering
\caption{\textbf{Multi-seed autonomous-inference results.}
Official-test performance is reported as mean $\pm$ sample standard deviation
over ten training-order runs. PATH-CONT and TWO-STAGE-END are continued from
matched validation-selected PATH checkpoints; their defining difference is
the inclusion of autonomous endpoint supervision in TWO-STAGE-END.
Lower output endpoint RMSE is better.}
\label{tab:appendix_B1_multiseed_main_results_lr10}
\begin{adjustbox}{max width=\textwidth}
\begin{tabular}{llrrrr}
\toprule
Dataset & Method & Top-1 (\%) & Top-2 (\%) & Agreement (\%) &
Output endpoint RMSE \\
\midrule

MNIST
& PATH
& 89.112 $\pm$ 0.020
& 93.542 $\pm$ 0.020
& 90.218 $\pm$ 0.020
& 0.357101 $\pm$ 0.000055 \\

MNIST
& PATH-CONT
& 89.112 $\pm$ 0.016
& 93.537 $\pm$ 0.019
& 90.218 $\pm$ 0.015
& 0.357086 $\pm$ 0.000045 \\

MNIST
& TWO-STAGE-END
& 96.711 $\pm$ 0.065
& 98.599 $\pm$ 0.079
& 96.995 $\pm$ 0.053
& 0.119756 $\pm$ 0.001362 \\

Fashion-MNIST
& PATH
& 80.172 $\pm$ 0.024
& 92.064 $\pm$ 0.011
& 86.464 $\pm$ 0.021
& 0.309501 $\pm$ 0.000062 \\

Fashion-MNIST
& PATH-CONT
& 80.171 $\pm$ 0.014
& 92.064 $\pm$ 0.010
& 86.465 $\pm$ 0.011
& 0.309489 $\pm$ 0.000049 \\

Fashion-MNIST
& TWO-STAGE-END
& 86.399 $\pm$ 0.184
& 95.209 $\pm$ 0.074
& 95.206 $\pm$ 0.156
& 0.101118 $\pm$ 0.002507 \\
\bottomrule
\end{tabular}
\end{adjustbox}
\end{table}

\begin{figure}[htbp!]
    \centering
    \includegraphics[width=\textwidth]
    {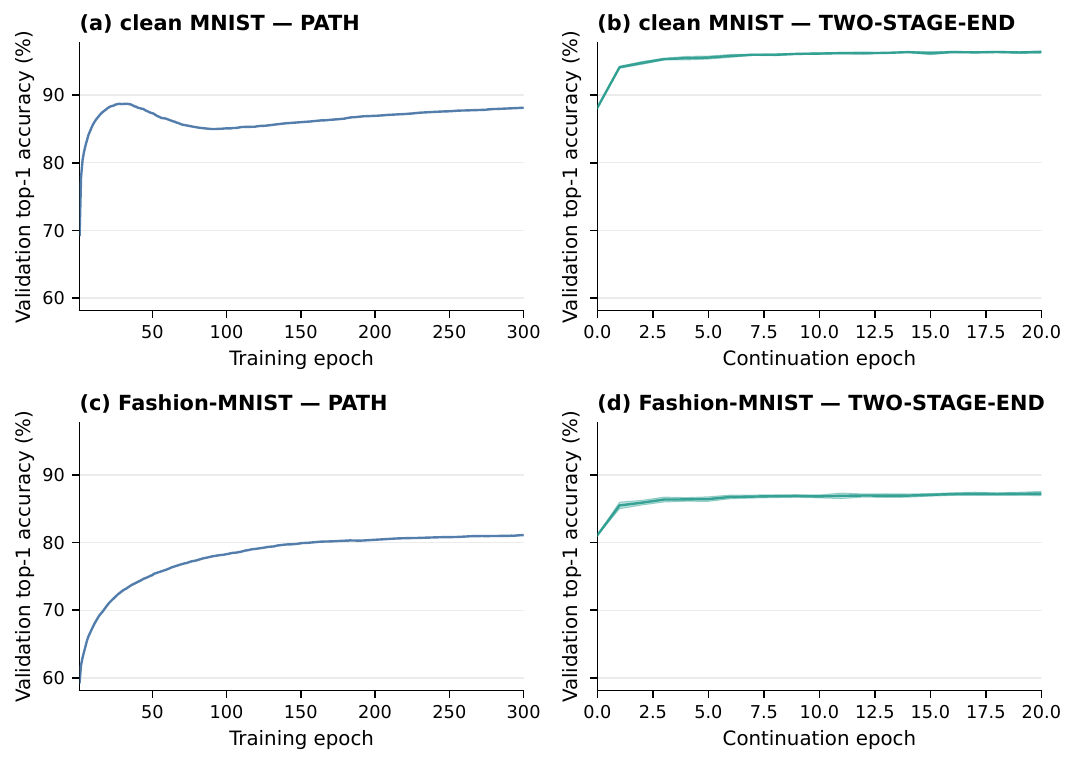}
    \caption{\textbf{Validation accuracy across training-order runs.}
    Validation top-1 accuracy during the $300$-epoch PATH stage and the
    subsequent $20$-epoch TWO-STAGE-END continuation on MNIST and
    Fashion-MNIST. For Stage II, continuation epoch $0$ denotes the
    validation-selected PATH checkpoint before endpoint fine-tuning.
    Solid curves show the mean over ten training-order runs and shaded regions
    show $\pm1$ sample standard deviation.}
    \label{fig:appendix_B1_validation_accuracy}
\end{figure}

Figure~\ref{fig:appendix_B1_validation_accuracy} shows the validation top-1
accuracy throughout Stage I and Stage II. The validation-selected PATH
checkpoint occurs at the final Stage-I epoch for all ten runs on both
datasets. Starting from these checkpoints, TWO-STAGE-END produces a rapid
increase in validation accuracy during the $20$-epoch continuation, with the
selected continuation checkpoints occurring between epochs $14$ and $20$ on
MNIST and between epochs $12$ and $20$ on Fashion-MNIST. The relatively
narrow across-run bands are consistent with the small run-to-run variability
observed in Table~\ref{tab:appendix_B1_multiseed_main_results_lr10}.

\begin{figure}[htbp!]
    \centering
    \includegraphics[width=0.82\textwidth]
    {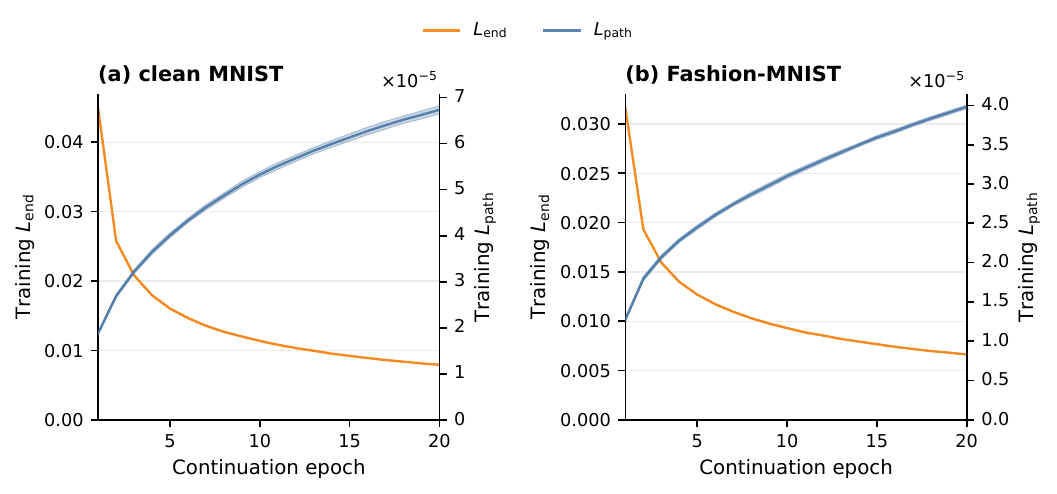}
    \caption{\textbf{Components of the TWO-STAGE-END objective.}
    Path loss $\mathcal L_{\rm path}$ and autonomous endpoint loss
    $\mathcal L_{\rm end}$ during the $20$-epoch Stage-II continuation on
    MNIST and Fashion-MNIST. Curves show means over ten training-order
    runs, with shaded regions denoting $\pm1$ sample standard deviation.}
    \label{fig:appendix_B1_stage2_components}
\end{figure}

To examine how the two forms of supervision evolve after the PATH
pretraining stage, Fig.~\ref{fig:appendix_B1_stage2_components} separates
$\mathcal L_{\rm path}$ and $\mathcal L_{\rm end}$ during TWO-STAGE-END
continuation. The endpoint term is evaluated from an autonomous rollout and
therefore directly penalizes the finite-time mismatch that remains after
teacher-forced path training. Its optimization substantially changes the
autonomous endpoint behavior even though continued minimization of
$\mathcal L_{\rm path}$ alone, as represented by PATH-CONT, produces almost no
change in the final inference metrics. The paired statistical significance and
run-by-run consistency of these differences are quantified in
Appendix~\ref{app:paired_improvements}.

\subsection{Paired Effects Across Training Runs}
\label{app:paired_improvements}

The multi-seed averages in Appendix~\ref{app:multiseed_main_results} establish
the overall performance differences among PATH, PATH-CONT, and
TWO-STAGE-END. We next exploit the paired experimental design to determine
whether these differences are reproduced consistently across changes in
training order. Paired effects, confidence intervals, and sign counts are
computed as defined in Appendix~\ref{app:evaluation_metrics}.

Table~\ref{tab:appendix_B2_paired_improvements_lr10} summarizes three
comparisons. TWO-STAGE-END versus PATH measures the total gain obtained after
introducing Stage-II endpoint supervision. TWO-STAGE-END versus PATH-CONT is
the stricter continuation control, because both methods start from the same
selected PATH checkpoint and undergo matched continuation optimization,
differing only in the inclusion of $\mathcal L_{\rm end}$. Finally,
PATH-CONT versus PATH measures the effect of additional path-only
optimization.

On MNIST, TWO-STAGE-END improves top-1 accuracy over PATH by
$7.599\pm0.059$ percentage points, with a two-sided $95\%$ Student-$t$
interval of $[7.557,\,7.641]$ percentage points. The corresponding
TWO-STAGE-END advantage over PATH-CONT is
$7.599\pm0.065$ percentage points. On Fashion-MNIST, the corresponding
improvements are $6.227\pm0.183$ and $6.228\pm0.181$ percentage points,
respectively. In both datasets, the top-1, top-2, teacher-agreement, and
endpoint-RMSE effects favor TWO-STAGE-END in all ten paired runs.

The continuous endpoint metric shows the same behavior. Relative to PATH,
TWO-STAGE-END reduces output endpoint RMSE by
$0.237345\pm0.001375$ on MNIST and
$0.208383\pm0.002519$ on Fashion-MNIST. Nearly identical reductions are
obtained relative to PATH-CONT. The paired bootstrap intervals closely agree
with the corresponding Student-$t$ intervals, indicating that the conclusion
does not depend on either interval construction.

In contrast, continued path-only optimization produces negligible changes.
The PATH-CONT minus PATH top-1 effects are
$0.000\pm0.012$ percentage points on MNIST and
$-0.001\pm0.011$ percentage points on Fashion-MNIST, while the corresponding
endpoint-RMSE reductions are only $1.5\times10^{-5}$ and
$1.3\times10^{-5}$. Some of these very small paired effects can be resolved
statistically because the across-run variability is also extremely small,
but their magnitudes are negligible relative to the changes produced by
TWO-STAGE-END.

Thus, under matched continuation conditions, the large improvement in
autonomous inference is associated with the introduction of endpoint
supervision rather than with additional optimization of the teacher-forced
path objective.

\begin{table}[htbp!]
\centering
\caption{\textbf{Paired effects across training-order runs.}
Effects are computed within each of ten matched runs before aggregation.
For accuracy and teacher agreement, $A-B$ denotes the metric of method $A$
minus that of method $B$, in percentage points. For endpoint RMSE, the sign is
reversed so that positive values denote a reduction in error for method $A$.
Confidence intervals use paired training runs as the sampling units.}
\label{tab:appendix_B2_paired_improvements_lr10}

\begin{adjustbox}{max width=\textwidth}
\begin{tabular}{lllcccc}
\toprule
Dataset & Comparison & Metric &
Mean $\pm$ SD &
95\% Student-$t$ CI &
95\% bootstrap CI &
$+/0/-$ \\
\midrule

MNIST
& TWO-STAGE-END $-$ PATH
& Top-1
& $+7.599 \pm 0.059$
& $[7.557,\,7.641]$
& $[7.564,\,7.633]$
& 10/0/0 \\

MNIST
& TWO-STAGE-END $-$ PATH
& Top-2
& $+5.057 \pm 0.079$
& $[5.001,\,5.113]$
& $[5.009,\,5.102]$
& 10/0/0 \\

MNIST
& TWO-STAGE-END $-$ PATH
& Teacher agreement
& $+6.777 \pm 0.049$
& $[6.742,\,6.812]$
& $[6.749,\,6.807]$
& 10/0/0 \\

MNIST
& TWO-STAGE-END $-$ PATH
& Output endpoint RMSE reduction
& $+0.237345 \pm 0.001375$
& $[0.236361,\,0.238328]$
& $[0.236579,\,0.238190]$
& 10/0/0 \\

\addlinespace

MNIST
& TWO-STAGE-END $-$ PATH-CONT
& Top-1
& $+7.599 \pm 0.065$
& $[7.553,\,7.645]$
& $[7.560,\,7.636]$
& 10/0/0 \\

MNIST
& TWO-STAGE-END $-$ PATH-CONT
& Top-2
& $+5.062 \pm 0.076$
& $[5.007,\,5.117]$
& $[5.016,\,5.106]$
& 10/0/0 \\

MNIST
& TWO-STAGE-END $-$ PATH-CONT
& Teacher agreement
& $+6.777 \pm 0.054$
& $[6.739,\,6.815]$
& $[6.745,\,6.808]$
& 10/0/0 \\

MNIST
& TWO-STAGE-END $-$ PATH-CONT
& Output endpoint RMSE reduction
& $+0.237330 \pm 0.001370$
& $[0.236350,\,0.238310]$
& $[0.236568,\,0.238164]$
& 10/0/0 \\

\addlinespace

MNIST
& PATH-CONT $-$ PATH
& Top-1
& $+0.000 \pm 0.012$
& $[-0.009,\,0.009]$
& $[-0.006,\,0.008]$
& 2/4/4 \\

MNIST
& PATH-CONT $-$ PATH
& Top-2
& $-0.005 \pm 0.005$
& $[-0.009,\,-0.001]$
& $[-0.008,\,-0.002]$
& 0/5/5 \\

MNIST
& PATH-CONT $-$ PATH
& Teacher agreement
& $+0.000 \pm 0.012$
& $[-0.009,\,0.009]$
& $[-0.006,\,0.008]$
& 2/4/4 \\

MNIST
& PATH-CONT $-$ PATH
& Output endpoint RMSE reduction
& $+0.000015 \pm 0.000012$
& $[0.000006,\,0.000023]$
& $[0.000008,\,0.000022]$
& 8/0/2 \\

\midrule

Fashion-MNIST
& TWO-STAGE-END $-$ PATH
& Top-1
& $+6.227 \pm 0.183$
& $[6.096,\,6.358]$
& $[6.113,\,6.327]$
& 10/0/0 \\

Fashion-MNIST
& TWO-STAGE-END $-$ PATH
& Top-2
& $+3.145 \pm 0.073$
& $[3.093,\,3.197]$
& $[3.103,\,3.190]$
& 10/0/0 \\

Fashion-MNIST
& TWO-STAGE-END $-$ PATH
& Teacher agreement
& $+8.742 \pm 0.169$
& $[8.621,\,8.863]$
& $[8.646,\,8.844]$
& 10/0/0 \\

Fashion-MNIST
& TWO-STAGE-END $-$ PATH
& Output endpoint RMSE reduction
& $+0.208383 \pm 0.002519$
& $[0.206581,\,0.210186]$
& $[0.206931,\,0.209887]$
& 10/0/0 \\

\addlinespace

Fashion-MNIST
& TWO-STAGE-END $-$ PATH-CONT
& Top-1
& $+6.228 \pm 0.181$
& $[6.098,\,6.358]$
& $[6.115,\,6.326]$
& 10/0/0 \\

Fashion-MNIST
& TWO-STAGE-END $-$ PATH-CONT
& Top-2
& $+3.145 \pm 0.070$
& $[3.095,\,3.195]$
& $[3.105,\,3.187]$
& 10/0/0 \\

Fashion-MNIST
& TWO-STAGE-END $-$ PATH-CONT
& Teacher agreement
& $+8.741 \pm 0.162$
& $[8.625,\,8.857]$
& $[8.649,\,8.838]$
& 10/0/0 \\

Fashion-MNIST
& TWO-STAGE-END $-$ PATH-CONT
& Output endpoint RMSE reduction
& $+0.208371 \pm 0.002519$
& $[0.206569,\,0.210172]$
& $[0.206922,\,0.209874]$
& 10/0/0 \\

\addlinespace

Fashion-MNIST
& PATH-CONT $-$ PATH
& Top-1
& $-0.001 \pm 0.011$
& $[-0.009,\,0.007]$
& $[-0.007,\,0.006]$
& 3/2/5 \\

Fashion-MNIST
& PATH-CONT $-$ PATH
& Top-2
& $+0.000 \pm 0.007$
& $[-0.005,\,0.005]$
& $[-0.004,\,0.004]$
& 2/6/2 \\

Fashion-MNIST
& PATH-CONT $-$ PATH
& Teacher agreement
& $+0.001 \pm 0.011$
& $[-0.007,\,0.009]$
& $[-0.005,\,0.008]$
& 4/2/4 \\

Fashion-MNIST
& PATH-CONT $-$ PATH
& Output endpoint RMSE reduction
& $+0.000013 \pm 0.000014$
& $[0.000002,\,0.000023]$
& $[0.000005,\,0.000022]$
& 7/0/3 \\

\bottomrule
\end{tabular}
\end{adjustbox}
\end{table}

\section{Dynamical and Trajectory Analysis}
\label{AppendixE}

\subsection{Output and Hidden Trajectory Reorganization}
\label{app:trajectory_reorganization}

We use the autonomous trajectory metrics defined in
Appendix~\ref{app:evaluation_metrics} to distinguish the evolution of the
output variables from that of the hidden physical representation. The
analysis uses the auxiliary single-run PATH and TWO-STAGE-END checkpoints
obtained with the learning-rate-$30$ continuation protocol. These checkpoints
are used here to examine dynamical behavior rather than to provide the
principal multi-run performance evidence of Appendix~\ref{AppendixB}.

For the $n$-th example, oscillator $i$, and integration time $t_k$, we compare
the physical readouts
\begin{equation*}
    x_{n,i}^{\rm S}(t_k)
    =
    \cos\theta_{n,i}^{\rm S}(t_k),
    \qquad
    x_{n,i}^{\rm TE}(t_k)
    =
    \cos\theta_{n,i}^{\rm TE}(t_k).
\end{equation*}
Trajectory RMSE is evaluated over the complete autonomous rollout and
separately over the hidden, output, and complete node sets. These quantities
measure autonomous trajectory fidelity in the physical activation
representation and are distinct from the teacher-forced phase-increment
objective $\mathcal L_{\rm path}$ used during Stage-I training.

Table~\ref{tab:trajectory_reorganization} reveals a clear separation between
the output and hidden trajectories. On MNIST, TWO-STAGE-END reduces the
output activation-trajectory RMSE from $0.355473$ to $0.182309$, a reduction
of approximately $48.7\%$. In contrast, the hidden activation-trajectory RMSE
increases from $0.183954$ to $0.612566$. The resulting all-node trajectory
RMSE therefore increases from $0.215272$ to $0.573603$.

Fashion-MNIST exhibits the same qualitative behavior. The output
activation-trajectory RMSE decreases from $0.297428$ for PATH to $0.160591$
for TWO-STAGE-END, corresponding to a reduction of approximately $46.0\%$,
whereas the hidden activation-trajectory RMSE increases from $0.137034$ to
$0.495577$. The all-node trajectory RMSE consequently increases from
$0.167915$ to $0.464642$.

Because the hidden, output, and all-node metrics use the same examples and
time points with equal scalar weighting, their squared errors satisfy the
exact decomposition
\begin{equation*}
    \left(
        R_{{\rm traj},x}^{\rm all}
    \right)^2
    =
    \frac{H}{N}
    \left(
        R_{{\rm traj},x}^{\HH}
    \right)^2
    +
    \frac{O}{N}
    \left(
        R_{{\rm traj},x}^{\OO}
    \right)^2,
\end{equation*}
with
\[
    H=64,\qquad O=10,\qquad N=74.
\]
The increase in the all-node error despite improved output alignment is
therefore explained by the substantially larger deviation of the hidden
activation trajectories from their prescribed teacher targets.

This separation reflects the different roles of the two training stages.
Stage I fits the Kuramoto vector field along the prescribed teacher trajectory
through the teacher-forced path objective. Stage II instead optimizes the
terminal output produced by a recursively evolving autonomous system. After
endpoint fine-tuning, improved finite-time inference and closer output
trajectories coexist with substantially greater departure of the hidden
activation trajectories from the transferred teacher representation.

The result therefore does not support an interpretation in which successful
autonomous inference requires reproduction of the complete internal teacher
trajectory. Rather, within the tested protocol, endpoint fine-tuning improves
the task-relevant output dynamics while allowing substantial reorganization
of the hidden physical representation. This observation does not imply that
the reorganized hidden dynamics themselves cause the improvement in
classification performance.

\begin{table}[htbp!]
    \centering
    \caption{\textbf{Autonomous output and hidden trajectory reorganization.}
    Activation-trajectory RMSE is evaluated over the complete
    $10{,}000$-example official test set and the deterministic $K=200$
    autonomous rollout. Output, hidden, and all-node metrics include $10$,
    $64$, and $74$ oscillators, respectively. Results use the auxiliary
    single-run continuation protocol. Lower values indicate closer agreement
    with the prescribed teacher activation trajectory.}
    \label{tab:trajectory_reorganization}
    \small
    \setlength{\tabcolsep}{5pt}
    \begin{tabular}{llrr}
        \toprule
        Dataset & Metric & PATH & TWO-STAGE-END \\
        \midrule

        MNIST
        & Output trajectory RMSE
        & $0.355473$ & $0.182309$ \\

        & Hidden trajectory RMSE
        & $0.183954$ & $0.612566$ \\

        & All-node trajectory RMSE
        & $0.215272$ & $0.573603$ \\

        \midrule

        Fashion-MNIST
        & Output trajectory RMSE
        & $0.297428$ & $0.160591$ \\

        & Hidden trajectory RMSE
        & $0.137034$ & $0.495577$ \\

        & All-node trajectory RMSE
        & $0.167915$ & $0.464642$ \\

        \bottomrule
    \end{tabular}
\end{table}

\subsection{Time-Resolved Autonomous Trajectory Errors}
\label{app:time_resolved_errors}

The aggregate trajectory errors in
Table~\ref{tab:trajectory_reorganization} summarize deviations over the
complete autonomous rollout but do not show how the separation between output
alignment and hidden-state reorganization develops in time. We therefore
evaluate the activation-space error independently at each integration time
point using the same auxiliary single-run checkpoints as in
Appendix~\ref{app:trajectory_reorganization}.

For a node subset $\mathcal I$, define the time-resolved activation error
\begin{equation*}
    e_x^{\mathcal I}(t_k)
    =
    \left[
        \frac{1}{M|\mathcal I|}
        \sum_{n=1}^{M}
        \sum_{i\in\mathcal I}
        \left(
            \cos\theta_{n,i}^{\rm S}(t_k)
            -
            \cos\theta_{n,i}^{\rm TE}(t_k)
        \right)^2
    \right]^{1/2},
\end{equation*}
where $M$ is the number of evaluation examples. We consider the hidden and
output subsets
\begin{equation*}
    \mathcal I^\HH
    =
    \{1,2\ldots,H\},
    \qquad
    \mathcal I^\OO
    =
    \{H+1,H+2\ldots,H+O\}.
\end{equation*}
The quantities $e_x^{\mathcal I^\HH}(t_k)$ and
$e_x^{\mathcal I^\OO}(t_k)$ therefore measure the instantaneous deviations
of the autonomous hidden and output activations from their prescribed teacher
trajectories.

\begin{figure*}[htbp!]
    \centering
    \includegraphics[width=0.75\textwidth]
    {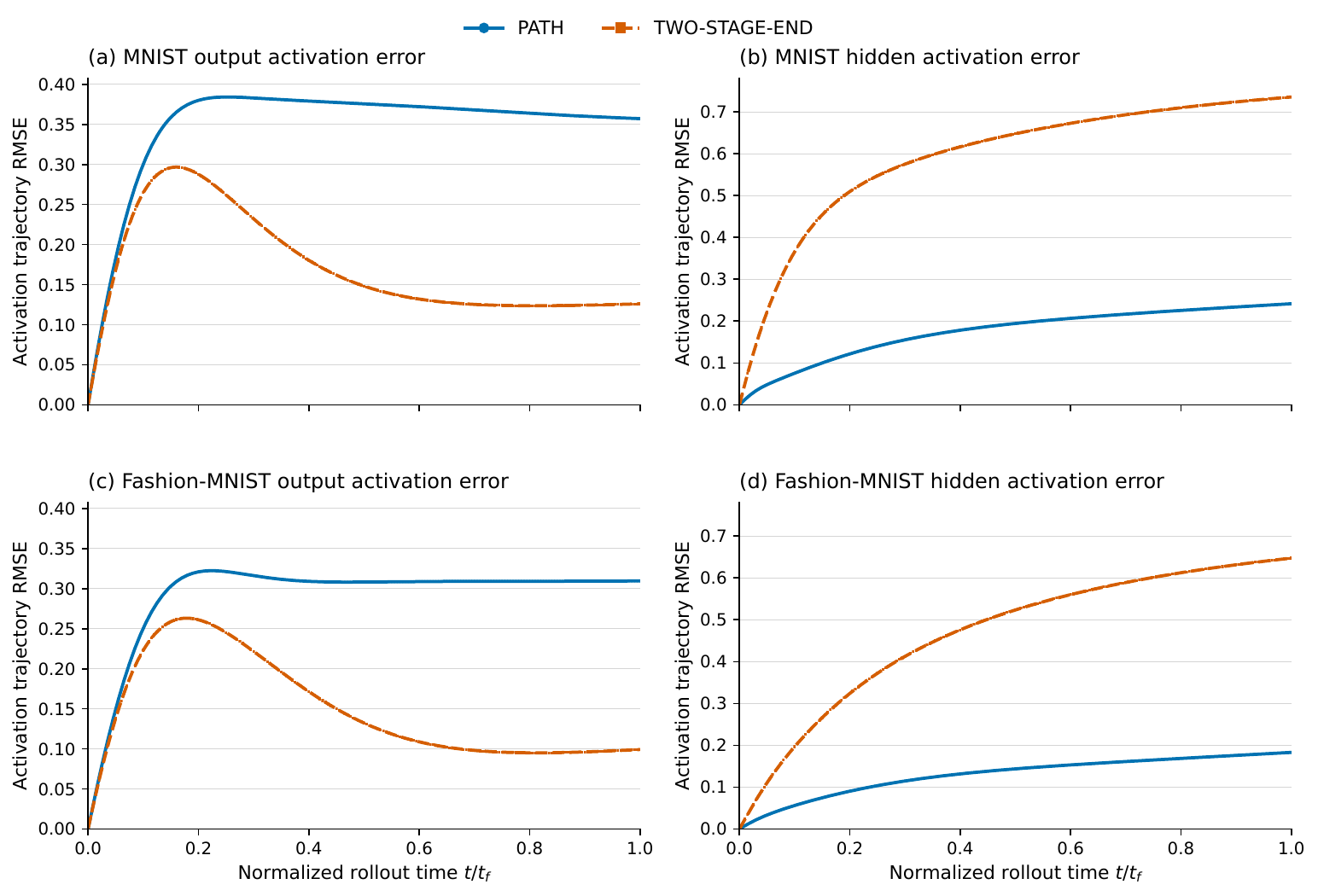}
    \caption{\textbf{Time-resolved autonomous trajectory errors.}
    Output and hidden activation-space RMSE as functions of normalized rollout
    time $t/t_{\rm f}$ for PATH and TWO-STAGE-END on MNIST and Fashion-MNIST.
    Curves are evaluated over the complete official test set and the indicated
    node subset. Throughout the postinitial rollout, TWO-STAGE-END remains
    closer to the prescribed teacher trajectory at the output nodes while
    deviating more strongly at the hidden nodes. Lower values indicate closer
    activation-space agreement with the teacher trajectory.}
    \label{fig:time_resolved_autonomous_errors}
\end{figure*}

Figure~\ref{fig:time_resolved_autonomous_errors} shows these errors over the
complete deterministic rollout as functions of normalized time $t/t_{\rm f}$.
Because the student and teacher share the initial condition
$\theta_i(0)=\pi/2$, both hidden and output activation errors vanish at
$t=0$.

The output trajectories show a consistent advantage for TWO-STAGE-END. At
every postinitial time point,
\begin{equation*}
    e_x^{\mathcal I^\OO,{\rm TWO}}(t_k)
    <
    e_x^{\mathcal I^\OO,{\rm PATH}}(t_k),
    \qquad
    k=1,2,\ldots,K.
\end{equation*}
Thus, the improved output alignment produced by endpoint fine-tuning is not
confined to the terminal readout; it is already present early in the
autonomous evolution and persists throughout the finite-time computation.

The hidden trajectories exhibit the opposite ordering:
\begin{equation*}
    e_x^{\mathcal I^\HH,{\rm TWO}}(t_k)
    >
    e_x^{\mathcal I^\HH,{\rm PATH}}(t_k),
    \qquad
    k=1,2,\ldots,K.
\end{equation*}
The greater departure of the hidden activations from the prescribed teacher
representation therefore also develops throughout the autonomous rollout
rather than appearing only near the terminal time.

The aggregate activation-trajectory RMSE reported in
Table~\ref{tab:trajectory_reorganization} is related to the time-resolved
error by
\begin{equation*}
    \left(
        R_{{\rm traj},x}^{\mathcal I}
    \right)^2
    =
    \frac{1}{K+1}
    \sum_{k=0}^{K}
    \left[
        e_x^{\mathcal I}(t_k)
    \right]^2.
\end{equation*}
Hence, the aggregate trajectory metric is the root-mean-square temporal
summary of the instantaneous deviations shown in
Fig.~\ref{fig:time_resolved_autonomous_errors}.

Together with Appendix~\ref{app:trajectory_reorganization}, these results show
that endpoint fine-tuning changes the autonomous computation throughout the
observation interval: the task-relevant output representation becomes more
closely aligned with the teacher trajectory, while the hidden representation
simultaneously departs farther from it. The improvement is therefore not
consistent with a picture in which Stage II modifies only the terminal readout
while otherwise preserving the Stage-I trajectory.

\subsection{Representative Autonomous Output Dynamics}
\label{app:representative_output_dynamics}

To visualize how class decisions emerge during finite-time autonomous
inference, we plot the ten output activations
\begin{equation*}
    x_{n,o}^{\rm S,\OO}(t)
    =
    \cos\theta_{n,o}^{\rm S,\OO}(t),
    \qquad
    o=1,2\ldots,O,
\end{equation*}
for representative correctly classified and misclassified test examples.
The figures use the same auxiliary single-run TWO-STAGE-END checkpoints as
the trajectory analyses in Appendices~\ref{app:trajectory_reorganization}
and \ref{app:time_resolved_errors}. Each panel also displays the corresponding
input image, and the predicted class is determined directly by the largest
output activation at $t_{\rm f}$, without an additional trainable readout.

Examples are chosen using fixed, class-balanced deterministic rules to avoid
selection based on visual inspection of the trajectories. For correctly
classified examples, candidates within each true class are ranked according
to how close their terminal classification margin is to the class-specific
median margin, where the margin is the difference between the largest and
second-largest terminal outputs. Twenty-five examples are selected from each
dataset with all ten classes represented.

For misclassified examples, we restrict attention to cases in which the true
class has the second-largest terminal output. Within each true class, an
example near the class-specific median separation between the predicted and
true-class terminal outputs is selected. This gives one representative
misclassified example per true class and ten examples per dataset. These
selection rules are used only for visualization and do not enter any
quantitative result reported in the paper.

\begin{figure*}[htbp!]
    \centering
    \includegraphics[width=\textwidth]
    {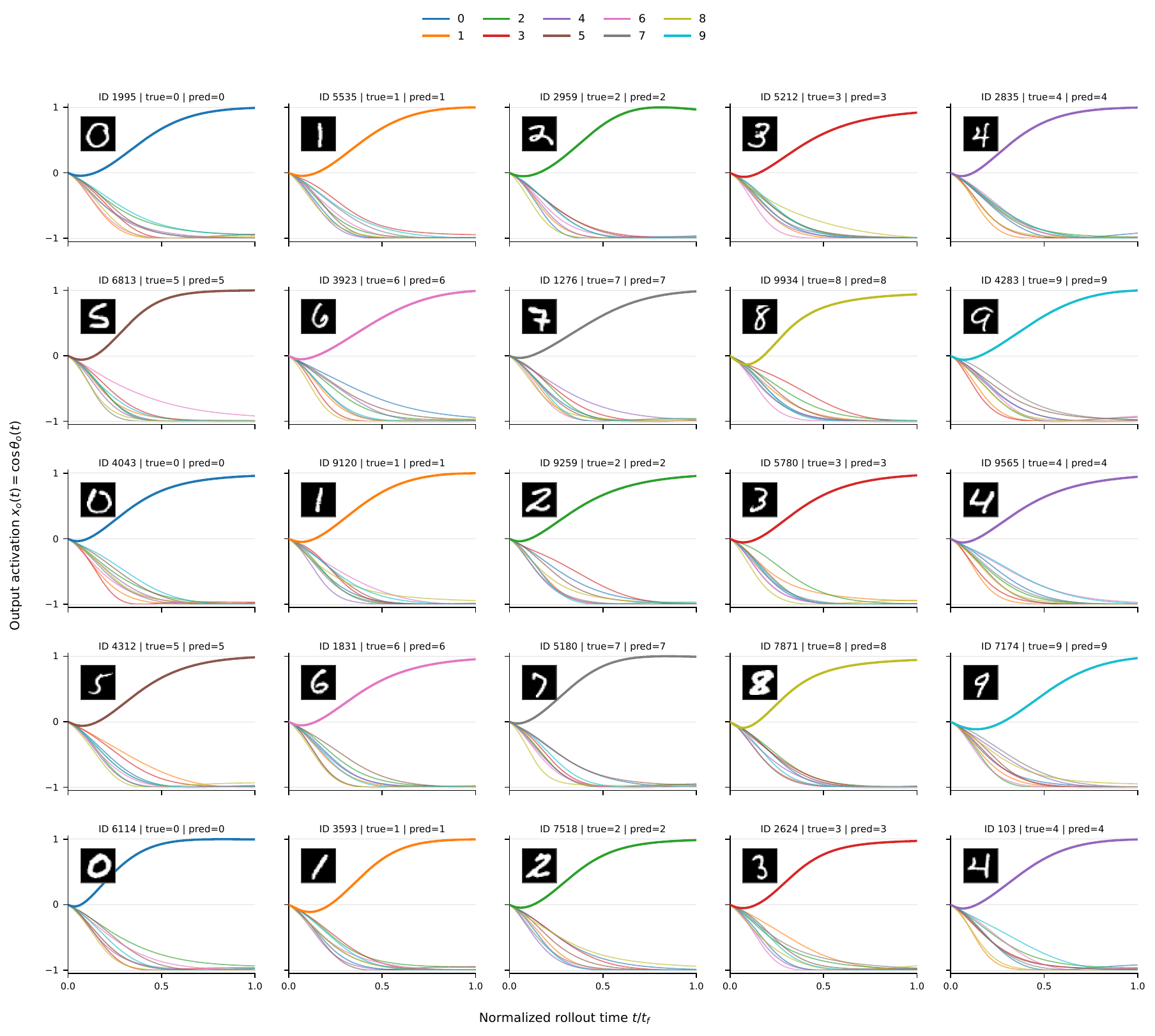}
    \caption{\textbf{Representative correctly classified MNIST output
    dynamics.}
    Twenty-five class-balanced test examples evaluated with TWO-STAGE-END.
    Each panel shows the input image and all ten autonomous output activations
    $\cos\theta_o(t)$ over normalized rollout time. Colors identify digit
    classes $0$--$9$ according to the shared legend. Examples are selected
    using the fixed class-balanced procedure described in the text.}
    \label{fig:mnist_correct_output_dynamics}
\end{figure*}

\begin{figure*}[htbp!]
    \centering
    \includegraphics[width=\textwidth]
    {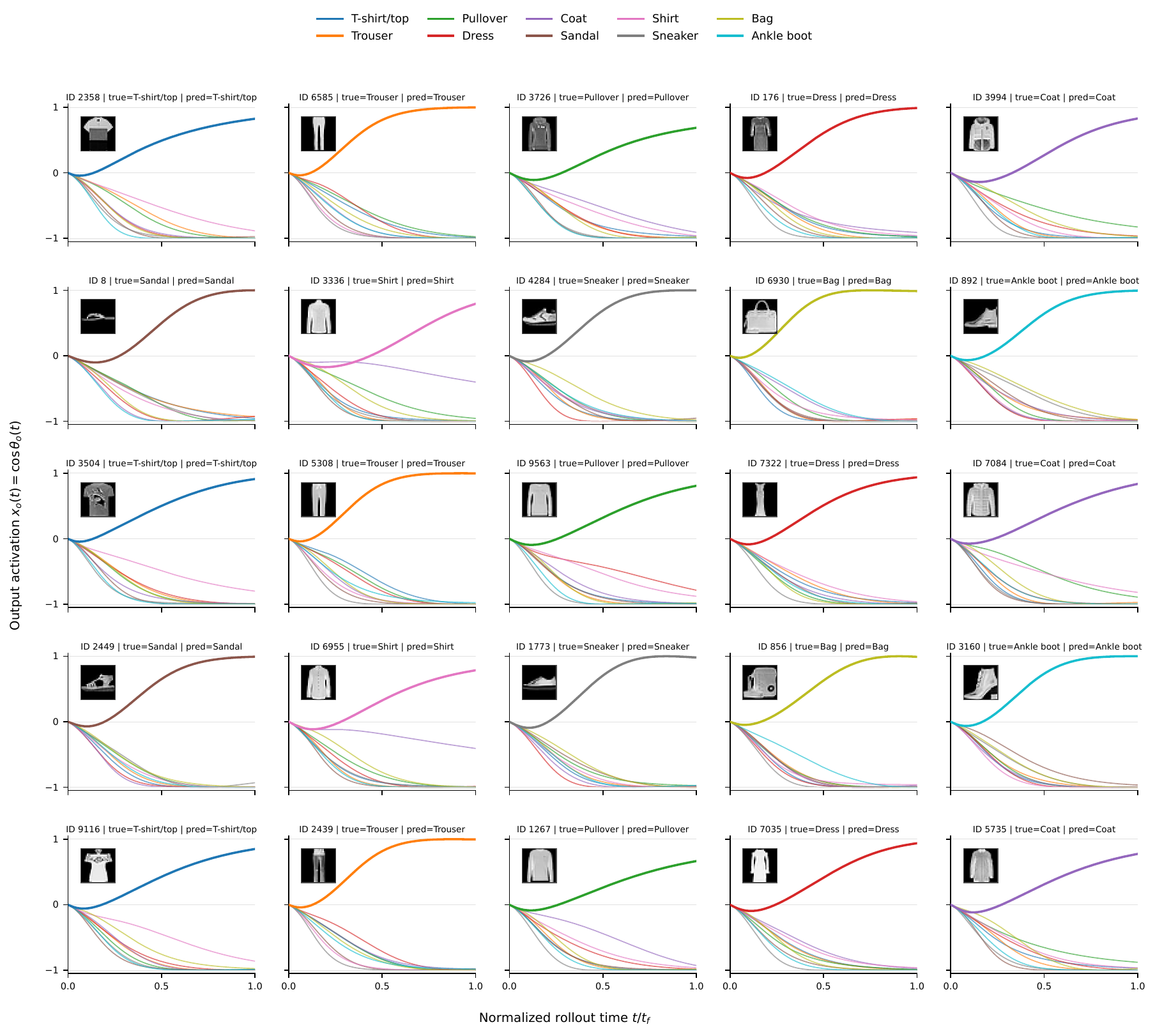}
    \caption{\textbf{Representative correctly classified Fashion-MNIST output
    dynamics.}
    Twenty-five class-balanced test examples evaluated with TWO-STAGE-END.
    Each panel shows the input image and all ten autonomous output
    activations over normalized rollout time. The shared legend identifies
    the ten Fashion-MNIST classes. Examples are selected using the same fixed
    class-balanced procedure as for MNIST.}
    \label{fig:fashion_correct_output_dynamics}
\end{figure*}

Figures~\ref{fig:mnist_correct_output_dynamics} and
\ref{fig:fashion_correct_output_dynamics} illustrate the formation of
finite-time class decisions for correctly classified examples. All output
activations begin from the common zero-readout state and evolve continuously
under the autonomous Kuramoto dynamics. During the rollout, the relative
ordering of the class outputs develops until the correct class attains the
largest activation at the observation time.

\begin{figure*}[htbp!]
    \centering
    \includegraphics[width=\textwidth]
    {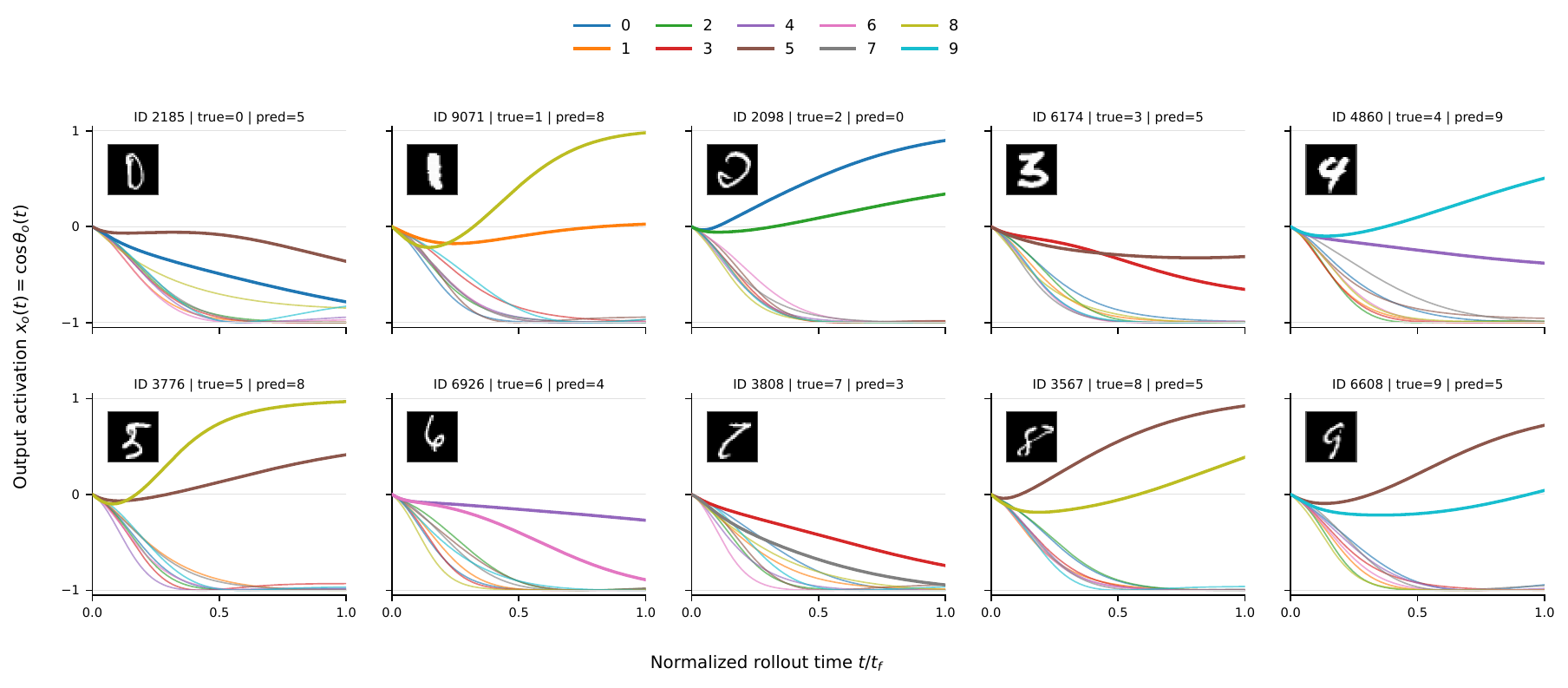}
    \caption{\textbf{Representative misclassified MNIST output
    dynamics.}
    Ten class-balanced test examples, one from each true digit class,
    evaluated with TWO-STAGE-END. Each panel shows the input image and all ten
    autonomous output trajectories; panel titles identify the true and
    predicted classes. The displayed cases are selected among examples for
    which the true class has the second-largest terminal output.}
    \label{fig:mnist_misclassified_output_dynamics}
\end{figure*}

\begin{figure*}[htbp!]
    \centering
    \includegraphics[width=\textwidth]
    {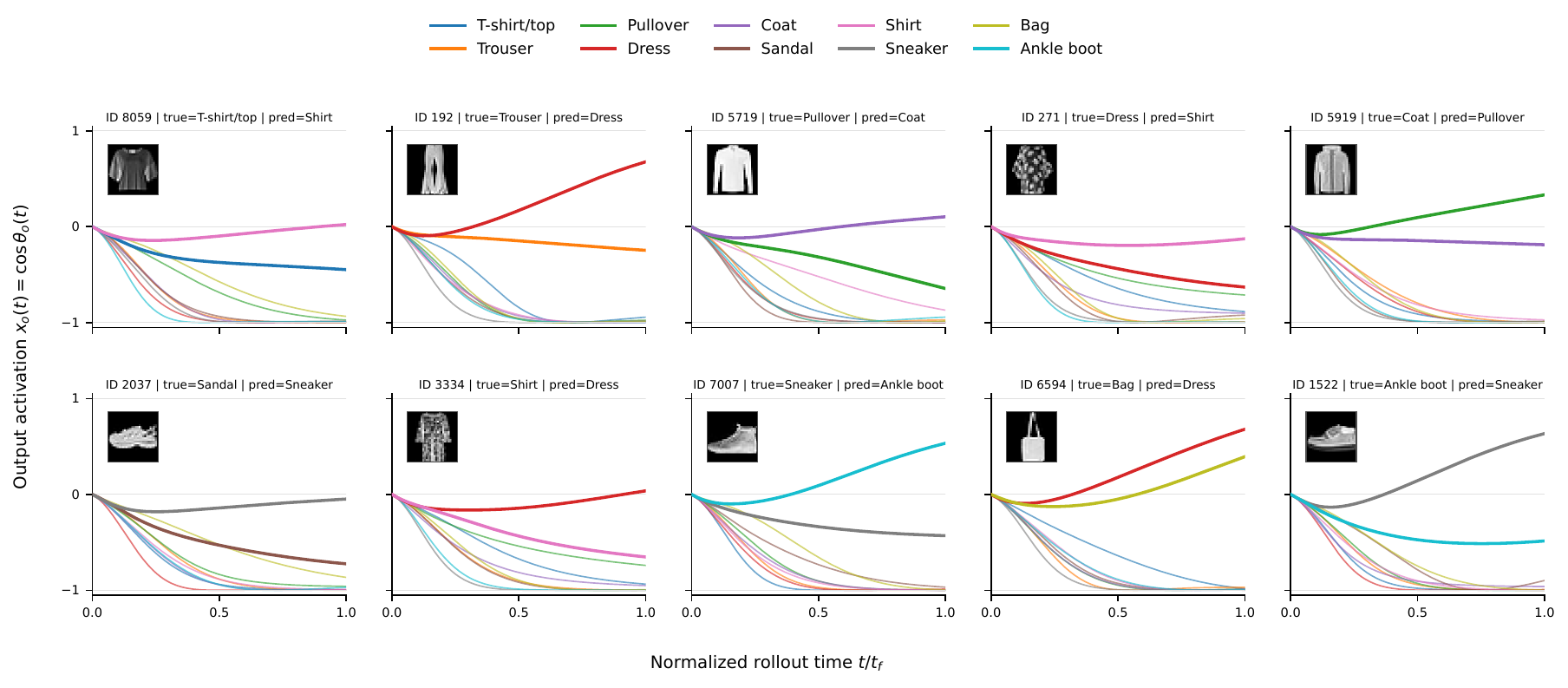}
    \caption{\textbf{Representative misclassified Fashion-MNIST output
    dynamics.}
    Ten class-balanced test examples, one from each true Fashion-MNIST class,
    evaluated with TWO-STAGE-END. Each panel shows the input image and all ten
    autonomous output trajectories; panel titles identify the true and
    predicted classes. As for MNIST, the displayed cases are selected
    among examples for which the true class has the second-largest terminal
    output.}
    \label{fig:fashion_misclassified_output_dynamics}
\end{figure*}

The misclassified examples in
Figs.~\ref{fig:mnist_misclassified_output_dynamics} and
\ref{fig:fashion_misclassified_output_dynamics} illustrate close finite-time
competition between output trajectories. In these selected cases, the true
class remains the second-largest terminal response, but a competing output
finishes higher at $t_{\rm f}$. The classification error is therefore determined by
the terminal ordering of continuously evolving class outputs, illustrating
the finite-time dynamical nature of the readout.

\section{Robustness to Neural-Teacher Architecture}
\label{AppendixC}
\subsection{Teacher-Architecture Sweep Design}
\label{app:fifteen_teacher_architectures}

To test whether the benefit of autonomous endpoint fine-tuning depends on a
particular neural-teacher representation, we perform a controlled
teacher-architecture sweep on Fashion-MNIST. The sweep contains fifteen fully
connected teachers with one to four hidden layers. For every architecture, the
total number of transferred hidden variables is fixed at
\begin{equation*}
    \sum_{\ell=1}^{L} h_\ell = 64,
\end{equation*}
where $L$ is the number of hidden layers and $h_\ell$ is the width of hidden
layer $\ell$. Each teacher therefore provides exactly $64$ hidden targets and
$10$ output targets to the same Kuramoto student with
\[
    H=64,\qquad O=10,\qquad N=74.
\]

All teachers use the tanh hidden activations and signed-softmax output
representation defined in Appendix~\ref{app:neural_teacher_training}. For a
teacher with hidden widths $(h_1,h_2\ldots,h_L)$, the transferred hidden state is
formed by concatenating the hidden-layer activations in network order,
\begin{equation*}
    a^{\rm TE,\HH}
    =
    \begin{bmatrix}
        h^{(1)}\\
        \vdots\\
        h^{(L)}
    \end{bmatrix}
    \in[-1,1]^{64}.
\end{equation*}
Together with the $10$ transferred output activations,
\begin{equation*}
    a^{\rm TE}
    =
    \begin{bmatrix}
        a^{\rm TE,\HH}\\
        a^{\rm TE,\OO}
    \end{bmatrix}
    \in[-1,1]^{74}.
\end{equation*}
Thus, every transferred hidden activation is assigned one-to-one to a hidden
oscillator. 

The fifteen architectures are listed in
Table~\ref{tab:fifteen_teacher_architectures}. Their labels indicate the
number and widths of the hidden layers: the prefixes \texttt{S}, \texttt{D},
\texttt{T}, and \texttt{Q} denote one, two, three, and four hidden layers,
respectively. For example, \texttt{D48\_16} denotes the architecture
$784$--$48$--$16$--$10$, whereas \texttt{T40\_08\_16} denotes
$784$--$40$--$8$--$16$--$10$.

The architecture set includes both changes in depth and contrasting
allocations of the same $64$ hidden variables. For example,
\texttt{D48\_16} and \texttt{D16\_48} reverse the width allocation of a
two-hidden-layer teacher, while \texttt{T40\_16\_08} and
\texttt{T40\_08\_16} exchange the widths of the final two hidden layers.
The sweep therefore changes the organization of the transferred neural
representation while keeping the number of physical hidden oscillators fixed.

A separate neural teacher is trained for each architecture using the common
Fashion-MNIST preprocessing and teacher-training procedures described in
Appendices~\ref{app:data_preprocessing} and
\ref{app:neural_teacher_training}. The corresponding Kuramoto students all use
the same $H=64$, $O=10$ physical architecture and numerical configuration
defined in Appendix~\ref{app:kuramoto_training}. A fixed training ordering is
used across the architecture sweep so that changes between configurations are
not intentionally accompanied by changes in minibatch order.

Each architecture is first trained with Stage-I PATH from exact-zero physical
parameters. TWO-STAGE-END then continues from the architecture-specific
validation-selected PATH checkpoint using
\begin{equation*}
    \mathcal L
    =
    \mathcal L_{\rm path}
    +
    \mathcal L_{\rm end}.
\end{equation*}
The architecture sweep uses the auxiliary single-run continuation protocol
with learning rate $30$, distinct from the learning-rate-$10$ multi-run
protocol used for the principal results in Appendix~\ref{AppendixB}.
Checkpoint selection follows Appendix~\ref{app:checkpoint_selection}.

The neural-teacher parameter count is not held fixed. In particular, changing
the first hidden-layer width changes the number of weights connecting the
$784$-dimensional input to the teacher representation. The sweep therefore
does not isolate teacher depth or parameter count as independent causal
variables. Rather, it tests whether the effect of endpoint fine-tuning
persists across substantially different teacher depths and layerwise
organizations while the physical Kuramoto student and the number of
transferred variables remain fixed.

The fifteen architectures are deliberately different teacher--student
configurations rather than repeated realizations of a single architecture.
Architecture-level changes and consistency across the tested set are
therefore interpreted descriptively rather than as independent statistical
replicates.

\begin{table*}[htbp!]
    \centering
    \caption{\textbf{Neural-teacher architectures in the Fashion-MNIST
    robustness sweep.}
    Every architecture transfers exactly $64$ hidden activations and $10$
    output activations to the same $74$-oscillator Kuramoto student. Teacher
    parameter counts include all affine weights and biases and are not held
    fixed across architectures.}
    \label{tab:fifteen_teacher_architectures}
    \small
    \begin{tabular}{llrrr}
        \toprule
        Family
        & Identifier
        & Hidden widths
        & Hidden depth
        & Teacher parameters \\
        \midrule

        Single
        & \texttt{S64}
        & $(64)$
        & $1$
        & $50{,}890$ \\

        \midrule
        Double
        & \texttt{D32\_32}
        & $(32,32)$
        & $2$
        & $26{,}506$ \\

        Double
        & \texttt{D48\_16}
        & $(48,16)$
        & $2$
        & $38{,}634$ \\

        Double
        & \texttt{D16\_48}
        & $(16,48)$
        & $2$
        & $13{,}866$ \\

        Double
        & \texttt{D40\_24}
        & $(40,24)$
        & $2$
        & $32{,}634$ \\

        Double
        & \texttt{D52\_12}
        & $(52,12)$
        & $2$
        & $41{,}586$ \\

        Double
        & \texttt{D56\_08}
        & $(56,8)$
        & $2$
        & $44{,}506$ \\

        \midrule
        Triple
        & \texttt{T22\_21\_21}
        & $(22,21,21)$
        & $3$
        & $18{,}435$ \\

        Triple
        & \texttt{T32\_16\_16}
        & $(32,16,16)$
        & $3$
        & $26{,}090$ \\

        Triple
        & \texttt{T40\_16\_08}
        & $(40,16,8)$
        & $3$
        & $32{,}282$ \\

        Triple
        & \texttt{T40\_08\_16}
        & $(40,8,16)$
        & $3$
        & $32{,}042$ \\

        Triple
        & \texttt{T32\_24\_08}
        & $(32,24,8)$
        & $3$
        & $26{,}202$ \\

        Triple
        & \texttt{T24\_24\_16}
        & $(24,24,16)$
        & $3$
        & $20{,}010$ \\

        Triple
        & \texttt{T16\_24\_24}
        & $(16,24,24)$
        & $3$
        & $13{,}818$ \\

        \midrule
        Quadruple
        & \texttt{Q16\_16\_16\_16}
        & $(16,16,16,16)$
        & $4$
        & $13{,}546$ \\

        \bottomrule
    \end{tabular}
\end{table*}

\subsection{Robustness Across Teacher Architectures}
\label{app:teacher_architecture_results}

\begin{figure*}[htbp!]
    \centering
    \includegraphics[width=\textwidth]
    {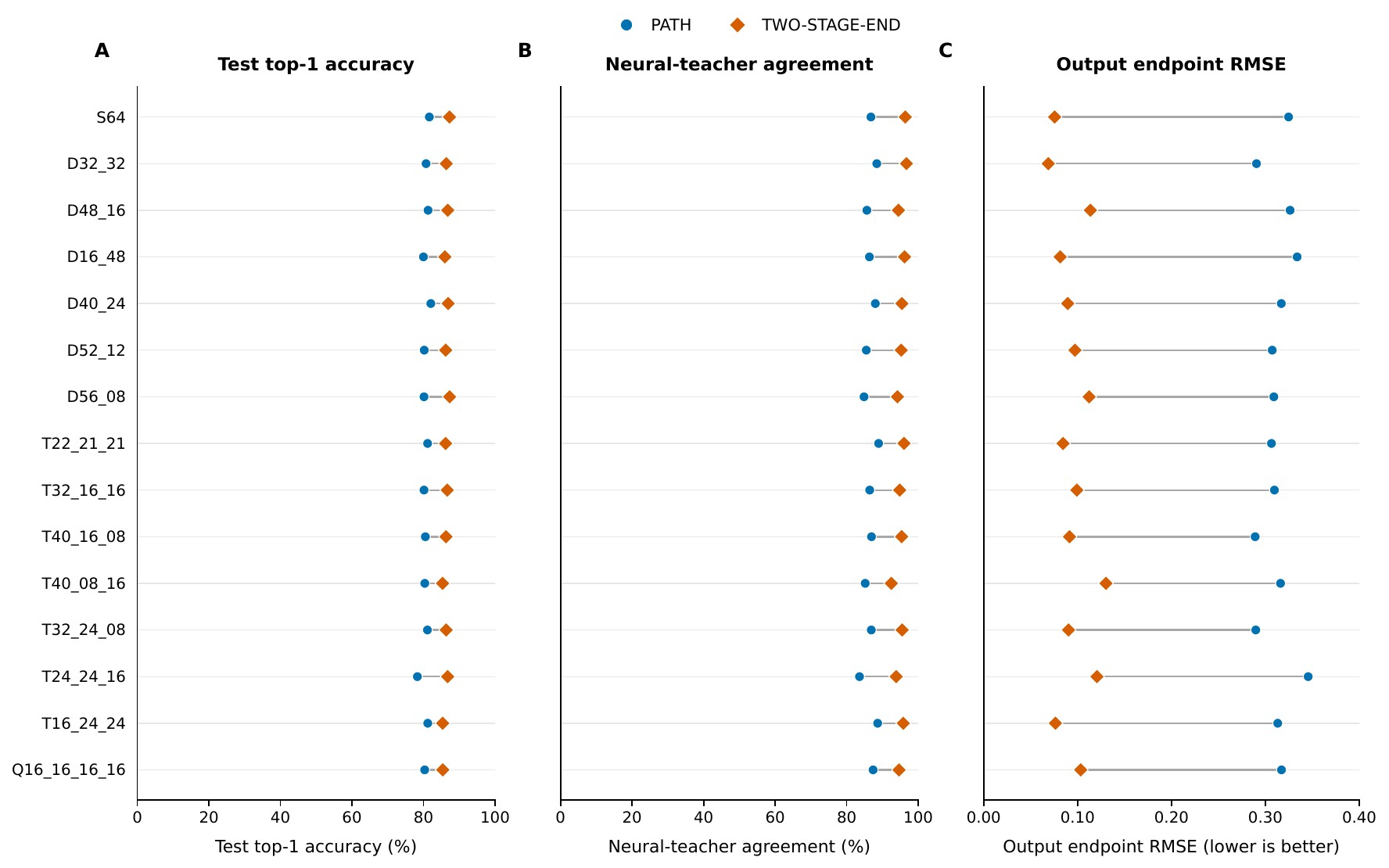}
    \caption{\textbf{Robustness of endpoint fine-tuning across neural-teacher
    architectures.}
    Paired PATH and TWO-STAGE-END results for fifteen Fashion-MNIST teacher
    architectures: (a) test top-1 accuracy, (b) neural-teacher agreement, and
    (c) output endpoint RMSE. Each line corresponds to one teacher
    architecture. TWO-STAGE-END improves top-1 accuracy and teacher agreement
    and reduces endpoint RMSE for all fifteen tested configurations.}
    \label{fig:teacher_architecture_complete_results}
\end{figure*}

Table~\ref{tab:appendix-c2-teacher-architecture-complete} and
Fig.~\ref{fig:teacher_architecture_complete_results} compare PATH and
TWO-STAGE-END across the fifteen Fashion-MNIST teacher architectures defined
in Appendix~\ref{app:fifteen_teacher_architectures}. Each architecture
constitutes a distinct teacher--student configuration, while the physical
Kuramoto student size and the number of transferred variables are held fixed.

The effect of endpoint fine-tuning is consistent across the complete
architecture set. TWO-STAGE-END improves test top-1 accuracy for all fifteen
teachers. The descriptive mean improvement is $5.73$ percentage points, with
architecture-level gains ranging from $4.16$ to $8.47$ percentage points.
Thus, the improvement is not restricted to the teacher architecture used in
the principal experiment, but persists across the tested changes in teacher
depth and hidden-layer allocation.

Neural-teacher agreement shows the same directional consistency:
TWO-STAGE-END increases agreement for all fifteen architectures, with a
descriptive mean gain of $8.50$ percentage points and individual gains ranging
from $7.10$ to $10.23$ percentage points. For every architecture, the increase
in teacher agreement exceeds the corresponding increase in ground-truth
top-1 accuracy. This is consistent with the different roles of the two
metrics: endpoint fine-tuning directly improves reproduction of the teacher
output, whereas agreement with the teacher does not necessarily translate
one-to-one into additional ground-truth accuracy.

Continuous endpoint fidelity also improves in every tested configuration.
For all fifteen architectures, $R_{\rm end}^{\OO,{\rm TWO}}<R_{\rm end}^{\OO,{\rm PATH}}$.

The simultaneous improvement in classification accuracy, teacher agreement,
and continuous endpoint matching indicates that the effect is not a
consequence of isolated changes in the final hard decision.


\begin{table*}[htbp!]
\centering
\caption{\textbf{Fashion-MNIST teacher-architecture robustness sweep.}
Results are from the common auxiliary single-run continuation protocol defined
in Appendix~\ref{app:fifteen_teacher_architectures}. $\Delta$ denotes
TWO-STAGE-END minus PATH for top-1 accuracy and neural-teacher agreement.
Lower output endpoint RMSE is better.}
\label{tab:appendix-c2-teacher-architecture-complete}
\small
\setlength{\tabcolsep}{3.0pt}
\resizebox{\textwidth}{!}{%
\begin{tabular}{lrrrrrrrrr}
\toprule
Architecture
& Teacher
& \multicolumn{3}{c}{Test top-1 accuracy}
& \multicolumn{3}{c}{Neural-teacher agreement}
& \multicolumn{2}{c}{Output endpoint RMSE} \\
\cmidrule(lr){2-2}
\cmidrule(lr){3-5}
\cmidrule(lr){6-8}
\cmidrule(lr){9-10}
& Top-1 (\%)
& PATH (\%)
& TWO-STAGE-END (\%)
& $\Delta$ (pp)
& PATH (\%)
& TWO-STAGE-END (\%)
& $\Delta$ (pp)
& PATH
& TWO-STAGE-END \\
\midrule

S64
& 87.29 & 81.62 & 87.24 & +5.62
& 86.77 & 96.39 & +9.62
& 0.324676 & 0.075511 \\

D32\_32
& 86.15 & 80.71 & 86.35 & +5.64
& 88.41 & 96.70 & +8.29
& 0.290514 & 0.068691 \\

D48\_16
& 86.71 & 81.27 & 86.75 & +5.48
& 85.64 & 94.47 & +8.83
& 0.326297 & 0.113679 \\

D16\_48
& 85.46 & 79.97 & 85.96 & +5.99
& 86.35 & 96.16 & +9.81
& 0.333781 & 0.081464 \\

D40\_24
& 86.69 & 82.04 & 86.88 & +4.84
& 88.00 & 95.42 & +7.42
& 0.316916 & 0.089550 \\

D52\_12
& 86.14 & 80.19 & 86.19 & +6.00
& 85.48 & 95.23 & +9.75
& 0.307263 & 0.097198 \\

D56\_08
& 86.69 & 80.13 & 87.30 & +7.17
& 84.85 & 94.18 & +9.33
& 0.308989 & 0.112328 \\

T22\_21\_21
& 85.64 & 81.14 & 86.17 & +5.03
& 88.92 & 96.02 & +7.10
& 0.306453 & 0.084495 \\

T32\_16\_16
& 85.79 & 80.13 & 86.63 & +6.50
& 86.43 & 94.81 & +8.38
& 0.309598 & 0.099203 \\

T40\_16\_08
& 85.97 & 80.51 & 86.28 & +5.77
& 86.93 & 95.36 & +8.43
& 0.289078 & 0.091387 \\

T40\_08\_16
& 84.62 & 80.35 & 85.32 & +4.97
& 85.17 & 92.46 & +7.29
& 0.316094 & 0.130139 \\

T32\_24\_08
& 86.10 & 81.09 & 86.34 & +5.25
& 86.87 & 95.49 & +8.62
& 0.289721 & 0.090320 \\

T24\_24\_16
& 86.09 & 78.28 & 86.75 & +8.47
& 83.58 & 93.81 & +10.23
& 0.345565 & 0.120578 \\

T16\_24\_24
& 84.91 & 81.18 & 85.34 & +4.16
& 88.65 & 95.80 & +7.15
& 0.313109 & 0.076402 \\

Q16\_16\_16\_16
& 84.32 & 80.32 & 85.36 & +5.04
& 87.38 & 94.62 & +7.24
& 0.317103 & 0.103331 \\

\bottomrule
\end{tabular}%
}
\end{table*}

\section{Training-Protocol Ablations}
\label{AppendixD}

\subsection{Roles of Path and Endpoint Supervision}
\label{app:full_protocol_ablation}

We use a six-method ablation to distinguish the roles of teacher-forced path
supervision, autonomous endpoint supervision, and staged initialization.
The methods are defined in Appendix~\ref{app:checkpoint_selection}.
This experiment uses the auxiliary single-run protocol with learning rate
$30$ for the continuation and direct-training branches, and is therefore
separate from the principal multi-run continuation study in
Appendix~\ref{AppendixB}. Table~\ref{tab:protocol_ablation} summarizes the
autonomous test performance on MNIST and Fashion-MNIST.

The first comparison asks whether the improvement observed after Stage I can
be explained simply by additional optimization. PATH-CONT continues
optimization of $\mathcal L_{\rm path}$ from the selected PATH checkpoint,
but produces essentially no change in autonomous inference on either dataset.
On MNIST, top-1 accuracy changes from $89.13\%$ for PATH to
$89.12\%$ for PATH-CONT; on Fashion-MNIST, both obtain $80.13\%$.
Their teacher agreement and output endpoint RMSE are likewise nearly
unchanged. Additional path-only optimization is therefore insufficient to
produce the large continuation-stage improvement.

The direct-training controls reveal a different role for path supervision
when optimization begins from exact-zero physical parameters.
ZERO-END-ONLY optimizes $\mathcal L_{\rm end}$ alone from zero, whereas
Z-END optimizes
$\mathcal L_{\rm path}+\mathcal L_{\rm end}$ under matched direct-training
conditions. On MNIST, ZERO-END-ONLY reaches only $20.88\%$ top-1
accuracy, compared with $95.84\%$ for Z-END. The same qualitative contrast
appears on Fashion-MNIST, where the corresponding accuracies are $18.65\%$
and $86.11\%$. Endpoint RMSE and neural-teacher agreement change consistently
with this large difference in task performance. Under the tested
initialization and optimization protocol, endpoint supervision alone therefore
does not effectively bootstrap the Kuramoto model from zero, whereas the path
objective provides strong optimization guidance.

Once Stage-I path training has produced a usable physical model, however, the
role of $\mathcal L_{\rm path}$ changes. PRETRAIN-END-ONLY and
TWO-STAGE-END begin from the same selected PATH checkpoint and differ in
whether the path loss remains active during continuation. Their resulting
performance is very similar on both datasets. On MNIST,
PRETRAIN-END-ONLY and TWO-STAGE-END reach $96.61\%$ and $96.47\%$ top-1
accuracy, respectively, while on Fashion-MNIST they reach $86.58\%$ and
$86.63\%$. Their endpoint RMSE values are likewise close. Within this
single-run ablation, there is therefore no clear evidence that retaining
$\mathcal L_{\rm path}$ during Stage II is necessary for the task-level gain
once path pretraining has already established a suitable initialization.

These comparisons distinguish two empirical roles of the training objectives.
Path supervision is important for constructing a trainable dynamical
initialization from exact-zero parameters, whereas autonomous endpoint
supervision is the principal driver of the subsequent improvement in
finite-time inference. Continued path-only training provides little additional
benefit after Stage I, and retaining the path objective alongside endpoint
supervision during Stage II produces no clear advantage in this auxiliary
single-run comparison.

\begin{table*}[htbp!]
    \centering
    \caption{\textbf{Training-protocol ablation on MNIST and
    Fashion-MNIST.}
    Results are from the auxiliary single-run protocol defined in
    Appendix~\ref{app:checkpoint_selection}. Continuation and direct-training
    ablation branches use learning rate $30$. All checkpoints are selected
    using validation data only. Agreement denotes neural-teacher agreement;
    lower output endpoint RMSE is better.}
    \label{tab:protocol_ablation}
    \small
    \setlength{\tabcolsep}{4.0pt}
    \begin{adjustbox}{max width=\textwidth}
    \begin{tabular}{llccrrr}
        \toprule
        Dataset
        & Method
        & Initialization
        & Objective
        & Top-1 (\%)
        & Agreement (\%)
        & Output endpoint RMSE \\
        \midrule

        MNIST
        & PATH
        & Exact zero
        & $\mathcal L_{\rm path}$
        & 89.13
        & 90.23
        & 0.357147 \\

        MNIST
        & PATH-CONT
        & PATH checkpoint
        & $\mathcal L_{\rm path}$
        & 89.12
        & 90.22
        & 0.357083 \\

        MNIST
        & Z-END
        & Exact zero
        & $\mathcal L_{\rm path}+\mathcal L_{\rm end}$
        & 95.84
        & 95.50
        & 0.148850 \\

        MNIST
        & ZERO-END-ONLY
        & Exact zero
        & $\mathcal L_{\rm end}$
        & 20.88
        & 20.98
        & 0.633421 \\

        MNIST
        & PRETRAIN-END-ONLY
        & PATH checkpoint
        & $\mathcal L_{\rm end}$
        & 96.61
        & 96.79
        & 0.123558 \\

        MNIST
        & TWO-STAGE-END
        & PATH checkpoint
        & $\mathcal L_{\rm path}+\mathcal L_{\rm end}$
        & 96.47
        & 96.60
        & 0.126015 \\

        \midrule

        Fashion-MNIST
        & PATH
        & Exact zero
        & $\mathcal L_{\rm path}$
        & 80.13
        & 86.43
        & 0.309598 \\

        Fashion-MNIST
        & PATH-CONT
        & PATH checkpoint
        & $\mathcal L_{\rm path}$
        & 80.13
        & 86.43
        & 0.309507 \\

        Fashion-MNIST
        & Z-END
        & Exact zero
        & $\mathcal L_{\rm path}+\mathcal L_{\rm end}$
        & 86.11
        & 93.10
        & 0.124091 \\

        Fashion-MNIST
        & ZERO-END-ONLY
        & Exact zero
        & $\mathcal L_{\rm end}$
        & 18.65
        & 16.46
        & 0.569114 \\

        Fashion-MNIST
        & PRETRAIN-END-ONLY
        & PATH checkpoint
        & $\mathcal L_{\rm end}$
        & 86.58
        & 94.91
        & 0.099167 \\

        Fashion-MNIST
        & TWO-STAGE-END
        & PATH checkpoint
        & $\mathcal L_{\rm path}+\mathcal L_{\rm end}$
        & 86.63
        & 94.81
        & 0.099203 \\

        \bottomrule
    \end{tabular}
    \end{adjustbox}
\end{table*}

\begin{figure*}[htbp!]
    \centering
    \includegraphics[width=0.75\textwidth]
    {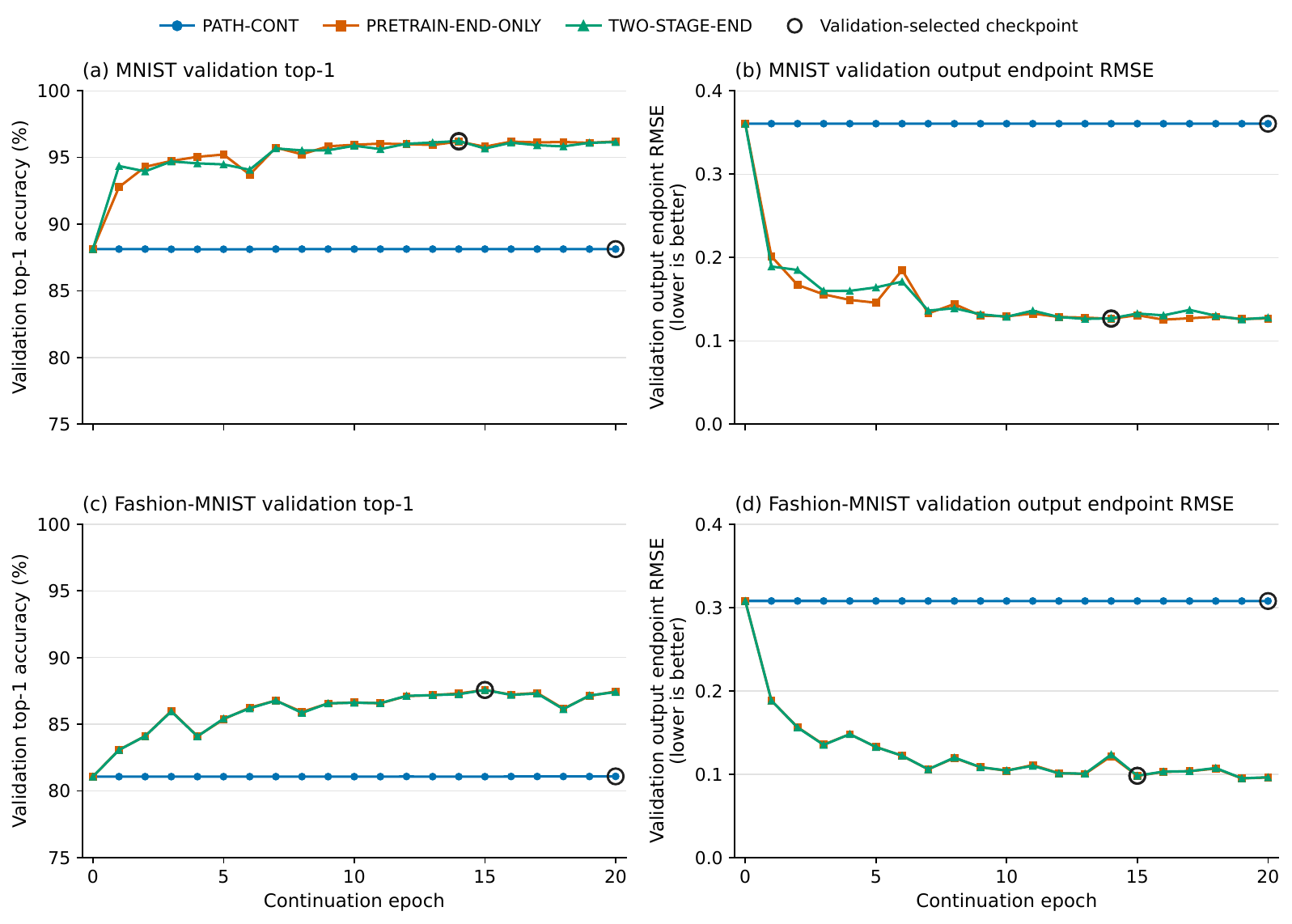}
    \caption{\textbf{Matched continuation behavior after PATH pretraining.}
    Validation top-1 accuracy and autonomous output endpoint RMSE for
    PATH-CONT, PRETRAIN-END-ONLY, and TWO-STAGE-END under the auxiliary
    single-run continuation protocol on MNIST and Fashion-MNIST.
    All three methods begin from the same dataset-specific PATH checkpoint
    under matched continuation conditions. PATH-CONT optimizes
    $\mathcal L_{\rm path}$, PRETRAIN-END-ONLY optimizes
    $\mathcal L_{\rm end}$, and TWO-STAGE-END optimizes
    $\mathcal L_{\rm path}+\mathcal L_{\rm end}$. Open circles mark the
    validation-selected continuation checkpoints. Lower endpoint RMSE is
    better.}
    \label{fig:matched_continuation_curves}
\end{figure*}

Figure~\ref{fig:matched_continuation_curves} shows the corresponding
validation behavior for the three continuation branches that share the same
PATH initialization. PATH-CONT remains close to the Stage-I starting point in
both validation accuracy and autonomous endpoint RMSE. By contrast,
PRETRAIN-END-ONLY and TWO-STAGE-END improve rapidly after endpoint supervision
is introduced and follow closely similar continuation trajectories. The
validation dynamics therefore support the same interpretation as the
autonomous test results: the principal change after PATH pretraining is
associated with autonomous endpoint supervision rather than with additional
path-only optimization.

\subsection{Endpoint-Weight Sensitivity}
\label{app:endpoint_weight_sensitivity}

We next examine the sensitivity of Stage-II continuation to the relative
weight of autonomous endpoint supervision. This auxiliary single-run study is
performed on MNIST using the same PATH initialization and matched
continuation conditions as the ablations in
Appendix~\ref{app:full_protocol_ablation}. The continuation objective is
\begin{equation*}
    \mathcal L
    =
    \mathcal L_{\rm path}
    +
    \lambda_{\rm end}\mathcal L_{\rm end},
\end{equation*}
with
\[
    \lambda_{\rm end}\in\{0,0.1,1,10\},
\]
where $\lambda_{\rm end}=0$ corresponds to PATH-CONT. The nonzero-weight
branches use the auxiliary continuation protocol with learning rate $30$.
All branches begin from the same validation-selected PATH checkpoint and
differ only in the weight assigned to $\mathcal L_{\rm end}$.

Table~\ref{tab:endpoint_weight_sensitivity} shows that introducing endpoint
supervision produces a large improvement for both moderate nonzero weights.
At $\lambda_{\rm end}=0.1$, top-1 accuracy increases from $89.12\%$ for the
path-only control to $96.43\%$, while the output endpoint RMSE decreases from
$0.357083$ to $0.124096$. At $\lambda_{\rm end}=1$, the corresponding values
are $96.47\%$ and $0.126015$. The small and metric-dependent differences
between these two settings indicate that the continuation-stage benefit is not
confined to a single narrowly tuned endpoint weight.

A larger endpoint weight produces qualitatively different behavior. At
$\lambda_{\rm end}=10$, the output endpoint RMSE remains lower than for
PATH-CONT, decreasing from $0.357083$ to $0.258149$, but top-1 accuracy falls
to $85.04\%$. Neural-teacher agreement shows the same degradation. Thus,
improving the average continuous endpoint match is not by itself sufficient
to improve the hard classification decision: excessive weighting of the
endpoint objective can reduce aggregate endpoint error while degrading
task-level inference.

The principal experiments use $\lambda_{\rm end}=1$ throughout. The present
sweep is reported as an auxiliary sensitivity analysis rather than as a
test-set model-selection procedure, and no corresponding multi-weight sweep is
performed on Fashion-MNIST. Within the tested range, the results support a
moderate endpoint contribution while also showing that overly strong endpoint
weighting can be detrimental.

\begin{table}[htbp!]
    \centering
    \caption{\textbf{Endpoint-weight sensitivity on MNIST.}
    All branches begin from the same validation-selected PATH checkpoint under
    matched continuation conditions. $\lambda_{\rm end}=0$ corresponds to
    PATH-CONT. The nonzero-weight branches use the auxiliary single-run
    continuation protocol with learning rate $30$. Lower output endpoint RMSE
    is better.}
    \label{tab:endpoint_weight_sensitivity}
    \small
    \setlength{\tabcolsep}{5pt}
    \begin{tabular}{lrrr}
        \toprule
        $\lambda_{\rm end}$
        & Top-1 (\%)
        & Agreement (\%)
        & Output endpoint RMSE \\
        \midrule

        $0$
        & $89.12$
        & $90.22$
        & $0.357083$ \\

        $0.1$
        & $96.43$
        & $97.09$
        & $0.124096$ \\

        $1$
        & $96.47$
        & $96.60$
        & $0.126015$ \\

        $10$
        & $85.04$
        & $84.85$
        & $0.258149$ \\

        \bottomrule
    \end{tabular}
\end{table}

\section{Hidden-Width Scaling}
\label{AppendixF}

\subsection{Matched Teacher--Kuramoto Scaling Design}
\label{app:matched_scaling_protocol}

The present path-matching construction associates each transferred
neural-teacher variable with one dynamical variable of the Kuramoto student.
Consequently, changing the number of hidden oscillators also changes the
dimension of the transferred teacher representation. We therefore scale the
neural teacher and Kuramoto student together within a common matched
architecture family.

For a teacher with three hidden layers of widths $(h_1,h_2,h_3)$, the
transferred hidden representation is formed by concatenating the hidden-layer
activations,
\begin{equation*}
    a^{\rm TE,\HH}
    =
    \begin{bmatrix}
        h^{(1)}\\
        h^{(2)}\\
        h^{(3)}
    \end{bmatrix},
\end{equation*}
with
$ h_1+h_2+h_3=H$, 
where $H$ is the number of hidden oscillators in the corresponding Kuramoto
student. Together with the $O=10$ signed-softmax output activations defined in
Appendix~\ref{app:neural_teacher_training}, the complete transferred state is
\begin{equation*}
    a^{\rm TE}
    =
    \begin{bmatrix}
        a^{\rm TE,\HH}\\
        a^{\rm TE,\OO}
    \end{bmatrix}
    \in[-1,1]^{H+O}.
\end{equation*}
Its dimension therefore exactly matches the $H$ hidden and $O$ output
oscillators of the physical student. No projection, padding, subsampling, or
learned adapter is introduced between the neural and physical
representations.

We use the matched three-hidden-layer family in
Table~\ref{tab:matched_scaling_teacher_family}. The teacher depth and
hidden-layer width ratio are fixed, with $h_1:h_2:h_3=2:1:1$,
while the total hidden dimension varies over
\[
    H\in\{16,32,64,128\}.
\]
The $H=64$ member, $784\text{-}32\text{-}16\text{-}16\text{-}10$, coincides with the neural-teacher architecture used in the principal
$74$-oscillator experiment. The other widths preserve the same architectural
family while changing the dimension of both the teacher representation and
the physical student.

\begin{table}[htbp!]
\centering
\caption{\textbf{Matched teacher--Kuramoto family used for hidden-width
scaling.}
For each $H$, the number of transferred neural-teacher hidden activations
equals the number of Kuramoto hidden oscillators. All teachers have three
hidden layers with a fixed $2{:}1{:}1$ width ratio, and all physical students
contain $O=10$ output oscillators.}
\label{tab:matched_scaling_teacher_family}
\begin{tabular}{ccc}
\toprule
$H$ & Neural-teacher architecture & Kuramoto size $N=H+O$ \\
\midrule
16  & $784$--$8$--$4$--$4$--$10$       & $26$ \\
32  & $784$--$16$--$8$--$8$--$10$      & $42$ \\
64  & $784$--$32$--$16$--$16$--$10$    & $74$ \\
128 & $784$--$64$--$32$--$32$--$10$    & $138$ \\
\bottomrule
\end{tabular}
\end{table}

A separate frozen neural teacher is used for each dataset and hidden width,
with checkpoint selection following the common validation-based
teacher-training procedure of Appendix~\ref{app:neural_teacher_training}.
The corresponding Kuramoto students use the same physical model and numerical
configuration as Appendix~\ref{app:kuramoto_training}, with only the hidden
dimension and the associated parameter dimensions changed.

For each width, PATH and TWO-STAGE-END are evaluated using the principal
training protocol. Stage I uses the path-matching objective, and Stage II
continues from the width- and run-matched validation-selected PATH checkpoint
with learning rate $10$ and autonomous endpoint supervision. Five matched
training-order runs are used at every width on both datasets, allowing
PATH and TWO-STAGE-END to be compared pairwise within each system size.


\subsection{Hidden-Width Scaling Results}
\label{app:hidden_width_scaling_results}

Table~\ref{tab:appendix_F2_hidden_width_scaling} summarizes the matched
teacher--Kuramoto scaling results defined in
Appendix~\ref{app:matched_scaling_protocol}. At each hidden dimension,
PATH and TWO-STAGE-END are evaluated over the same five paired training-order
runs, while the teacher entry corresponds to the frozen neural teacher
associated with that matched system size.

The neural teachers become progressively more accurate as the matched hidden
dimension increases, but PATH does not systematically benefit from the larger
teacher--student systems. On MNIST, teacher top-1 accuracy increases
from $90.08\%$ at $H=16$ to $97.47\%$ at $H=128$. PATH instead reaches its
highest mean accuracy at $H=32$ and subsequently decreases from
$90.780\pm0.007\%$ to $86.846\pm0.013\%$ at $H=128$. TWO-STAGE-END shows a
different scaling trend, increasing from $91.182\pm0.070\%$ at $H=16$ to
$96.752\pm0.210\%$ at $H=128$.

Fashion-MNIST exhibits the same qualitative distinction. Teacher accuracy
increases from $82.60\%$ to $86.48\%$ across the tested range. PATH again
reaches its highest mean accuracy at $H=32$ and does not improve
systematically at larger widths, whereas TWO-STAGE-END increases from
$83.148\pm0.112\%$ at $H=16$ to $86.764\pm0.088\%$ at $H=128$.

At every tested width on both datasets, TWO-STAGE-END outperforms PATH in all
five paired training-order runs. On MNIST, the mean paired top-1
advantage increases from $2.982\pm0.058$ percentage points at $H=16$ to
$9.906\pm0.209$ percentage points at $H=128$. On Fashion-MNIST, the
corresponding advantages are $2.692\pm0.135$, $3.416\pm0.086$,
$6.296\pm0.061$, and $6.110\pm0.082$ percentage points for
$H=16,32,64,$ and $128$, respectively.

Continuous endpoint fidelity shows the same separation. TWO-STAGE-END has
lower output endpoint RMSE than PATH at every tested width on both datasets.
For PATH, endpoint fidelity does not improve systematically as the matched
system becomes larger and deteriorates at the larger widths. TWO-STAGE-END,
in contrast, maintains substantially closer alignment with the corresponding
teacher outputs throughout the tested range. The simultaneous improvements in
classification and continuous endpoint fidelity indicate that the scaling
advantage is not restricted to changes in the final hard-label decision.

The incremental accuracy benefit of increasing system size becomes smaller at
the largest tested widths. From $H=64$ to $H=128$, mean TWO-STAGE-END top-1
accuracy increases by only $0.030$ percentage points on MNIST and
$0.308$ percentage points on Fashion-MNIST, compared with substantially
larger gains at smaller widths. These observations indicate diminishing
returns over the tested range, but do not establish an asymptotic performance
plateau.

The same results can also be viewed relative to the accuracy level of the
matched neural teachers. On MNIST, the mean teacher-minus-PATH accuracy
gap increases from $1.880$ percentage points at $H=16$ to $10.624$
percentage points at $H=128$; on Fashion-MNIST it increases from $2.144$ to
$5.826$ percentage points. TWO-STAGE-END remains much closer to the matched
teacher accuracy throughout the tested range, with an absolute mean gap no
larger than $1.128$ percentage points across all dataset--width combinations.
Negative gaps at several widths simply indicate that the student attains
slightly higher aggregate test accuracy than its matched teacher and should
not be interpreted as sample-wise reproduction of the teacher classifier.

Taken together, increasing the matched teacher--Kuramoto system size does not
by itself guarantee improved autonomous inference. Although the matched
teachers become progressively more accurate, the PATH student does not
consistently exploit the additional representational capacity and increasingly
departs from the teacher accuracy level at larger widths. TWO-STAGE-END
substantially reduces this mismatch, improves autonomous accuracy across the
tested widths, and maintains lower endpoint error. As emphasized in
Appendix~\ref{app:matched_scaling_protocol}, these results describe matched
teacher--student system-size scaling and should not be interpreted as an
isolated causal effect of Kuramoto hidden width.

\begin{table*}[htbp!]
\centering
\caption{\textbf{Matched teacher--Kuramoto hidden-width scaling results.}
Teacher top-1 entries correspond to the frozen neural teacher associated with
each width. Student quantities are reported as mean $\pm$ sample standard
deviation over five paired training-order runs.
TWO-STAGE-END$-$PATH denotes the paired top-1 improvement in percentage
points. Endpoint RMSE compares autonomous terminal output activations with the
corresponding neural-teacher targets; lower values are better.}
\label{tab:appendix_F2_hidden_width_scaling}

\begin{adjustbox}{max width=\textwidth}
\begin{tabular}{llcccccc}
\toprule
Dataset
& $H$
& Teacher Top-1 (\%)
& PATH Top-1 (\%)
& TWO-STAGE-END Top-1 (\%)
& TWO-STAGE-END$-$PATH (pp)
& PATH endpoint RMSE
& TWO-STAGE-END endpoint RMSE \\
\midrule

MNIST
& 16
& 90.08
& 88.200 $\pm$ 0.023
& 91.182 $\pm$ 0.070
& 2.982 $\pm$ 0.058
& 0.341461 $\pm$ 0.000115
& 0.126897 $\pm$ 0.001712 \\

& 32
& 93.70
& 90.780 $\pm$ 0.007
& 94.828 $\pm$ 0.116
& 4.048 $\pm$ 0.117
& 0.291821 $\pm$ 0.000158
& 0.125116 $\pm$ 0.001537 \\

& 64
& 95.90
& 89.116 $\pm$ 0.011
& 96.722 $\pm$ 0.072
& 7.606 $\pm$ 0.079
& 0.357096 $\pm$ 0.000054
& 0.120144 $\pm$ 0.000891 \\

& 128
& 97.47
& 86.846 $\pm$ 0.013
& 96.752 $\pm$ 0.210
& 9.906 $\pm$ 0.209
& 0.445772 $\pm$ 0.000054
& 0.127378 $\pm$ 0.004596 \\

\midrule

Fashion-MNIST
& 16
& 82.60
& 80.456 $\pm$ 0.034
& 83.148 $\pm$ 0.112
& 2.692 $\pm$ 0.135
& 0.291675 $\pm$ 0.000275
& 0.121808 $\pm$ 0.002344 \\

& 32
& 84.87
& 82.036 $\pm$ 0.015
& 85.452 $\pm$ 0.077
& 3.416 $\pm$ 0.086
& 0.262271 $\pm$ 0.000231
& 0.101992 $\pm$ 0.001652 \\

& 64
& 85.79
& 80.160 $\pm$ 0.027
& 86.456 $\pm$ 0.036
& 6.296 $\pm$ 0.061
& 0.309541 $\pm$ 0.000059
& 0.100330 $\pm$ 0.002459 \\

& 128
& 86.48
& 80.654 $\pm$ 0.013
& 86.764 $\pm$ 0.088
& 6.110 $\pm$ 0.082
& 0.339279 $\pm$ 0.000025
& 0.079827 $\pm$ 0.003762 \\

\bottomrule
\end{tabular}
\end{adjustbox}
\end{table*}

\section{Stochastic Inference and Ensemble Averaging}
\label{AppendixG}
\subsection{Stochastic Dynamics and Ensemble Readout}
\label{app:stochastic_protocol}

To evaluate inference under thermal fluctuations, we introduce stochasticity
only during autonomous inference and leave the trained Kuramoto parameters
unchanged. PATH and TWO-STAGE-END are evaluated using five paired
training-order runs from the principal training protocol on both MNIST
and Fashion-MNIST; no model is retrained or adapted in the presence of noise.

For test example $n$ and stochastic replicate $m$, the phase dynamics are
\begin{equation*}
    {\rm d}\theta_{n,i,m}^{\rm S}
    =
    f_{\Theta,i}
    \left(
        \theta_{n,m}^{\rm S},
        u_n
    \right)
    {\rm d}t
    +
    \sqrt{2\mu k_{\rm B}T}\,
    {\rm d}W_{n,i,m},
\end{equation*}
where $f_{\Theta,i}$ is the deterministic Kuramoto vector field defined in
Appendix~\ref{app:kuramoto_training}, $k_{\rm B}T$ controls the thermal-noise
strength, and $W_{n,i,m}$ is a standard Wiener process. Independent noise acts
on every oscillator, and the deterministic dynamics are recovered at
$k_{\rm B}T=0$.

The stochastic dynamics are integrated using Euler--Maruyama on the same
observation interval and nominal time grid as deterministic inference,
\[
t_{\rm f}=0.2,\qquad K=200,\qquad \Delta t=10^{-3}.
\]
For each step, we first compute the unconstrained phase update
\begin{equation*}
\widetilde{\theta}^{\rm S}_{n,i,k+1,m}
=
\theta^{\rm S}_{n,i,k,m}
+
\Delta t\,
f_{\Theta,i}
\left(
\theta^{\rm S}_{n,k,m},u_n
\right)
+
\sqrt{2\mu k_{\rm B}T\Delta t}\,
\eta_{n,i,k,m},
\end{equation*}
where
\[
\eta_{n,i,k,m}\sim\mathcal N(0,1).
\]
The numerical phase representative is then reduced to the principal circular
branch,
\begin{equation*}
\theta^{\rm S}_{n,i,k+1,m}
=
\operatorname{atan2}
\left(
\sin\widetilde{\theta}^{\rm S}_{n,i,k+1,m},
\cos\widetilde{\theta}^{\rm S}_{n,i,k+1,m}
\right).
\end{equation*}
This wrapping selects an equivalent numerical representative of the periodic
phase state; the Kuramoto vector field and cosine readout are both
$2\pi$-periodic.

\paragraph{Matched stochastic forcing.}

Common random numbers are used for comparisons between trained checkpoints.
For a fixed input, temperature, stochastic replicate, integration step, and
oscillator, the compared PATH and TWO-STAGE-END models receive the same
Gaussian increment. The same prescribed noise realization is also used across
the five matched training-order runs, so differences among trained models are
not confounded by independently sampled stochastic forcing. Distinct
stochastic replicates use distinct Gaussian fields, and different
temperatures are evaluated with distinct prescribed noise realizations.

\paragraph{Single-trajectory and ensemble inference.}

For stochastic replicate $m$, the terminal score of output oscillator $o$ is
\begin{equation*}
    s_{n,o,m}^{\rm S,\OO}
    =
    \cos
    \theta_{n,o,m,K}^{\rm S,\OO},
    \qquad
    o=1,2,\ldots,O.
\end{equation*}
Each trained checkpoint is evaluated with $M_{\rm ens}=10$
stochastic trajectories per input. We distinguish two inference modes.

For mean single-trajectory performance, each stochastic replicate is
treated as a separate noisy evaluation of the complete test set. The
classification metric is computed independently for each replicate and then
averaged over the $M_{\rm ens}$ replicates. This quantity characterizes the
performance of a single noisy physical realization.

For ensemble inference, the continuous terminal outputs are instead
averaged before classification,
\begin{equation*}
    \overline{s}_{n,o}^{\rm S,\OO}
    =
    \frac{1}{M_{\rm ens}}
    \sum_{m=1}^{M_{\rm ens}}
    s_{n,o,m}^{\rm S,\OO},
\end{equation*}
followed by
\begin{equation*}
    \widehat y_n^{\rm S,ens}
    =
    \arg\max_o
    \overline{s}_{n,o}^{\rm S,\OO}.
\end{equation*}
Thus, ensemble inference averages the continuous physical outputs rather than
performing majority voting over trajectory-level hard predictions. Each noisy
trajectory contributes through the magnitudes of all terminal output
activations. At $k_{\rm B}T=0$, all replicates coincide with the deterministic
rollout, and the single-trajectory and ensemble procedures reduce to the same
prediction.

\paragraph{Statistical aggregation.}

Stochastic trajectories are Monte Carlo replicates nested within a trained
Kuramoto model and are not treated as independent training runs. Single-
trajectory and ensemble metrics are first computed within each of the five
trained checkpoints. Reported means and sample standard deviations are then
computed across these five training-order runs, following
Appendix~\ref{app:evaluation_metrics}.

\subsection{Temperature-Dependent Robustness}
\label{app:temperature_robustness}

We evaluate PATH and TWO-STAGE-END under inference-time thermal noise over
\begin{equation*}
    k_{\rm B}T
    \in
    \{0,0.1,0.25,0.5,1\}.
\end{equation*}
No model is retrained or adapted under noise. For each dataset, method,
training-order run, and temperature, we evaluate ten stochastic trajectories
using the matched-noise protocol of Appendix~\ref{app:stochastic_protocol}.
Results are reported for both mean single-trajectory performance and
$M_{\rm ens}=10$ continuous-output ensemble inference. Means and sample
standard deviations are computed across the five trained runs.

Table~\ref{tab:appendix_G2_temperature_top1} summarizes the top-1 results.
Single-trajectory accuracy decreases monotonically with temperature for both
methods and datasets. At the largest tested temperature,
$k_{\rm B}T=1$, MNIST accuracy falls to
$67.117\pm0.003\%$ for PATH and $82.191\pm0.318\%$ for TWO-STAGE-END,
compared with deterministic accuracies of $89.116\pm0.011\%$ and
$96.722\pm0.072\%$, respectively. Fashion-MNIST shows the same behavior,
with single-trajectory accuracies decreasing to
$56.649\pm0.009\%$ for PATH and $71.985\pm0.267\%$ for
TWO-STAGE-END. Thus, sufficiently strong thermal fluctuations substantially
degrade individual physical realizations, although TWO-STAGE-END remains more
robust than PATH.

Continuous-output ensemble averaging recovers much of the lost performance.
At $k_{\rm B}T=1$, averaging ten trajectories increases MNIST accuracy
from $67.117\%$ to $88.884\%$ for PATH and from $82.191\%$ to
$96.142\%$ for TWO-STAGE-END. On Fashion-MNIST, the corresponding increases
are from $56.649\%$ to $76.684\%$ and from $71.985\%$ to $84.958\%$.
For TWO-STAGE-END, the $M_{\rm ens}=10$ accuracy at the largest tested
temperature therefore remains within $0.580$ percentage points of its
deterministic value on MNIST and within $1.498$ percentage points on
Fashion-MNIST.

The advantage of endpoint fine-tuning persists throughout the temperature
sweep. At every nonzero temperature on both datasets, the
$M_{\rm ens}=10$ accuracy of TWO-STAGE-END exceeds that of PATH in all five
paired training-order runs. At $k_{\rm B}T=1$, the paired
TWO-STAGE-END minus PATH improvements are
$7.258\pm0.069$ percentage points on MNIST and
$8.274\pm0.190$ percentage points on Fashion-MNIST, with two-sided $95\%$
paired Student-$t$ intervals of $[7.172,7.344]$ and
$[8.038,8.510]$ percentage points, respectively.

A distinct nonmonotonic effect is observed for MNIST PATH under ensemble
inference. Its $M_{\rm ens}=10$ top-1 accuracy increases from
$89.116\pm0.011\%$ at $k_{\rm B}T=0$ to
$90.624\pm0.018\%$ at $k_{\rm B}T=0.1$ and
$91.124\pm0.018\%$ at $k_{\rm B}T=0.25$, before decreasing at larger
temperatures. The paired improvement at $k_{\rm B}T=0.25$ is
$2.008\pm0.019$ percentage points relative to deterministic inference and is
positive in all five training-order runs. We therefore describe this as a
reproducible low-temperature accuracy enhancement for MNIST PATH under
$M_{\rm ens}=10$ ensemble inference. This observation is specific to the
tested ensemble protocol and is not interpreted as evidence of stochastic
resonance.

Continuous endpoint fidelity is consistent with the classification results.
For every tested temperature on both datasets, TWO-STAGE-END has lower
ensemble output endpoint RMSE than PATH. At $k_{\rm B}T=1$, the mean endpoint
RMSE values are $0.386359$ for PATH and $0.252062$ for TWO-STAGE-END on
MNIST, and $0.357304$ and $0.233505$, respectively, on Fashion-MNIST.
Thus, the stochastic robustness advantage is also present in the continuous
teacher--student output representation rather than only in the final hard
classification decision.

Together, these results identify two complementary mechanisms of
inference-time robustness. Endpoint fine-tuning produces individual noisy
trajectories that are more robust than those obtained after path training
alone, while repeated stochastic evaluation followed by continuous-output
averaging suppresses much of the remaining inference variability. The former
is a property of the trained dynamics, whereas the latter provides an
additional inference strategy for operating the same physical model under
thermal fluctuations.

\begin{table*}[htbp!]
    \centering
    \caption{\textbf{Temperature-dependent top-1 accuracy under stochastic
    inference.}
    Mean-single evaluates each of ten stochastic trajectories independently
    and averages the resulting accuracies within each trained model.
    $M_{\rm ens}=10$ instead averages the ten continuous terminal outputs
    before classification. Entries are mean $\pm$ sample standard deviation
    across five training-order runs. At $k_{\rm B}T=0$, all stochastic
    trajectories reduce to the deterministic rollout.}
    \label{tab:appendix_G2_temperature_top1}
    \small
    \setlength{\tabcolsep}{4.5pt}
    \begin{adjustbox}{max width=\textwidth}
    \begin{tabular}{llrrr}
        \toprule
        Dataset
        & Method
        & $k_{\rm B}T$
        & Mean-single top-1 (\%)
        & $M_{\rm ens}=10$ top-1 (\%) \\
        \midrule

        MNIST & PATH & $0$
        & $89.1160\pm0.0114$ & $89.1160\pm0.0114$ \\
        & PATH & $0.1$
        & $86.1694\pm0.0124$ & $90.6240\pm0.0182$ \\
        & PATH & $0.25$
        & $82.2876\pm0.0040$ & $91.1240\pm0.0182$ \\
        & PATH & $0.5$
        & $76.1984\pm0.0104$ & $90.1240\pm0.0114$ \\
        & PATH & $1$
        & $67.1172\pm0.0026$ & $88.8840\pm0.0207$ \\

        \addlinespace

        MNIST & TWO-STAGE-END & $0$
        & $96.7220\pm0.0716$ & $96.7220\pm0.0716$ \\
        & TWO-STAGE-END & $0.1$
        & $96.0428\pm0.0285$ & $96.6900\pm0.0771$ \\
        & TWO-STAGE-END & $0.25$
        & $94.4832\pm0.0675$ & $96.6380\pm0.0642$ \\
        & TWO-STAGE-END & $0.5$
        & $90.8326\pm0.1603$ & $96.5540\pm0.0789$ \\
        & TWO-STAGE-END & $1$
        & $82.1908\pm0.3178$ & $96.1420\pm0.0753$ \\

        \midrule

        Fashion-MNIST & PATH & $0$
        & $80.1600\pm0.0274$ & $80.1600\pm0.0274$ \\
        & PATH & $0.1$
        & $76.5070\pm0.0122$ & $80.0280\pm0.0239$ \\
        & PATH & $0.25$
        & $71.9750\pm0.0074$ & $79.6800\pm0.0100$ \\
        & PATH & $0.5$
        & $65.6258\pm0.0071$ & $78.5280\pm0.0179$ \\
        & PATH & $1$
        & $56.6492\pm0.0087$ & $76.6840\pm0.0055$ \\

        \addlinespace

        Fashion-MNIST & TWO-STAGE-END & $0$
        & $86.4560\pm0.0358$ & $86.4560\pm0.0358$ \\
        & TWO-STAGE-END & $0.1$
        & $85.4040\pm0.0900$ & $86.4060\pm0.1193$ \\
        & TWO-STAGE-END & $0.25$
        & $83.8114\pm0.1088$ & $86.1380\pm0.1256$ \\
        & TWO-STAGE-END & $0.5$
        & $80.1554\pm0.1514$ & $85.4600\pm0.0972$ \\
        & TWO-STAGE-END & $1$
        & $71.9850\pm0.2665$ & $84.9580\pm0.1871$ \\

        \bottomrule
    \end{tabular}
    \end{adjustbox}
\end{table*}

\subsection{Representative Stochastic Output Dynamics}
\label{app:representative_stochastic_output_dynamics}

To illustrate stochastic finite-time inference under the conditions analyzed in Appendix~\ref{app:temperature_robustness}, we visualize
time-resolved output trajectories for four representative Fashion-MNIST
examples at $k_{\rm B}T=1.$
The examples are evaluated using one prespecified training-order run with
$M_{\rm ens}=10$ stochastic trajectories per input. PATH and TWO-STAGE-END
receive matched Gaussian perturbations for each stochastic replicate according
to the common-random-number protocol of
Appendix~\ref{app:stochastic_protocol}. 

\begin{figure*}[htbp!]
    \centering
    \includegraphics[width=\textwidth]
    {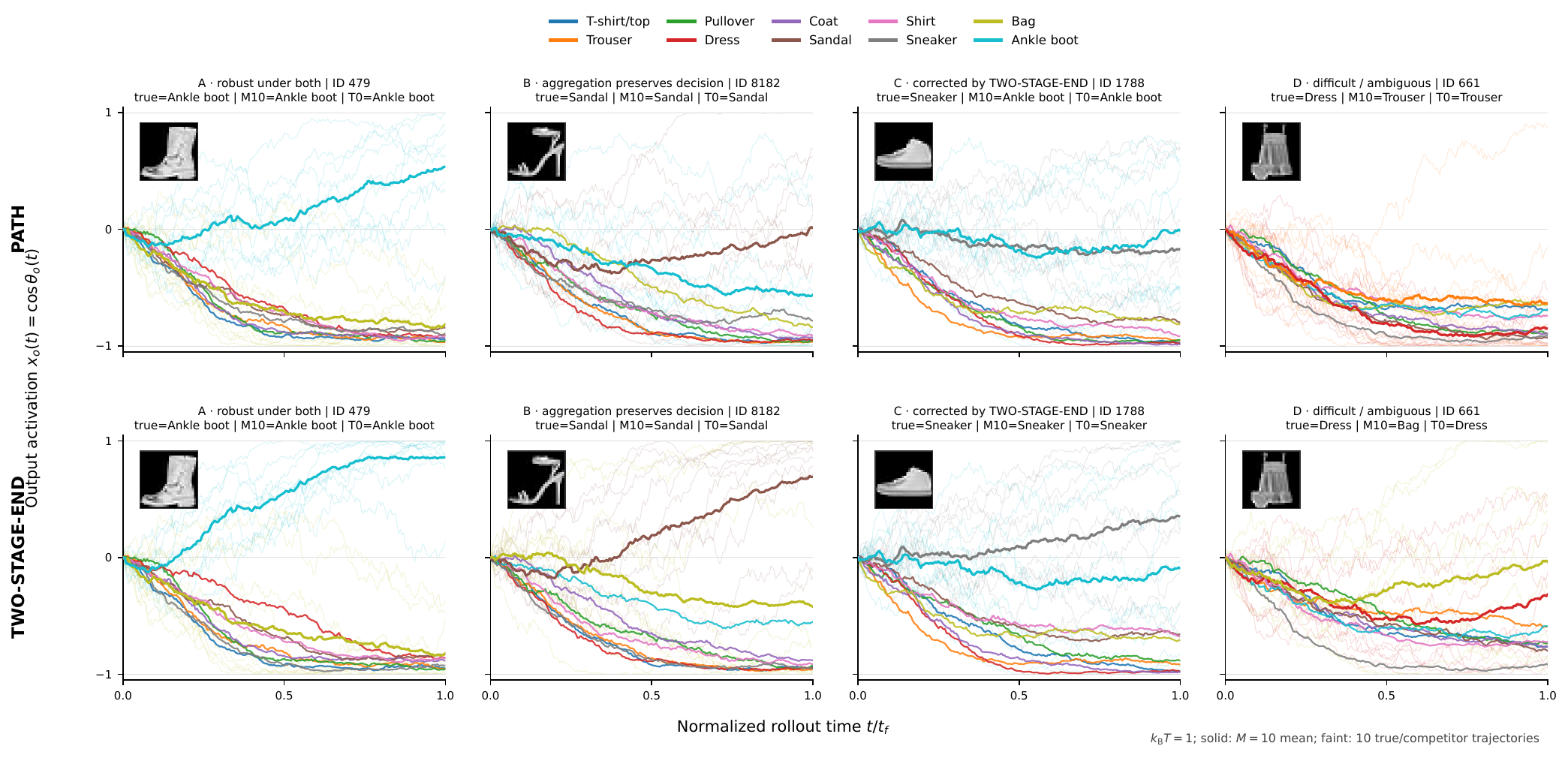}
    \caption{\textbf{Representative stochastic output dynamics at
    $k_{\rm B}T=1$.}
    Fashion-MNIST examples evaluated with $M_{\rm ens}=10$ stochastic
    trajectories per input. The top row shows PATH and the bottom row shows
    TWO-STAGE-END under matched stochastic forcing. Solid curves denote the
    ensemble-mean output activations of all ten classes; faint curves show the
    individual stochastic trajectories of the true class and the strongest
    competing class.
    \textbf{(A)} Both methods remain correct across all individual
    trajectories.
    \textbf{(B)} Continuous-output averaging preserves the correct decision
    despite errors in some individual trajectories.
    \textbf{(C)} PATH remains misclassified after stochastic averaging,
    whereas TWO-STAGE-END retains the correct class.
    \textbf{(D)} A difficult example in which stochastic perturbations also
    overturn the TWO-STAGE-END ensemble decision.}
    \label{fig:appendix_G3_stochastic_output_dynamics}
\end{figure*}

For stochastic replicate $r$, the output activation of class $o$ is
\begin{equation*}
    x_{n,o,r}^{\rm S,\OO}(t_k)
    =
    \cos\theta_{n,o,r}^{\rm S,\OO}(t_k),
\end{equation*}
and the ensemble-mean trajectory is
\begin{equation*}
    \overline{x}_{n,o}^{\rm S,\OO}(t_k)
    =
    \frac{1}{M_{\rm ens}}
    \sum_{r=1}^{M_{\rm ens}}
    x_{n,o,r}^{\rm S,\OO}(t_k).
\end{equation*}
In Fig.~\ref{fig:appendix_G3_stochastic_output_dynamics}, solid curves show
the ensemble-mean trajectories of all ten output classes. Faint curves show
the individual stochastic trajectories of the true class and of the strongest
competing class, defined for each method as the non-true class with the
largest ensemble-mean terminal output.

These examples provide qualitative illustrations of several stochastic
inference regimes underlying the quantitative robustness results in
Appendix~\ref{app:temperature_robustness}.

\section{Numerical Sensitivity}
\label{AppendixH}

The principal deterministic evaluations use explicit Euler integration with
\[
    t_{\rm f}=0.2,
    \qquad
    K=200,
    \qquad
    \Delta t=10^{-3}.
\]
To assess sensitivity to this discretization, we re-evaluate the same trained
PATH and TWO-STAGE-END checkpoints using a twofold finer grid,
\[
    K=400,
    \qquad
    \Delta t=5\times10^{-4},
\]
while keeping the observation time, initial condition, model parameters, and
evaluation set unchanged. No model is retrained, fine-tuned, or reselected at
the refined resolution. Five paired training-order runs are evaluated for
each method on both MNIST and Fashion-MNIST.

We compare the two resolutions at the task, decision, and continuous-output
levels. In addition to the change in top-1 accuracy, the hard-prediction
disagreement is
\begin{equation*}
    D_{200,400}
    =
    \frac{1}{M}
    \sum_{n=1}^{M}
    \mathbf 1
    \left[
        \widehat y_n^{(200)}
        \neq
        \widehat y_n^{(400)}
    \right],
\end{equation*}
and the RMS change in the terminal continuous outputs is
\begin{equation*}
    \Delta_{\rm RMS}^{\OO}
    =
    \left[
        \frac{1}{MO}
        \sum_{n=1}^{M}
        \sum_{o=1}^{O}
        \left(
            s_{n,o}^{(200)}
            -
            s_{n,o}^{(400)}
        \right)^2
    \right]^{1/2}.
\end{equation*}

Table~\ref{tab:appendix_H_numerical_refinement} summarizes the refinement
results. Across the four dataset--method combinations, the largest mean
top-1 change is $0.096$ percentage points, observed for MNIST PATH.
TWO-STAGE-END changes by only $+0.002\pm0.004$ percentage points on 
MNIST and $-0.006\pm0.011$ percentage points on Fashion-MNIST. The large
accuracy separation between PATH and TWO-STAGE-END is therefore preserved
under the refined integration grid.

Hard predictions are likewise stable. The mean fractions of test examples
whose predicted class changes between $K=200$ and $K=400$ are $0.132\%$ for
MNIST PATH, $0.002\%$ for MNIST TWO-STAGE-END, $0.050\%$ for
Fashion-MNIST PATH, and $0.020\%$ for Fashion-MNIST TWO-STAGE-END.
Continuous terminal outputs also change only slightly, with
$\Delta_{\rm RMS}^{\OO}$ of order $10^{-3}$ in all four cases. The largest
elementwise terminal-output difference observed across all evaluated models is
\begin{equation*}
    \Delta_{\max}^{\OO}
    =
    \max_{n,o}
    \left|
        s_{n,o}^{(200)}
        -
        s_{n,o}^{(400)}
    \right|
    =
    0.0304694.
\end{equation*}

\begin{table*}[htbp!]
    \centering
    \caption{\textbf{Sensitivity to a twofold integration-grid refinement.}
    The same trained checkpoints are evaluated with $K=200$ and $K=400$
    explicit-Euler steps at fixed $t_{\rm f}=0.2$, without retraining or checkpoint
    reselection. Top-1 values and paired changes are reported as
    mean $\pm$ sample standard deviation over five training-order runs.
    Prediction disagreement is the fraction of test examples whose top-1
    prediction changes between the two resolutions, and
    $\Delta_{\rm RMS}^{\OO}$ measures the RMS change in the continuous
    terminal outputs.}
    \label{tab:appendix_H_numerical_refinement}
    \small
    \setlength{\tabcolsep}{4.0pt}
    \begin{adjustbox}{max width=\textwidth}
    \begin{tabular}{llrrrrr}
        \toprule
        Dataset
        & Method
        & Top-1 $K=200$ (\%)
        & Top-1 $K=400$ (\%)
        & $K400-K200$ (pp)
        & Prediction disagreement
        & $\Delta_{\rm RMS}^{\OO}$ \\
        \midrule

        MNIST
        & PATH
        & $89.116\pm0.011$
        & $89.020\pm0.045$
        & $-0.096\pm0.038$
        & $0.132\%$
        & $0.000853\pm0.0000003$ \\

        MNIST
        & TWO-STAGE-END
        & $96.722\pm0.072$
        & $96.724\pm0.073$
        & $+0.002\pm0.004$
        & $0.002\%$
        & $0.001057\pm0.000029$ \\

        \midrule

        Fashion-MNIST
        & PATH
        & $80.160\pm0.027$
        & $80.148\pm0.022$
        & $-0.012\pm0.013$
        & $0.050\%$
        & $0.000855\pm0.0000002$ \\

        Fashion-MNIST
        & TWO-STAGE-END
        & $86.456\pm0.036$
        & $86.450\pm0.043$
        & $-0.006\pm0.011$
        & $0.020\%$
        & $0.001165\pm0.000028$ \\

        \bottomrule
    \end{tabular}
    \end{adjustbox}
\end{table*}

The local integration behavior is also consistent with the grid refinement.
Across all dataset--method groups, the maximum observed single-step phase
increment decreases approximately by a factor of two when $\Delta t$ is
halved. The largest such increment occurs for MNIST TWO-STAGE-END and
changes from $0.204291$ at $K=200$ to $0.102146$ at $K=400$.
All evaluated model parameters, simulated phase states, and terminal outputs
remain finite at both resolutions.

For the deterministic rollouts considered in this numerical-refinement
analysis, the phase coordinates are integrated directly in their continuous
numerical representation: no clipping, modulo-$2\pi$ reduction, or circular
wrapping is applied. Their physical periodicity enters
through the periodic Kuramoto vector field and the output readout
$\cos\theta$. Thus, the refinement comparison tests the same underlying phase
dynamics at two temporal resolutions rather than introducing a
resolution-dependent phase-processing operation.

Taken together, the small task-level changes, rare prediction changes, small
continuous-output differences, and approximately halved single-step phase
increments show that the reported PATH--TWO-STAGE-END comparison is stable
under a twofold refinement of the integration grid. This analysis is a
discretization-sensitivity test over the two resolutions considered here. It
does not constitute a proof of numerical convergence, establish a formal
convergence order for explicit Euler, or demonstrate convergence of the
$K=400$ trajectories to the exact continuous-time solution.

\section{Cross-Platform Resource and Protocol Comparison}
\label{AppendixI}

\subsection{Comparison Scope, Counting Conventions, and Evidence Provenance}
\label{app:cross_platform_scope}

Tables~\ref{tab:cross_platform_resources} and
\ref{tab:cross_platform_training} compare the proposed method with selected
trainable physical dynamical systems based on coupled oscillators, quantum
annealing, and coherent-Ising-machine dynamics. The comparison is intended to
clarify differences in model-level resources, implementation status, training
procedure, and inference protocol. It is not a controlled ranking of
classification accuracy, hardware efficiency, latency, or energy consumption.

The cited studies use different datasets, input encodings, dynamical
architectures, training procedures, and implementation platforms. In
particular, full MNIST, Fashion-MNIST, MNIST/100, and the $8\times8$ Digits
dataset are distinct classification tasks. Reported accuracies therefore serve
only to identify the operating point associated with each architecture and
should not be interpreted as results under a common benchmark protocol.
Likewise, logical resource counts, numerical simulations, and physical-hardware
measurements represent different forms of evidence. The following conventions
are used to keep these distinctions explicit.

\paragraph{Logical dynamical variables.}

We count hidden and output dynamical variables separately and define
\begin{equation*}
    N_{\rm logical}
    =
    N_{\rm hidden}
    +
    N_{\rm output}.
    \label{eq:cross_platform_logical_nodes}
\end{equation*}
A logical dynamical variable is counted once at the model level, irrespective
of how many circuit, photonic, spin, or auxiliary components may be required
to realize it physically. Fixed or externally imposed input variables that do
not evolve as hidden or output states are excluded from $N_{\rm logical}$ and
reported separately as input resources.

Logical variables are not identified with implementation-level device counts.
A single logical oscillator or spin may require multiple physical components,
and practical implementations may additionally require circuitry for input
generation, coupling, coefficient storage, biasing, control, and readout.
Implementation-level physical counts are therefore reported only when they are
explicitly supported by the cited source; otherwise the corresponding entry is
reported as NR.

\paragraph{Input resources.}

We distinguish the input dimension $D$, the number of programmable input
coefficients, and any additional physical or dynamical input sources. These
quantities characterize different aspects of an input architecture and need
not coincide.

For the proposed model,
\begin{equation*}
    b^{\HH}(u)
    =
    W^{\HH}u+c^{\HH},
    \qquad
    W^{\HH}\in\mathbb R^{H\times D},
\end{equation*}
so the number of programmable input-to-hidden coefficients is
\begin{equation*}
    N_{\rm coeff}^{\rm input}
    =
    DH.
    \label{eq:cross_platform_input_weights}
\end{equation*}
At the principal operating point,
\[
    D=784,
    \qquad
    H=64,
\]
giving
\[
    N_{\rm coeff}^{\rm input}
    =
    50{,}176.
\]
This is a coefficient count, not a count of independent physical input
sources. Architectures that explicitly instantiate fixed or phase-coded input
oscillators are therefore reported separately from architectures in which
weighted inputs generate local fields.

\paragraph{Dynamical couplers.}

A dynamical coupler denotes an independently programmable interaction between
evolving dynamical variables. For symmetric hidden and output variables, the
possible pairwise blocks contain
\begin{equation*}
     N_{\rm coup}^{\HH\HH}=
    \frac{H(H-1)}{2},\; N_{\rm coup}^{\HH\OO}=
    HO,\; N_{\rm coup}^{\OO\OO}=
    \frac{O(O-1)}{2}.
\end{equation*}
Only interaction blocks present in the stated architecture are counted.
Input coefficients, local fields, fixed input encodings, and readout
operations are not included as oscillator-to-oscillator dynamical couplers.

The proposed Kuramoto topology contains symmetric hidden--hidden and
hidden--output interactions, with no self-coupling and no output--output
coupling. Hence
\begin{equation*}
    N_{\rm coup}
    =
    \frac{H(H-1)}{2}
    +
    HO.
    \label{eq:cross_platform_ours_couplers}
\end{equation*}
For $H=64$ and $O=10$,
\[
    N_{\rm coup}
    =
    2{,}656.
\]
For the comparison studies, the same model-level convention is applied to the
independently programmed interactions specified by each source. These logical
coupler counts are not interpreted as implementation-level wire, switch, or
embedded-hardware counts.

\paragraph{Training and inference stages.}

We distinguish an autonomous dynamical evolution from a digital evaluation of
a state or vector field. A dynamical evolution is a relaxation, anneal, or
finite-time rollout used to generate a physical or modeled dynamical state;
evaluating a vector field digitally on a prescribed trajectory is not counted
as such an evolution.

In the proposed method, Stage-I path training evaluates the Kuramoto vector
field along the prescribed teacher trajectory and therefore does not require
an autonomous rollout. Stage II differentiates through one autonomous
finite-time rollout, while the principal deterministic inference protocol uses
one autonomous rollout to the fixed observation time $t_{\rm f}$. The corresponding
training and inference conventions for the comparison methods are reported in
Table~\ref{tab:cross_platform_training}. Counts of dynamical evolutions are
used only to describe protocol structure and are not treated as common units of
physical cost across different technologies.

\paragraph{Evidence provenance.}

Superscripts in Tables~\ref{tab:cross_platform_resources} and
\ref{tab:cross_platform_training} indicate the provenance of individual
entries:
\begin{description}
    \item[\textbf{H}]
    a result or quantity obtained from a reported physical-hardware
    experiment;

    \item[\textbf{S}]
    a result obtained from numerical or dynamical simulation;

    \item[\textbf{D}]
    an architecture, parameter, resource value, or protocol stated directly
    in the cited source; and

    \item[\textbf{A}]
    an exact analytical count obtained from an explicitly specified
    architecture.
\end{description}
NR denotes a quantity that is not reported with sufficient information to be
recovered without additional assumptions.

The provenance codes do not represent a ranking of evidence quality. In
particular, an analytical logical-resource count does not imply a demonstrated
physical implementation, and a simulated classification result does not imply
hardware-measured performance. The resource and protocol comparisons below are
therefore interpreted within the task definitions, architectural assumptions,
and implementation status reported for each study.

\subsection{Resource and Implementation Comparison}
\label{app:cross_platform_resources_section}

Table~\ref{tab:cross_platform_resources} compares selected trainable physical
dynamical systems in terms of logical dynamical-state size, input resources,
programmable interactions, and implementation evidence. All entries follow
the counting conventions and provenance labels defined in
Appendix~\ref{app:cross_platform_scope}. Because the studies use different
tasks, input architectures, training procedures, and implementation
platforms, the reported accuracies identify representative operating points
rather than forming a controlled performance ranking.

\begin{table*}[htbp!]
\centering
\caption{\textbf{Resource and implementation comparison across selected
trainable physical dynamical systems.}
Logical dynamical variables, input parameterization, and dynamical couplings
are reported separately from implementation evidence. Reported task accuracies
refer to different benchmark settings and are not directly comparable.
Superscripts denote evidence provenance as defined in
Appendix~\ref{app:cross_platform_scope}; NR denotes insufficient information
for the common counting convention.}
\label{tab:cross_platform_resources}

\setlength{\tabcolsep}{2.8pt}
\renewcommand{\arraystretch}{1.18}
\scriptsize

\begin{adjustbox}{max width=\textwidth}
\begin{tabularx}{\textwidth}{
    >{\raggedright\arraybackslash}p{2.55cm}
    *{6}{>{\raggedright\arraybackslash}X}}
\toprule
Metric
& This work
& \cite{rageau2025training}
& \cite{gower2025train}
& \cite{wang2024training}
& \cite{laydevant2024training}
& \cite{fan2025equilibrium} \\
\midrule

Platform
& Kuramoto network
& Kuramoto network$^{\mathrm D}$
& Oscillator Ising machine$^{\mathrm D}$
& XY/Kuramoto network$^{\mathrm D}$
& D-Wave Ising machine$^{\mathrm D}$
& Coherent-Ising-machine$^{\mathrm D}$ \\

Task / dataset
& MNIST / Fashion-MNIST
& MNIST$^{\mathrm D}$
& MNIST / Fashion-MNIST$^{\mathrm D}$
& $8\times8$ Digits$^{\mathrm D}$
& MNIST/100$^{\mathrm D}$
& MNIST$^{\mathrm D}$ \\

Reported task accuracy
& {$96.711\pm0.065\%^{\mathrm S}$ /
   $86.399\pm0.184\%^{\mathrm S}$}
& $97.77\pm0.08\%^{\mathrm S}$
& {$97.2\pm0.1\%^{\mathrm S}$ /
   $88.0\pm0.1\%^{\mathrm S}$}
& $94.1\%^{\mathrm S}$
& $88.8\pm1.5\%^{\mathrm H}$
& $88\%^{\mathrm S}$ / $87\%^{\mathrm S}$ \\

Hidden dynamical variables
& $64$
& $500^{\mathrm D}$
& $500^{\mathrm D}$
& $200^{\mathrm D}$
& $120^{\mathrm D}$
& $120^{\mathrm D}$ / $256^{\mathrm D}$ \\

Output dynamical variables
& $10$
& $10^{\mathrm D}$
& $10^{\mathrm D}$
& $10^{\mathrm D}$
& $40^{\mathrm D}$
& NR \\

Total logical dynamical variables
& $74$
& $510^{\mathrm A}$
& $510^{\mathrm A}$
& $210^{\mathrm A}$
& $160^{\mathrm A}$
& NR \\

Programmable input-to-hidden coefficients
& $50{,}176$
& $392{,}000^{\mathrm A}$
& $392{,}000^{\mathrm A}$
& $12{,}800^{\mathrm A}$
& $94{,}080^{\mathrm A}$
& NR \\

Input coupling / encoding
& {Input-dependent hidden local fields
   $b^{\mathrm H}(u)=W^{\mathrm H}u+c^{\mathrm H}$}
& {Trainable source--hidden couplings from
   phase-coded input sources$^{\mathrm D}$}
& {Input-weighted sums generate hidden
   bias fields$^{\mathrm D}$}
& {Trainable couplings from fixed-phase
   input units$^{\mathrm D}$}
& {Digital matrix--vector product generates
   hidden bias fields$^{\mathrm D}$}
& NR \\

Dynamical couplers
& $2{,}656$
& $5{,}000^{\mathrm A}$
& $5{,}000^{\mathrm A}$
& $2{,}000^{\mathrm A}$
& $4{,}800$ logical$^{\mathrm A}$
& NR \\

Implementation evidence
& {System-level numerical simulation;
   fabricated classifier footprint NR}
& {System-level numerical simulation$^{\mathrm S}$;
   classifier footprint NR}
& {System-level numerical simulation$^{\mathrm S}$;
   classifier footprint NR}
& {System-level numerical simulation$^{\mathrm S}$;
   classifier footprint NR}
& {Physical D-Wave experiment$^{\mathrm H}$;
   approximately six physical spins per logical neuron reported$^{\mathrm D}$;
   exact footprint NR}
& {Numerical CIM simulation$^{\mathrm S}$;
   classifier footprint NR} \\

\bottomrule
\end{tabularx}
\end{adjustbox}
\end{table*}

At the principal operating point, the proposed model contains $64$ hidden and
$10$ output oscillators, giving $N_{\rm logical}=74$. Among the
cross-platform operating points in
Table~\ref{tab:cross_platform_resources} for which complete state counts are
available, this is the smallest logical dynamical-state count. This refers
specifically to the evolving logical state and is not a hardware-footprint or
parameter-count claim.

\begin{figure*}[htbp!]
    \centering
    \includegraphics[width=0.58\textwidth]
    {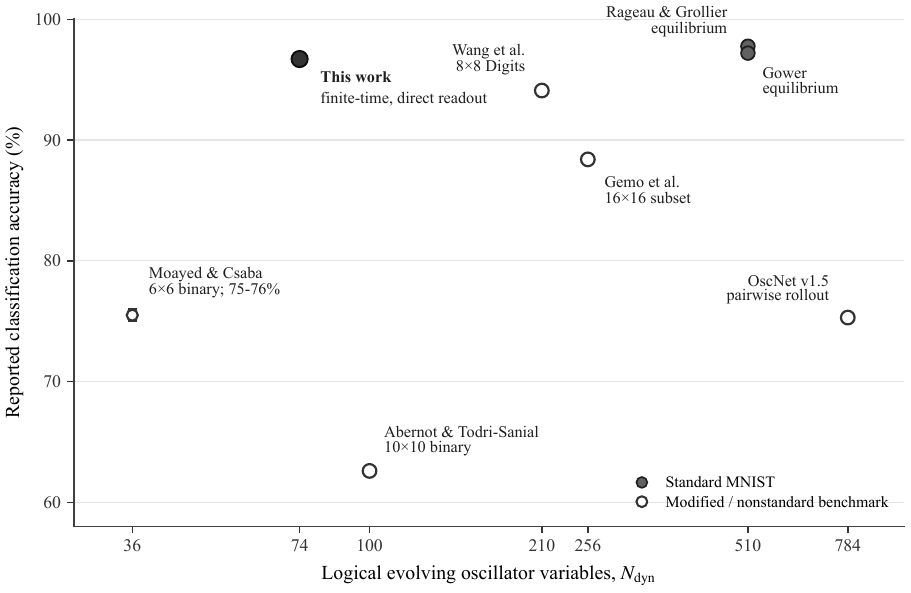}
    \caption{\textbf{Reported handwritten-digit classification accuracy
    versus logical oscillator-state scale for selected oscillator networks.}
    The horizontal axis reports the number of logical oscillator variables
    that evolve during inference under the counting convention of
    Appendix~\ref{app:cross_platform_scope}. Filled markers denote standard
    MNIST operating points, whereas open markers denote modified or
    nonstandard handwritten-digit benchmarks or inference protocols.
    The comparison contextualizes dynamical-state scale rather than defining
    a controlled accuracy or hardware-efficiency ranking. For OscNet v1.5,
    $N_{\rm dyn}=784$ denotes the phase variables evolved in each pairwise
    classification rollout.}
    \label{fig:oscillator_digit_scale}
\end{figure*}

Figure~\ref{fig:oscillator_digit_scale} provides complementary context within
the more closely related oscillator-network literature. In addition to the
standard-MNIST operating points of this work, Rageau and Grollier, and Gower
\citep{rageau2025training,gower2025train}, it includes smaller oscillator
systems evaluated under modified or nonstandard handwritten-digit protocols
\citep{moayed2026design,abernot2023training,wang2024training,
gemo2026leveraging,cai2025oscnet}.

The figure makes clear that small oscillator systems have previously been
studied on simplified or modified digit-classification tasks. Among the
standard-MNIST points shown, however, the present $74$-variable system reaches
$96.711\%$ accuracy, compared with $97.77\%$ and $97.2\%$ for the two
$510$-variable oscillator systems. This comparison is descriptive: the
methods differ in preprocessing, interaction structure, training procedure,
and inference protocol.

Resource counts must also be interpreted separately. The proposed model has
$50{,}176$ programmable input coefficients and $2{,}656$ logical dynamical
couplers, but these are not counts of independent physical input sources or
implementation-level interconnects. Likewise, the studies in
Table~\ref{tab:cross_platform_resources} differ substantially in input
generation and implementation evidence. Most reported classifier results are
numerical, whereas \citet{laydevant2024training} use physical D-Wave hardware;
for \citet{fan2025equilibrium}, incomplete classifier-level state information
prevents recovery of a common $N_{\rm logical}$.

Overall, the comparison places the proposed classifier at a compact logical
operating point while emphasizing that state dimension is only one component
of physical-computing complexity. A controlled hardware comparison would
additionally require common assumptions about implementation, precision,
latency, and energy.

\subsection{Training and Inference Protocol Comparison}
\label{app:cross_platform_training_section}

Table~\ref{tab:cross_platform_training} compares how the selected methods use
their dynamical systems during training and inference. The comparison focuses
on the states used to construct parameter updates, the number and type of
dynamical evaluations required during training, the role of physical hardware,
and the state used for prediction. These protocol counts describe
computational structure only; they are not treated as common measures of
latency, energy, or hardware cost across different physical platforms.

\begin{table*}[htbp!]
\centering
\caption{\textbf{Comparison of training and inference protocols.}
A dynamical evolution denotes an autonomous relaxation, anneal, or finite-time
rollout of the modeled physical state. Digital evaluation of a vector field on
a prescribed trajectory is not counted as an autonomous dynamical evolution.
Superscripts denote evidence provenance as defined in
Appendix~\ref{app:cross_platform_scope}. NR denotes information not reported
with sufficient detail for the common comparison.}
\label{tab:cross_platform_training}

\setlength{\tabcolsep}{2.8pt}
\renewcommand{\arraystretch}{1.18}
\scriptsize

\begin{adjustbox}{max width=\textwidth}
\begin{tabularx}{\textwidth}{
    >{\raggedright\arraybackslash}p{2.35cm}
    *{6}{>{\raggedright\arraybackslash}X}}
\toprule
Metric
& \centering This work\arraybackslash
& \centering \cite{rageau2025training}\arraybackslash
& \centering \cite{gower2025train}\arraybackslash
& \centering \cite{wang2024training}\arraybackslash
& \centering \cite{laydevant2024training}\arraybackslash
& \centering \cite{fan2025equilibrium}\arraybackslash \\
\midrule

Training rule
& {Teacher-path pretraining followed by autonomous endpoint
   fine-tuning}
& {Three-phase equilibrium propagation$^{\mathrm D}$}
& {Centered / symmetric equilibrium propagation$^{\mathrm D}$}
& {Two-phase equilibrium propagation$^{\mathrm D}$}
& {Equilibrium propagation with quantum annealing$^{\mathrm D}$}
& {Equilibrium propagation with numerical CIM state search$^{\mathrm D}$}
\\

State generation for training
& {Stage I: vector-field evaluation on a prescribed teacher trajectory;
   Stage II: autonomous finite-time rollout}
& {Free equilibrium and positive/negative nudged equilibria$^{\mathrm D}$}
& {Free equilibrium and positive/negative nudged equilibria$^{\mathrm D}$}
& {Free and nudged equilibria from randomized phase
   initializations$^{\mathrm D}$}
& {Forward-annealed free state and reverse-annealed nudged
   state$^{\mathrm D}$}
& {Free and nudged stable states from numerical CIM search$^{\mathrm D}$}
\\

Dynamical evaluations / sampling per update
& {Stage I: no autonomous rollout;
   Stage II: one autonomous rollout;
   no stochastic sampling in principal training}
& {Three simulated relaxations$^{\mathrm D}$}
& {Three simulated relaxations$^{\mathrm D}$}
& {Two relaxations per sampled initial configuration;
   gradients averaged over initial configurations$^{\mathrm D}$}
& {Two annealing stages; ten hardware samples per phase,
   with the lowest-energy sample retained$^{\mathrm D}$}
& {Two numerical CIM state searches;
   repeated-search count NR$^{\mathrm D}$}
\\

Hardware in reported training
& {Numerical simulation}
& {Numerical simulation$^{\mathrm S}$}
& {Numerical simulation$^{\mathrm S}$}
& {Numerical simulation$^{\mathrm S}$}
& {D-Wave hardware$^{\mathrm H}$}
& {Numerical CIM simulation$^{\mathrm S}$}
\\

Parameter-update computation
& {Digital parameter optimization}
& {Numerical Adam update$^{\mathrm D}$}
& {Numerical parameter update$^{\mathrm D}$}
& {Numerical parameter update$^{\mathrm D}$}
& {Classical SGD using hardware-generated states$^{\mathrm D}$}
& {Numerical EP weight update; optimizer details NR$^{\mathrm D}$}
\\

Inference state
& {Autonomous state at fixed observation time}
& {Free relaxed oscillator state$^{\mathrm D}$}
& {Free relaxed oscillator state$^{\mathrm D}$}
& {Free equilibrium$^{\mathrm D}$}
& {Annealed spin state$^{\mathrm D}$}
& {Stable CIM state$^{\mathrm D}$}
\\

Principal inference evolution
& {One autonomous finite-time rollout;
   simulated in the present study}
& {One free relaxation$^{\mathrm D}$}
& {One free relaxation$^{\mathrm D}$}
& {One free relaxation$^{\mathrm D}$}
& {Single post-training D-Wave call reported$^{\mathrm D}$}
& {NR}
\\

\bottomrule
\end{tabularx}
\end{adjustbox}
\end{table*}

\subsection{Broader Context Among Dynamical Classifiers}
\label{app:broader_dynamical_context}

To place the dynamical-state scale of our principal operating point in broader
context, we include representative energy-based, equilibrium, neural-ODE,
reservoir, analog-computing, and spiking classifiers evaluated on MNIST and/or
Fashion-MNIST.

Because these systems implement computation differently, the comparison is
descriptive rather than a controlled efficiency ranking. For continuous-state
models, $N_{\rm dyn}$ counts logical scalar variables that evolve during
inference, excluding clamped inputs and purely digital input/output
projections. For the representative spiking neural network, we instead report
the number of non-input spiking neurons. Sequential-input recurrent models,
such as sequential-MNIST LSTMs, are excluded because pixels are presented over
successive recurrent steps rather than defining a single static-input
dynamical inference process.

Table~\ref{tab:broader_dynamical_context} therefore compares state scale
together with dynamics, input protocol, and readout. Reported accuracies are
not directly comparable when dataset splits or inference protocols differ.

\begin{table*}[htbp!]
\centering
\scriptsize
\renewcommand{\arraystretch}{1.22}
\setlength{\tabcolsep}{3.0pt}

\caption{\textbf{Broader context among dynamical classifiers on MNIST and
Fashion-MNIST.}
Representative operating points are shown for context rather than as a
controlled efficiency ranking. For continuous-state systems, $N_{\rm dyn}$
counts logical scalar variables evolving during inference, excluding clamped
inputs and purely digital projections; for the SNN, we report the number of
non-input spiking neurons. NR denotes not reported.
}

\label{tab:broader_dynamical_context}

\resizebox{\textwidth}{!}{%
\begin{tabular}{lcccccccc}
\toprule
Metric &
\textbf{This work} &
\shortstack{\textbf{Whitelam}\\\citep{whitelam2026training}} &
\shortstack{\textbf{Equilibrium}\\\textbf{Propagation}\\
\citep{scellier2016equilibrium}} &
\shortstack{\textbf{SubDEQ}\\
\citep{sittoni2024subhomogeneous}} &
\shortstack{\textbf{AOC}\\
\citep{kalinin2025analog}} &
\shortstack{\textbf{nmODE}\\
\citep{yi2023nmode}} &
\shortstack{\textbf{MRESN}\\
\citep{lopezortiz2024exploring}} &
\shortstack{\textbf{SNN}\\
\citep{bacho2023exploring}}
\\
\midrule

Dynamical substrate &
\shortstack{Kuramoto\\phase network} &
\shortstack{Thermodynamic\\Langevin network} &
\shortstack{Continuous\\energy-based network} &
\shortstack{Deep-equilibrium\\network} &
\shortstack{Analog optical\\equilibrium model} &
\shortstack{Neural-memory\\ODE} &
\shortstack{Echo-state\\reservoirs} &
\shortstack{Spiking\\neural network}
\\

Representative architecture &
\shortstack{$784\rightarrow64+10$\\oscillators} &
\shortstack{$784\rightarrow64+10$\\physical variables} &
$784$--$500$--$10$ &
\shortstack{$784\rightarrow87$\\latent state} &
\shortstack{$784\rightarrow16\rightarrow10$\\
digital IP / AOC / OP} &
\shortstack{$784\rightarrow2048\rightarrow10$} &
\shortstack{$4\times500$\\reservoir nodes} &
\shortstack{MNIST: $800$--$10$\\
Fashion: $400$--$400$--$10$}
\\

Dynamical-state scale &
\textbf{$N_{\rm dyn}=74$} &
$N_{\rm dyn}=74$ &
$N_{\rm dyn}=510$ &
$N_{\rm dyn}=87$ &
\textbf{$N_{\rm dyn}=16$} &
$N_{\rm dyn}=2048$ &
$N_{\rm dyn}=2000$ &
\shortstack{810 non-input\\spiking neurons$^{\S}$}
\\

Input protocol &
\shortstack{Static image;\\hidden local fields} &
\shortstack{Static image;\\trainable input couplings} &
\shortstack{Static clamped\\input units} &
\shortstack{Static image;\\learned input map} &
\shortstack{Static image;\\learned digital IP} &
\shortstack{Static image;\\learned input map} &
\shortstack{Static image;\\reservoir input weights} &
\shortstack{Static image encoded\\as input spike times}
\\

Inference regime &
\shortstack{Fixed finite-time\\transient} &
\shortstack{Finite-time stochastic\\trajectory} &
\shortstack{Free-phase\\equilibrium} &
\shortstack{Implicit\\fixed point} &
\shortstack{Analog\\fixed point} &
\shortstack{Finite-time ODE\\integration} &
\shortstack{Reservoir\\transient} &
\shortstack{Finite-time\\spike dynamics}
\\

Energy / gradient structure &
\shortstack{Strict deterministic\\gradient flow} &
\shortstack{Potential-gradient\\Langevin drift} &
\shortstack{Energy-based\\gradient dynamics} &
\shortstack{Generic constrained\\fixed-point map} &
\shortstack{Generic nonlinear\\fixed-point map} &
\shortstack{Decoupled globally\\attracting ODEs} &
\shortstack{Generic recurrent\\reservoir dynamics} &
\shortstack{Hybrid threshold/reset\\dynamics}
\\

Internal dynamical
interaction structure &
\shortstack{Symmetric pairwise\\oscillator couplings} &
\shortstack{Potential-derived\\interactions} &
\shortstack{Symmetric\\recurrent weights} &
\shortstack{Learned latent\\interaction map} &
\shortstack{Learned $16\times16$ map;\\no symmetry constraint} &
\shortstack{No memory--memory\\coupling} &
\shortstack{Fixed random\\recurrent weights} &
\shortstack{Feedforward\\synaptic connections}
\\

Trainable post-dynamical
readout &
None &
None &
None &
\shortstack{Yes; learned\\classifier} &
\shortstack{Yes; digital\\output projection} &
\shortstack{Yes; learned\\decision layer} &
\shortstack{Yes; trained\\output layer} &
\shortstack{None; spiking\\output layer}
\\

Prediction state &
\shortstack{Output oscillators\\directly at $t_{\rm f}$} &
\shortstack{Output variables\\directly at $t_{\rm f}$} &
\shortstack{Output units at\\free fixed point} &
\shortstack{Classifier on\\equilibrium state} &
\shortstack{Digital projection of\\analog fixed point} &
\shortstack{Decision layer on\\terminal ODE state} &
\shortstack{Readout of\\reservoir state} &
\shortstack{Output-neuron\\spike response}
\\

MNIST top-1 &
\textbf{$96.711\pm0.065\%$} &
$92.0\%^{\dagger}$ &
$\sim97\%$ &
$98.080\pm0.102\%^{\ddagger}$ &
$\sim95\%^{\P}$ &
$99.52$--$99.59\%^{\|}$ &
$96.64\pm0.001\%$ &
$98.88\pm0.02\%$
\\

Fashion-MNIST top-1 &
\textbf{$86.399\pm0.184\%$} &
$\approx77.3\%^{\#}$ &
NR &
NR &
$\sim86\%^{\P}$ &
NR &
$86.77\pm0.13\%$ &
$90.19\pm0.12\%$
\\

Implementation evidence &
\shortstack{System-level\\numerical simulation} &
\shortstack{System-level\\numerical simulation} &
\shortstack{Numerical\\simulation} &
\shortstack{Numerical\\simulation} &
\shortstack{Physical analog\\hardware} &
\shortstack{Numerical\\simulation} &
\shortstack{Numerical\\implementation} &
\shortstack{Numerical\\simulation}
\\

\bottomrule
\end{tabular}%
}

\vspace{1mm}
\begin{minipage}{0.98\textwidth}
\footnotesize

$^{\dagger}$Whitelam reports $92.0\%$ MNIST top-1 accuracy after averaging
the continuous outputs of ten stochastic trajectories; the corresponding
single-trajectory accuracy is $91.7\%$.

$^{\ddagger}$The dense Tanh SubDEQ reports a test error of
$1.92\pm0.102\%$, corresponding to $98.080\pm0.102\%$ accuracy.
The study uses a $71\%/14.5\%/14.5\%$ train/validation/test hold-out split
rather than the standard official MNIST train/test split.

$^{\P}$AOC values denote the reported 256-weight, 16-state physical hardware
operating point and are read approximately from the reported classification
results. The 16 evolving variables refer only to the analog equilibrium core;
the learned digital input and output projections are not included in
$N_{\rm dyn}$.

$^{\|}$Five independently trained 2048-memory-neuron nmODE models achieve
$99.52$--$99.59\%$ MNIST accuracy individually. The reported $99.71\%$
five-model voting result is not used here.

$^{\#}$Whitelam reports a Fashion-MNIST neural-teacher accuracy of $84.9\%$
and a thermodynamic-computer accuracy equal to $91\%$ of the teacher value,
giving approximately $77.3\%$.

$^{\S}$For the representative fully connected SNN, both the MNIST
$800$--$10$ and Fashion-MNIST $400$--$400$--$10$ architectures contain
810 non-input spiking neurons. This neuron count is not treated as directly
equivalent to 810 scalar continuous state variables because each neuron has
hybrid continuous and event-driven dynamics.

\end{minipage}
\end{table*}

Two observations are useful for interpreting
Table~\ref{tab:broader_dynamical_context}. First, a small dynamical state
space is not unique to the present model. For example, SubDEQ operates with
an 87-dimensional latent equilibrium state, and the AOC demonstrates a
16-dimensional physical fixed-point core. These systems, however, employ
different dynamical classes and learned input/output interfaces. Conversely,
representative neural-ODE, reservoir, and spiking classifiers operate at
substantially larger dynamical scales while achieving higher or comparable
classification accuracy.

The distinguishing feature of the present operating point is therefore not
state dimension alone. Rather, the 74-variable classifier combines a compact
dynamical state with strict gradient-flow dynamics, symmetric pairwise
oscillator interactions, restricted connectivity, no trainable
post-dynamical classifier, and classification directly from the oscillator
state at a prescribed finite time. The heterogeneous comparison above is
intended to make these simultaneous constraints explicit rather than to
define a universal measure of dynamical or hardware efficiency.

\end{document}